\documentclass[a4paper,11pt]{article}
\pdfoutput=1
\usepackage{jheppub}
\usepackage[T1]{fontenc}
\usepackage{graphicx}
\usepackage{amsmath}
\usepackage{amsmath}
\usepackage{xspace}
\usepackage{subfig}
\usepackage{xcolor}
\usepackage{mathtools}
\usepackage{braket}
\usepackage{multirow}

\newcommand{\search}[1]{\textbf{#1}}

\title{Neutrinoless double beta decay versus \\ other probes of heavy sterile neutrinos}
\author[a, b]{Patrick D. Bolton,}
\author[a, c]{Frank F. Deppisch}
\author[d]{and P. S. Bhupal Dev}

\affiliation[a]{Department of Physics and Astronomy, University College London, \\Gower Street, London WC1E 6BT, UK}
\affiliation[b]{Wright Laboratory, Department of Physics, Yale University,\\New Haven, Connecticut 06520-8120, USA}
\affiliation[c]{Institut f{\"u}r Hochenergiephysik, {\"O}sterreichische Akademie der Wissenschaften,\\Nikolsdorfer Gasse 18, 1050 Wien, Austria}
\affiliation[d]{Department of Physics and McDonnell Center for the Space Sciences, Washington University,\\St. Louis, MO 63130, USA}

\emailAdd{patrick.bolton.17@ucl.ac.uk}
\emailAdd{f.deppisch@ucl.ac.uk}
\emailAdd{bdev@wustl.edu}

\abstract{We make a comparative study of the neutrinoless double beta decay constraints on heavy sterile neutrinos versus other direct and indirect constraints from both lepton number conserving and violating processes, as a sensitive probe of the extent of lepton number violation and possible interference effects in the sterile sector. We introduce a phenomenological parametrisation of the simplified one-generation seesaw model with one active and two sterile neutrino states in terms of experimentally measurable quantities, such as active-sterile neutrino mixing angles, CP phases, masses and mass splittings. This simple parametrisation enables us to analytically derive a spectrum of possible scenarios between the canonical seesaw with purely Majorana heavy neutrinos and inverse seesaw with pseudo-Dirac ones. We then go on to constrain the simplified parameters of this model from various experiments at the energy, intensity and cosmic frontiers. We emphasise that the constraints from lepton number violating processes strongly depend on the mass splitting between the two sterile states and the relative CP phase between them. This is particularly relevant for neutrinoless double beta decay, which is weakened for small mass splitting and opposite CP parities between the sterile states. On the other hand, neutrinoless double beta decay is especially sensitive for Majorana sterile neutrinos with masses around $0.1-10$ GeV.}
	

\begin{document}	
\maketitle
\flushbottom

%%%%%%%%%%%%%%%%%%%%%%%%%%%%%%%%%%%%%%%%%%%%%%%%%%%%%%%%%%%%%%%%%%%%%%%%%%%%%%%%%
\section{Introduction}
%%%%%%%%%%%%%%%%%%%%%%%%%%%%%%%%%%%%%%%%%%%%%%%%%%%%%%%%%%%%%%%%%%%%%%%%%%%%%%%%%

The observation of neutrino oscillations in solar, atmospheric, reactor and accelerator neutrino data~\cite{Tanabashi:2018oca} implies that at least two of the three active neutrinos have a small but non-zero mass and that individual lepton flavour is violated. In the Standard Model (SM), neutrinos have only one helicity state $\nu_L$, and therefore, cannot acquire a Dirac mass through the Higgs mechanism, unlike the charged fermions. A Majorana mass term of the form $\bar{\nu}_{L}^C\nu_{L}$ (where $\nu_L^C\equiv \nu_L^TC^{-1}$, $C$ being the charge conjugation matrix) is also forbidden in the SM due to its gauge structure and particle content. Specifically, the SM does not contain an $SU(2)_L$ triplet Higgs which could give rise to the $\bar{\nu}_{L}^C\nu_{L}$ term. By adding a SM-singlet, right-handed (RH) neutrino field $\nu_R$ per generation to the SM, one could in principle generate a Dirac mass term; however, to get sub-eV left-handed (LH) neutrino masses as required by the neutrino oscillation data, one needs the Dirac Yukawa couplings to be $\lesssim 10^{-12}$. While theoretically allowed,  such a scenario  would be rather uninteresting from an experimental point of view. A more appealing choice is to break the accidental $(B-L)$~symmetry of the SM to generate a Majorana neutrino mass at tree or loop-level. 

The simplest tree-level realisation of the $(B-L)$-breaking is through the effective dimension-5 Weinberg operator $(L^TH)(L^TH)/\Lambda$, where $L$ and $H$ are the $SU(2)_L$ lepton and Higgs doublets of the SM and $\Lambda$ is the scale of new physics that induces $(B-L)$-breaking~\cite{Weinberg:1979sa}. Here the intermediate heavy particles integrated out in the low-energy theory are SM-singlet fermions, identified as the RH Majorana neutrinos $\nu_{R,i}$ (with $i=1,2,\cdots$) with mass $m_{N_i}$. This is widely known as the {\em type-I seesaw} mechanism~\cite{Minkowski:1977sc, Mohapatra:1979ia, Gellmann:1980vs, Yanagida:1979as, Schechter:1980gr}. In the minimal type-I seesaw extension of the SM, the RH Majorana neutrinos, also known as the sterile neutrino states (or heavy neutral leptons in some literature), being SM gauge-singlets, can only interact with the SM particles through their mixing with the active neutrino states.  In the traditional `vanilla' seesaw mechanism, this active-sterile neutrino mixing is given by  
\begin{align}
\label{eq:vanilla}
	V_{\ell N} \ \simeq \ \sqrt{\frac{m_\nu}{m_N}} 
	\ \lesssim \ 10^{-6} \sqrt{\frac{100~\text{GeV}}{m_N}}\,,
\end{align}
due to the smallness of the light neutrino mass $m_\nu \lesssim 0.1~{\rm eV}$, as inferred from neutrino oscillation data~\cite{Tanabashi:2018oca}, as well as the cosmological limit on the sum of active neutrino masses~\cite{Aghanim:2018eyx}. Thus for a low seesaw scale in the sub-TeV range, the experimental effects of the sterile neutrino are expected to be small, unless they have additional interactions, e.g. when they are charged under an extra gauge $U(1)_{B-L}$. However, there exists a class of minimal SM plus low-scale type-I seesaw scenarios~\cite{Pilaftsis:1991ug, Buchmuller:1991ce, Gluza:2002vs, Pilaftsis:2004xx, Kersten:2007vk, Xing:2009in, Gavela:2009cd, He:2009ua, Adhikari:2010yt, Ibarra:2010xw, Deppisch:2010fr, Ibarra:2011xn, Mitra:2011qr}, where $V_{\ell N}$ can be sizeable while still satisfying the light neutrino data. This is made possible by assigning specific textures to the Dirac and Majorana mass matrices. The stability of these textures can in principle be guaranteed by enforcing extra symmetries in the lepton sector~\cite{Pilaftsis:2004xx, Shaposhnikov:2006nn, Kersten:2007vk, Deppisch:2010fr, Dev:2013oxa, Chattopadhyay:2017zvs}. We will generically assume this to be the case for our subsequent discussion and freely vary the active-sterile mixing up to ${\cal O}(1)$, without referring to any particular texture or model-building aspects.  
	 
For the conventional seesaw scenarios mentioned above, the active neutrino masses are inversely proportional to the lepton number violating (LNV)  Majorana mass scale $m_N$ (hence the name `seesaw'). There exist important variations, such as {\em inverse seesaw}~\cite{mohapatra:1986aw, Nandi:1985uh, mohapatra:1986bd}, {\it linear seesaw}~\cite{wyler:1983dd, akhmedov:1995ip, akhmedov:1995vm, malinsky:2005bi} and generalised inverse seesaw~\cite{Ma:2009du, Dev:2012sg, Dev:2012bd}, where the active neutrino masses are {\em directly} proportional to the lepton-number breaking scale. In such scenarios, large active-sterile neutrino mixing can be achieved rather naturally, irrespective of the sterile neutrino mass spectrum. Experimentally, the main distinguishing feature of these variants from pure type-I seesaw is the {\em pseudo-Dirac} nature of the sterile neutrinos, which suppresses the LNV signals, in contrast with the purely {\em Majorana} nature of the sterile neutrinos in the type-I seesaw scenario.  However, this distinction may not be as clear in the presence of additional CP phases in the sterile sector, depending on the sterile neutrino mass spectrum~\cite{Bray:2007ru, Gluza:2016qqv, Das:2017hmg}, with important implications for collider physics and leptogenesis~\cite{Dev:2019rxh}. 

The aim of this paper is to show how both purely \textit{Majorana} and \textit{pseudo-Dirac} limits (and the spectrum of possible cases in between) can be understood from a simple phenomenological parametrisation in terms of the {\it a priori} measurable mixing angles, CP phases and mass eigenvalues involving the sterile neutrinos. In order to show our results analytically as much as possible, we will work in a simplified single-generational picture (involving only the electron flavour) and introduce the generalised inverse seesaw with two SM-singlet Weyl fermions $\nu_{R,1}$ and $\nu_{R,2}$. In this case, the unitary matrix $V$ that diagonalises the full neutrino mass matrix $\mathcal{M}_{\nu}$ is a $3\times 3$ matrix containing three mixing angles ($\vartheta_{e1}, \vartheta_{e2}, \vartheta_{12}$, with the first two being the mixing angles of the active neutrino with the two sterile states, while the last one being the mixing angle in the sterile sector) and three CP phases ($\delta, \phi_1, \phi_2$, with $\delta$ being the Dirac CP phase and $\phi_1, \phi_2$ being the Majorana phases). We will apply this general parametrisation to identify the regions of parameter space allowed by consistency relations among the neutrino mass matrix elements. We will look at how the \textit{Majorana} and \textit{pseudo-Dirac} limits of the sterile state pair depend on the phases $\phi_1$ and $\phi'_2=\phi_2-2\delta$ and how these in turn are completely determined by the active-sterile squared mixing strengths $s^2_{e1}\equiv \sin^2\vartheta_{e1}$ and $s^2_{e2}\equiv \sin^2\vartheta_{e2}$ as a result of the $(1, 1)$ element of $\mathcal{M}_{\nu}$ being zero. We will also use the $(1, 3)$ element, which can be set to zero by a particular rotation and therefore parametrisation of $\mathcal{M}_{\nu}$, to constrain the sterile-sterile squared mixing strength $s_{12}^2\equiv\sin^2\vartheta_{12}$ and linear combination of phases $\delta'=2\phi_2+\delta$. The angle $\vartheta_{12}$ is an unobservable parameter in the SM because it is not contained in the mixing strengths $V_{eN_i}$. Nonetheless, the solution for $s_{12}^2$ in the chosen parametrisation gives a parametrisation-independent value for the light active neutrino mass at one-loop, which we enforce to be considerably smaller than the tree-level mass.

In order to put the neutrinoless double beta ($0\nu\beta\beta$) decay sensitivity  into context, we review the  experimental constraints on the sterile neutrino sector from both lepton number conserving (LNC) and violating channels from high-energy collider searches, high-intensity beam dump and meson decay experiments, beta decays and other nuclear processes, active-sterile neutrino oscillation experiments, electroweak precision data and other indirect laboratory searches, as well as cosmological and astrophysical observations. We pay particular attention to possible interference effects from two sterile states on the LNV constraints as a function of their mass splitting. We give special emphasis on the $0\nu\beta\beta$ decay constraint, which has been argued to be the most stringent one for active-sterile neutrino mixing in the electron flavour; see e.g. summary plots in Refs.~\cite{Atre:2009rg, Deppisch:2015qwa, deGouvea:2015euy, Chrzaszcz:2019inj}.  We re-evaluate the $0\nu\beta\beta$ decay constraints using the general parametrisation discussed above and show how these are affected by the sterile neutrino mass splitting and CP phases, in comparison to other laboratory constraints. In fact, under certain conditions, we find the $0\nu\beta\beta$ decay constraints to be {\it weaker} than the direct search limits from colliders, thus reinforcing the importance of independent direct searches for sterile neutrinos in all flavours. On the other hand, for the theoretically interesting mass regime $m_{N}\approx 0.1 - 10$ GeV, $0\nu\beta\beta$ decay is comparable to current and future direct searches. As we work in a simplified one-generation framework containing a single active neutrino state which we identify as the electron neutrino, we cannot model the coherent contribution of the other two light states to $0\nu\beta\beta$ decay. Our main focus is on the constraints on sterile neutrinos in a simplified yet consistent seesaw picture but we will comment on the omission of the other light states (see the discussion at the end of Sec.~\ref{sec:0vbbcontribs}).

The rest of this paper is organised as follows. In Sec.~\ref{sec:model} we introduce the generalised inverse seesaw for the neutrino mass matrix $\mathcal{M}_{\nu}$, which reduces to the type-I, inverse and linear seesaw scenarios under different limits. We also investigate the masses at tree-level and at one-loop for the light mostly-active neutrinos as a function of the model parameters. In Sec.~\ref{sec:parametrization} we introduce a phenomenological parametrisation of the unitary matrix $V$ that diagonalises $\mathcal{M}_{\nu}$ in terms of three mixing angles ($\vartheta_{e1}, \vartheta_{e2}, \vartheta_{12}$) and three CP phases ($\delta, \phi_1, \phi_2$), and identify the regions of parameter space allowed by the consistency relations implied by $(\mathcal{M}_{\nu})_{ 11}=0$. In  Sec.~\ref{sec:searches} we review the current upper limits and future sensitivities on the active-sterile mixing strength $|V_{eN}|^2$ as a function of the sterile mass $m_{N}$ from 1 eV to 10 TeV. In Sec.~\ref{sec:0vbb}, we re-evaluate the $0\nu\beta\beta$ decay constraints in the generalised inverse seesaw, particularly for different values of the splitting $\Delta m_{N}$, and make a comparison with other constraints discussed earlier. We conclude in Sec.~\ref{sec:conclusions}. For completeness, the summary plots for constraints on $|V_{\mu N}|^2$ and $|V_{\tau N}|^2$ are given in Appendix~\ref{sec:app}.

%%%%%%%%%%%%%%%%%%%%%%%%%%%%%%%%%%%%%%%%%%%%%%%%%%%%%%%%%%%%%%%%%%%%%%%%%%%%%%%%%
\section{Generalised seesaw and neutrino mass spectrum}
\label{sec:model}
%%%%%%%%%%%%%%%%%%%%%%%%%%%%%%%%%%%%%%%%%%%%%%%%%%%%%%%%%%%%%%%%%%%%%%%%%%%%%%%%%
\subsection{Model setup}
\label{sec:model-setup}
%\frank{Define setup stringently.} 
We consider the addition of two SM-singlet Weyl fermions $\nu_{R,1}$ and $\nu_{R,2}$ to the SM particle content. We restrict ourselves to the first generation of SM fermions, which is the most relevant for $0\nu\beta\beta$ decay, and also allows us to present the gist of our results analytically. The SM Lagrangian is then extended to
\begin{align}
\label{eq:lagrangian}
\mathcal{L} \ = \ \mathcal{L}_{SM}
- y_{e i}\bar L_e \widetilde{H} \nu_{R,i} 
- \frac{1}{2} (\mathcal{M}_S)_{ij} \bar{\nu}^C_{R,i} \nu^{\phantom{c}}_{R,j} 
+ \text{H.c.}\,.
\end{align}
Here, $L_e  = (\nu_{L,e}, e_L)^T$ is the first-generation SM lepton doublet, $\widetilde{H} = i\sigma_2 H^*$ with $H=(H^0, H^-)^T$ being the SM Higgs doublet and $\sigma_2$ being the second Pauli matrix, $\mathcal{M}_S$ is the Majorana mass term for the sterile states, and a summation over the sterile states is assumed (with $i=1,2$). After electroweak symmetry breaking by the vacuum expectation value $\langle H^0\rangle =v\simeq 174$ GeV, we obtain the neutrino Dirac mass terms $(\mathcal{M}_D)_i = y_{e i} v$, and the Lagrangian~\eqref{eq:lagrangian} gives rise to the following neutrino mass matrix in the basis $\left(\nu_{L,e}^C, \nu_{R,1}, \nu_{R,2}\right)$: 
\begin{align}
\label{eq:massmatrix}
\mathcal{M}_\nu \ = \  
\begin{pmatrix}
0       & (\mathcal{M}_D)_1    & (\mathcal{M}_D)_2      \\
(\mathcal{M}_D)_1 & (\mathcal{M}_S)_{11} & (\mathcal{M}_S)_{12} \\
(\mathcal{M}_D)_2 & (\mathcal{M}_S)_{12} & (\mathcal{M}_S)_{22} 
\end{pmatrix}
\ \equiv \  
\begin{pmatrix}
0     & \mathcal{M}_D \\ 
\mathcal{M}_D^T & \mathcal{M}_S 
\end{pmatrix} .
\end{align}
The above mass matrix can be diagonalised by a $3\times 3$ unitary matrix $V$ such that 
\begin{align}
\label{eq:diagonalization}
V^\dagger \cdot \mathcal{M}_\nu \cdot V^* 
\ = \ \text{diag}\left(m_\nu, m_{N_1}, m_{N_2}\right)\,,
\end{align}
giving rise to three mass eigenvalues $m_\nu$, $m_{N_1}$, $m_{N_2}$ which can be chosen to be real and non-negative. We have denoted the mass eigenvalues suggestively for the case we will focus on, with one light, dominantly active, state ($m_\nu \lesssim 1$~eV) and two much heavier, dominantly sterile, states ($m_{N_1}, \,m_{N_2} \gg m_\nu$). Accordingly, we conventionally order the mass eigenstates $\nu^{\prime C}_{L,e}$, $\nu^\prime_{R,1}$, $\nu^\prime_{R,2}$ defined by
\begin{align}
\begin{pmatrix}
\nu^{\prime C}_{L,e} \\ \nu^\prime_{R,1} \\ \nu^\prime_{R,2}
\end{pmatrix} 
\ = \ V \cdot \begin{pmatrix}\nu_{L,e}^C \\ \nu_{R,1} \\ \nu_{R,2} \end{pmatrix},
\end{align}
by increasing mass, $m_\nu \ll m_{N_1} \leq m_{N_2}$. The corresponding Majorana states are then defined as 
\begin{gather}
\nu_e = \nu^\prime_{L,e} + \nu^{\prime C}_{L,e}\, ,\quad 
N_1   = \nu^\prime_{R,1} + \nu^{\prime C}_{R,1}\, ,\quad
N_2   = \nu^\prime_{R,2} + \nu^{\prime C}_{R,2}\, .
\end{gather}

The above minimal first-generation extension of the SM incorporates simplified versions of various seesaw scenarios:

\paragraph{The type-I seesaw}~\cite{Minkowski:1977sc, Mohapatra:1979ia, Gellmann:1980vs, Yanagida:1979as, Schechter:1980gr} is realised for $||\mathcal{M}_D|| \ll ||\mathcal{M}_S||$ (where $||{\cal M}||\equiv \sqrt{{\rm Tr}({\cal M}^\dag {\cal M})}$ is the norm of matrix ${\cal M}$). In fact, only one sterile state is minimally required to give mass to the one active neutrino considered here, i.e.
\begin{align}
\mathcal{M}_\nu \ = \ 
\begin{pmatrix}
0   & m_D \\
m_D & m_N    
\end{pmatrix},
\end{align}
with $m_D \ll m_N$. The light neutrino mass in this case is given by 
\begin{align}
\label{eq:type-I}
m_\nu \ \approx \ -\frac{m_D^2}{m_N} \, .
\end{align}

\paragraph{The minimal inverse seesaw}~\cite{mohapatra:1986aw, Nandi:1985uh, mohapatra:1986bd} incorporates $(\mathcal{M}_D)_2=0$ and $(\mathcal{M}_S)_{11}=0$, so the neutrino mass matrix~\eqref{eq:massmatrix} becomes 
\begin{align}
\mathcal{M}_\nu \ = \ 
\begin{pmatrix}
0 & m_D & 0     \\
m_D & 0   & m_S   \\
0 & m_S & \mu_S 
\end{pmatrix},
\end{align}
with $\mu_S$, $m_D \ll m_S$. The light neutrino mass in this limit is given by 
\begin{align}
\label{eq: mnu_minimalinverse}
m_\nu \ \approx \ -\mu_S \frac{m_D^2}{m_S^2} \, .
\end{align}

\paragraph{The generalised inverse seesaw}~\cite{Dev:2012sg, Dev:2012bd}  incorporates $(\mathcal{M}_D)_2=0$, but $(\mathcal{M}_S)_{11} = \mu_R \neq 0$, so the neutrino mass matrix~\eqref{eq:massmatrix} becomes 
\begin{align}
\label{eq:generalized}
\mathcal{M}_\nu \ = \  
\begin{pmatrix}
0 & m_D   & 0     \\
m_D & \mu_R & m_S   \\
0 & m_S  & \mu_S 
\end{pmatrix},
\end{align}
with $\mu_S$, $m_D \ll m_S$. This does not affect the mass of the light neutrino given by Eq.~\eqref{eq: mnu_minimalinverse} at tree-level, but will generate a one-loop correction~\cite{Pilaftsis:1991ug, Dev:2012sg} as discussed in Sec.~\ref{sec:loopcorrections}.

\paragraph{The minimal linear seesaw}~\cite{wyler:1983dd, akhmedov:1995ip, akhmedov:1995vm, malinsky:2005bi} has $(\mathcal{M}_D)_2 =\mu_F \neq 0$, but $(\mathcal{M}_S)_{11} =(\mathcal{M}_S)_{22}=0$: 
\begin{align}
\label{eq:linear}
\mathcal{M}_\nu = 
\begin{pmatrix}
0     & m_D & \mu_F \\
m_D   & 0   & m_S   \\
\mu_F & m_S & 0 
\end{pmatrix},
\end{align}
with $\mu_F$, $m_D \ll m_S$. The light neutrino mass in this case is given by 
\begin{align}
\label{eq: mnu_minimallinear}
m_\nu \ \approx \ -\mu_F \frac{m_D^2}{m_S^2}\,.
\end{align}
Note that the mass matrix~\eqref{eq:linear} can always be rotated to the form given by Eq.~\eqref{eq:generalized} with appropriately defined $\mu_R$ and $\mu_S$~\cite{Ma:2009du}. We will take advantage of this fact later to simplify our analysis, without loss of generality. 

In the above scenarios we have not specified the source of LNV. Whether any one of the terms in Eq.~\eqref{eq:massmatrix} violates lepton number will depend on the $L$ assignment for the two sterile neutrinos $\nu_{R,1}$,  $\nu_{R,2}$. For example, making the choice $L(\nu_{R,1}) = L(\nu_{R,2}) = L(\nu_{L,e}) = +1$, suggested by treating the sterile neutrinos as RH counterparts to the LH active neutrinos, will mean that both terms in $\mathcal{M}_D$ conserve $L$ whereas all terms in $\mathcal{M}_S$ violate $L$ by two units. On the other hand, if $L(\nu_{R,1}) = L(\nu_{L,e}) = +1, \,L(\nu_{R,2}) = -1$, the LNV terms are $(\mathcal{M}_D)_2$ and $(\mathcal{M}_S)_{12} = (\mathcal{M}_S)_{21}$. While the choice of the origin of LNV is crucial to describe the underlying model, from a phenomenological point of view, the lepton number assignment does not need to be fixed. Also, any observable LNV effect crucially depends on the relative CP phase between the two sterile eigenstates, as we will see below.  
In any case, the smallness of the parameters $\mu_{R, S, F}$ in the three seesaw variants discussed above is technically natural in the 't Hooft sense~\cite{thooft:1979}, i.e. in the limit of $\mu_{R,S,F}\to 0$, lepton number symmetry is restored and the light neutrino $\nu_{L,e}$ is exactly massless to all orders in perturbation theory, as in the SM.

\subsection{Radiative corrections to the neutrino mass}
\label{sec:loopcorrections}
The light neutrino mass acquires a one-loop radiative correction from the self-energy diagrams involving the SM gauge and Higgs bosons~\cite{Pilaftsis:1991ug, Grimus:2002nk, Fernandez-Martinez:2015hxa}, induced by the Lagrangian~\eqref{eq:lagrangian}. In terms of the $1\times 2$ matrix $\mathcal{M}_D$ and the $2\times 2$ matrix $\mathcal{M}_S$ as defined through Eq.~\eqref{eq:massmatrix}, the finite loop contribution in our single-generation case can be written as~\cite{Dev:2012sg}
\begin{align}
\label{eq:oneloopexact}
\delta m_\nu^\text{1-loop}  =  
\frac{\alpha_W\mathcal{M}_D \mathcal{M}_S}{16\pi m_W^2}
\bigg[  \frac{m_H^2}{\mathcal{M}_S^2 - m_H^2\mathbf{1}}\ln\left(\frac{\mathcal{M}_S^2}{m_H^2}\right)+ \frac{3m_Z^2}{\mathcal{M}_S^2 - m_Z^2\mathbf{1}}\,
\ln\left(\frac{\mathcal{M}_S^2}{m_Z^2}\right)
\bigg]\mathcal{M}_D^T\,.
\end{align}
Here, $\alpha_W = g^2/4\pi$ is the weak fine structure constant, $m_H = 125$~GeV, $m_W=80.4$ GeV and $m_Z = 91.2$~GeV are the SM Higgs, $W$ and $Z$ boson masses respectively, and $\mathbf{1}$ is the $2\times 2$ identity matrix. To a very good approximation, the expression Eq.~\eqref{eq:oneloopexact} can be simplified to~\cite{Dev:2012sg}
\begin{align}
\label{eq:oneloopapprox}
\delta m_\nu^\text{1-loop} \ \approx \ \frac{\alpha_{W}m^2_{D}\mu_{R}}{16 \pi m_{W}^2}\left[\frac{m^2_{H}}{m^2_{S}-m^2_{H}} \ln \left(\frac{m^2_{S}}{m^2_{H}}\right)+\frac{3 m^2_{Z}}{m^2_{S}-m^2_{Z}} \ln \left(\frac{m^2_{S}}{m^2_{Z}}\right)\right]\,,
\end{align}
in the limit $\mu_{R,S}\ll |m_S|$ of the generalised inverse seesaw mass matrix Eq.~\eqref{eq:generalized}.

In our analysis, we will require that the one-loop corrections are subdominant to the tree-level mass, using a 10\% contribution as the limit,
\begin{align}
\label{eq:loop-tree}
\delta m_\nu^\text{1-loop} 
\ \leq \ 0.1 m_\nu \, .
\end{align}
Using different loop-to-tree contribution ratios will not change our results qualitatively.

%%%%%%%%%%%%%%%%%%%%%%%%%%%%%%%%%%%%%%%%%%%%%%%%%%%%%%%%%%%%%%%%%%%%%%%%%%%%%%%%%
\section{Phenomenological parametrisation of the mixing matrix}
\label{sec:parametrization}
%%%%%%%%%%%%%%%%%%%%%%%%%%%%%%%%%%%%%%%%%%%%%%%%%%%%%%%%%%%%%%%%%%%%%%%%%%%%%%%%%
As noted before, we will neglect the flavour structure of the lepton sector and work in a single-generation picture with only an electron flavour active neutrino field and two sterile fields; $\nu_{L,e}^C$, $\nu_{R,1}$ and $\nu_{R,2}$. In this case the general neutrino mass matrix $\mathcal{M}_\nu$ in Eq.~\eqref{eq:massmatrix} can be diagonalised by a $3\times 3$ unitary matrix $V$ as described in Eq.~\eqref{eq:diagonalization}. It is simple to reverse this diagonalisation in order to express the mass matrix in terms of the \textit{a priori} measurable mixing angles, CP phases and mass eigenvalues, 
\begin{align}
	\mathcal{M}_\nu \ = \  
	V \cdot \text{diag}\left(m_\nu, m_{N_1}, m_{N_2}\right) \cdot V^T.
\end{align} 

We can first consider a parametrisation of $V$ analogous to that of the Pontecorvo-Maki–Nakagawa–Sakata (PMNS) mixing matrix accompanying charged currents in the SM,
\begin{align}
\label{eq:Vmatrix}
	V \ = \ \begin{pmatrix}
	U_{\nu} &  V_{eN}  \\
	V_{Ne} & V_{N}
	\end{pmatrix}&= 
	\begin{pmatrix}
		1 &                          0 &               0 \\
		0 & \phantom{-}c_{12} & s_{12} \\
		0 &          - s_{12} & c_{12}
	\end{pmatrix} \cdot
	\begin{pmatrix}
		c_{e 2} & 0 & \phantom{-}s_{e 2}e^{-i\delta} \\
	                             0 & 1 & 0                          \\
	  - s_{e 2}e^{i\delta} & 0 & c_{e 2}           
	\end{pmatrix} \cdot
	\begin{pmatrix}
		\phantom{-}c_{e 1} & s_{e 1} & 0 \\
		         - s_{e 1} & c_{e 1} & 0 \\
		                         0 &               0 & 1
 	\end{pmatrix} \cdot D \nonumber\\
	& \ = \ \begin{pmatrix}
		           c_{e 1}c_{e 2} & s_{e 1}c_{e 2} & s_{e 2}e^{-i\delta}            \\
		         - s_{e 1}c_{12} - c_{e 1}s_{e 2}s_{12}e^{i\delta} & 
		\phantom{-}c_{e 1}c_{12} - s_{e 1}s_{e 2}s_{12}e^{i\delta} & c_{e 2}s_{12} \\
		\phantom{-}s_{e 1}s_{12} - c_{e 1}s_{e 2}c_{12}e^{i\delta} &
		         - c_{e 1}s_{12} - s_{e 1}s_{e 2}c_{12}e^{i\delta} & c_{e 2}c_{12}
	\end{pmatrix} \cdot D \\
	& \ \approx \ \begin{pmatrix}
		                  1 & \phantom{-} s_{e 1} & \phantom{-}s_{e 2}e^{-i\delta}            \\
		     - s_{e 1}c_{12} - s_{e 2}s_{12}e^{i\delta} & 
	\phantom{-}c_{12} & s_{12} \\
	\phantom{-}s_{e 1}s_{12} - s_{e 2}c_{12}e^{i\delta} &
	         - s_{12} & c_{12}
	\end{pmatrix} \cdot D + \mathcal{O}\left(s_{e i}^2\right)\,,\nonumber
\end{align} 
in terms of the cosine $c_{ij} \equiv \cos\vartheta_{ij}$ and sine $s_{ij} \equiv \sin\vartheta_{ij}$ of the three mixing angles $\vartheta_{e 1}$, $\vartheta_{e 2}$ and $\vartheta_{12}$. They describe, respectively, the mixing between the mostly-active light neutrino mass eigenstate $\nu_{e}$ and the first mostly-sterile mass eigenstate $N_1$, $\nu_{e}$ and the second mostly-sterile mass eigenstate $N_2$ and finally between $N_{1}$ and $N_{2}$. The angles can in principle lie in the range $\vartheta_{ij} \in [0, \pi/2]$ and the equivalent of the Dirac CP phase in the range  $\delta \in [0,2\pi]$. $D$ is a diagonal matrix containing the remaining two Majorana phases $\phi_{1,2} \in [0,2\pi]$,
\begin{align}
	D \ = \ \begin{pmatrix}
	1 & 0 & 0 \\
	0 & e^{i\phi_1/2} & 0 \\
	0 & 0           &  e^{i\phi_2/2}
	\end{pmatrix}.
\end{align}
As for the light active neutrino PMNS mixing matrix, only two physical Majorana phases survive because an overall phase can be rotated away. 

Rather than this phenomenological approach we can instead write $V$ in a form explicitly imposing existing constraints from neutrino oscillations. A convenient way to do this is the so-called Casas-Ibarra parametrisation~\cite{Casas:2001sr}, which has been generalised in Ref.~\cite{Donini:2012tt} to include the complete parameter space of sterile neutrino masses and mixings. Here, in the three-generation picture and for two sterile states the active-sterile mixings are related to the light active neutrino masses $m_{i}$ (assuming $m_{1}=0$), heavy neutrino masses $m_{N_{i}}$ and PMNS mixing matrix elements by
\begin{align}
V_{e N_{i}} \ = \ i(U_{\text{PMNS}})_{e k}H_{kj}\sqrt{\frac{m_{j}}{m_{N_{i}}}}\mathcal{R}_{ij}^*\,,
\end{align}
where $\mathcal{R}$ is an arbitrary $2\times 2$ orthogonal matrix parametrised by a complex mixing angle $\vartheta_{45}+i\gamma_{45}$ and $H$ is a hermitian matrix encoding deviations from unitarity in the light neutrino sector. For fixed values of $m_{i}$ and $m_{N_{i}}$ the size of mixings $V_{eN_i}$ depend on $\vartheta_{45}$ and $\gamma_{45}$. In the phenomenological single-generation picture this translates to choices of the CP phases $\phi_1$, $\phi_2$ and $\delta$. We will proceed with our phenomenological approach because it is not our immediate goal to reproduce the observed light neutrino data, which is an implicit input to the Casas-Ibarra parametrisation. Our goal is to investigate in the most direct way the phenomenology of active-sterile mixing in the generalised inverse seesaw.

\subsection{Consistency relations}

We will now apply this general parametrisation to the seesaw scenarios discussed in Sec.~\ref{sec:model-setup}. Without a triplet Higgs extending the SM field content, the active neutrinos cannot acquire a mass of the form $\bar\nu_L^C \nu_L$ and thus the $(1,1)$ entry of $\mathcal{M}_\nu$ in Eq.~\eqref{eq:massmatrix} is strictly zero at tree-level. This requirement must be satisfied irrespective of the remaining mass matrix structure (i.e. type-I, inverse or linear seesaw). Written in terms of the phenomenological parameters this condition may be written as
\begin{align}
\label{eq:cond11}
	(\mathcal{M}_\nu)_{11} \ = \ 0 \quad\Rightarrow\quad
	  c_{e1}^2 c_{e2}^2 \,\frac{m_\nu}{m_{N_1}} 
	+ s_{e1}^2 c_{e2}^2 \, e^{i\phi_1}
	+ s_{e2}^2 \,\frac{m_{N_2}}{m_{N_1}} \,e^{i(\phi_2-2\delta)} \ = \ 0\,,
\end{align} 
where we have divided the sum by the heavy neutrino mass $m_{N_1}$. We note first that this constraint has no dependence on the sterile-sterile mixing angle $\vartheta_{12}$. It can also be seen that such a constraint is equivalent to the vanishing of the effective $0\nu\beta\beta$ decay mass $m_{\beta\beta} = \sum_i (U_\text{PMNS})^2_{ei} m_{i}$, where in that case the summation is over the three light neutrino mass eigenstates. While this would be an accidental cancellation -- possible for a normally ordered light neutrino spectrum with specific values of the Majorana phases in $U_\text{PMNS}$ (as opposed to $V$) -- the condition in Eq.~\eqref{eq:cond11} must always be satisfied at tree-level, putting requirements on the values of the three masses, three mixing angles and three CP phases. Instead of the parameter $m_{N_{2}}$ it is equally valid to use the mass splitting $\Delta m_{N}=m_{N_{2}}-m_{N_{1}}$, which will be of importance later.

\begin{figure}[t!]
	\centering
	\includegraphics[width=0.5\textwidth]{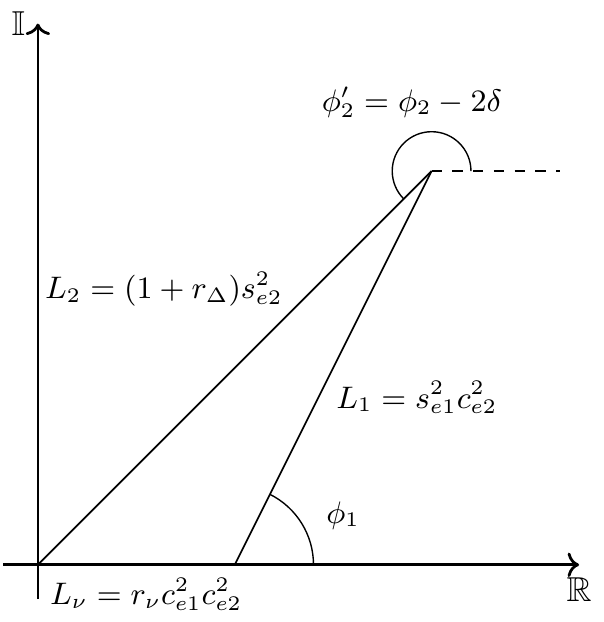}
	\caption{Visualisation of the $(\mathcal{M}_\nu)_{11} = 0$ constraint in Eq.~\eqref{eq:cond11} in the complex plane. The sides are given in terms of the dimensionless ratios $r_{\nu} = m_\nu/m_{N_1}$ and $r_{\Delta} = \Delta m_{N}/m_{N_1}$ along with squared sines and cosines of the active-sterile neutrino mixing angles $\vartheta_{e 1}$, $\vartheta_{e2}$.}
	\label{fig:triangle_diagram}
\end{figure}

As illustrated in Fig.~\ref{fig:triangle_diagram}, the condition in Eq.~\eqref{eq:cond11} can be visualised as a triangle in the complex plane, formed by three sides with lengths $L_\nu = c_{e1}^2 c_{e2}^2 m_\nu/m_{N_1}$, $L_1 = s_{e1}^2 c_{e2}^2$ and $L_2 = s_{e2}^2 m_{N_2}/m_{N_1}=s_{e2}^2 (m_{N_1}+\Delta m_{N})/m_{N_1}$. The angles between these sides are determined by the phase $\phi_1$ and the linear combination $\phi_2 - 2\delta$, which we label $\phi'_2$ for convenience. Not all combinations of the masses and mixings allow a triangle to be formed with side lengths $L_\nu$, $L_1$ and $L_2$. Specifically, the triangle can only be closed (for some values of $\phi_1$ and $\phi_2'$) if the longest length is smaller (or equal) to the sum of the shorter lengths, 
\begin{align}
	\text{max}(L_\nu, L_1, L_2) 
	\ \leq \ \text{min}(L_\nu, L_1, L_2) + \text{med}(L_\nu, L_1, L_2)\,.
\end{align}

\begin{figure}[t!]
	\centering
	\includegraphics[width=0.49\textwidth]{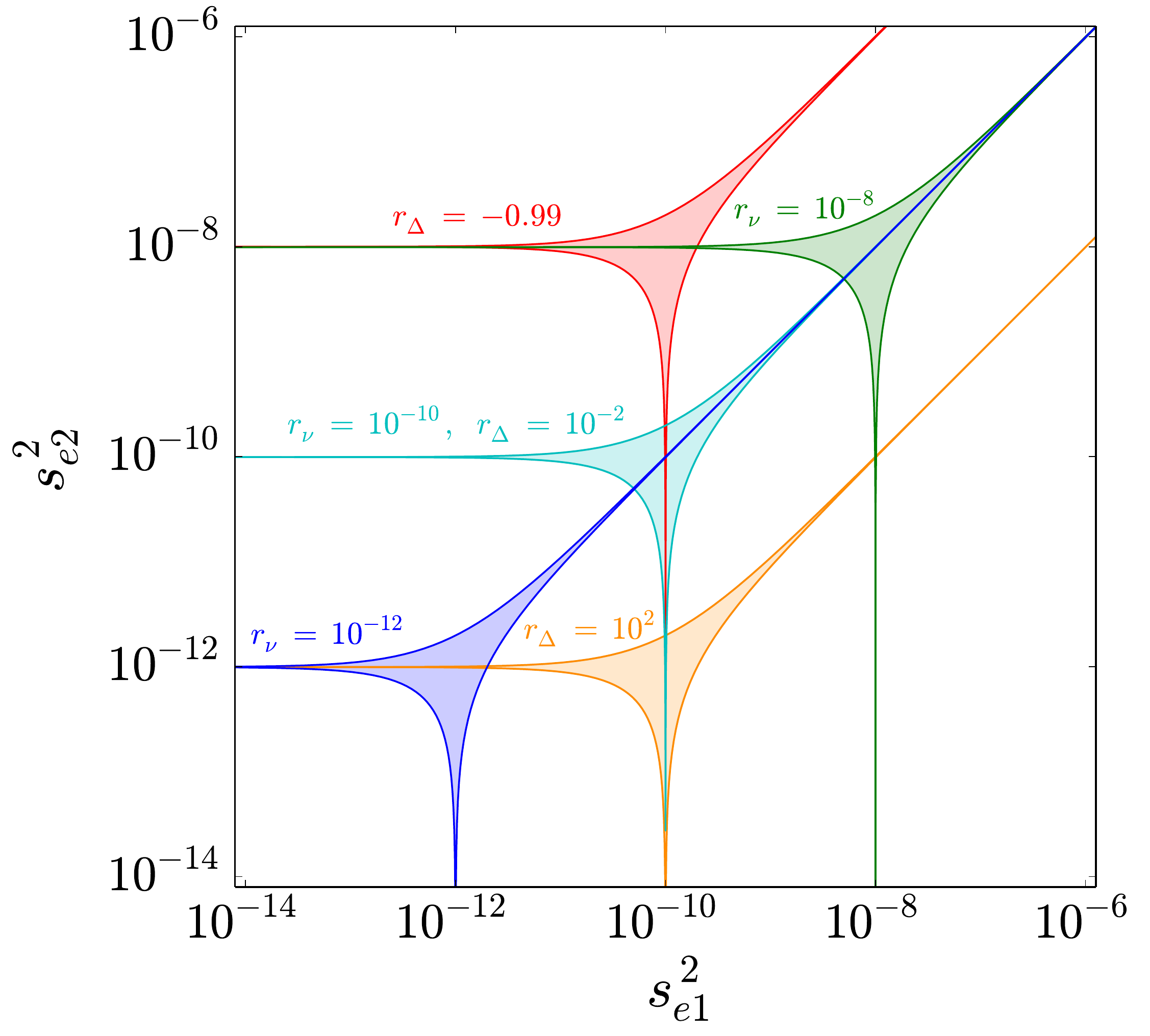}
	\includegraphics[width=0.49\textwidth]{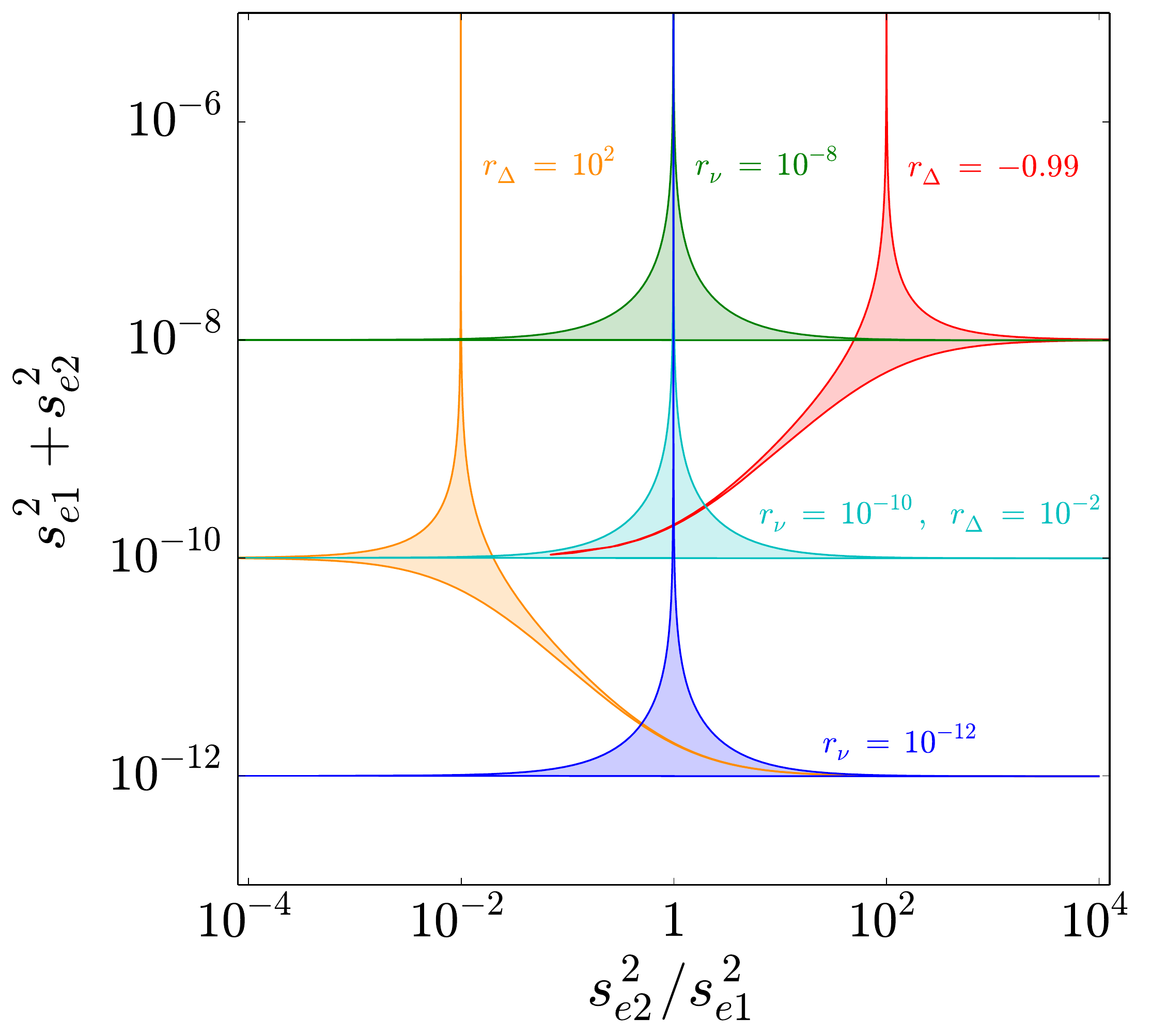}
	\caption{{\it Left:} Values of the squared active-sterile mixing strengths $s_{e1}^2$ and $s_{e2}^2$ satisfying the tree-level condition $(\mathcal{M}_\nu)_{11} = 0$ in Eq.~\eqref{eq:cond11} for different combinations of the light and heavy neutrino masses in the ratios $r_{\nu} = m_\nu / m_{N_1}$, and $r_{\Delta} = \Delta m_{N}/m_{N_1}$, as shown by the shaded regions. {\it Right:} Equivalent regions in the $s_{e2}^2/s_{e1}^2$ and $s_{e1}^2 + s_{e2}^2$ parameter space.}
	\label{fig:mixingangles}
\end{figure}

The allowed regions for the squared active-sterile mixing strengths $s_{e1}^2$ and $s_{e2}^2$ are shown in Fig.~\ref{fig:mixingangles} (left) for different choices of the light and heavy neutrino masses. The centre shape (light blue) corresponds to the choice $m_\nu / m_{N_1} = 10^{-10}$ and $\Delta m_{N}/m_{N_1} = 10^{-2}$. This for example could correspond to a light neutrino mass $m_\nu = 10^{-3}$ eV and heavy neutrino masses $m_{N_1} = 10$ MeV and $m_{N_2} = 10.01$ MeV. The allowed mixing strengths form a region centred around $s_{e1}^2 \approx s_{e2}^2 \approx m_\nu / m_{N_1}$. Thin, virtually line-like extensions to large $s_{e1}^2 = s_{e2}^2$, small $s_{e1}^2$ and small $s_{e2}^2$ are also possible. As can be seen from the dark blue and green regions, increasing (decreasing) $r_{\nu} = m_{\nu}/m_{N_{1}}$ will move the bulk of the region along the diagonal to higher (smaller) mixing. As can be seen from the yellow region, increasing the splitting $\Delta m_{N}$ shifts the allowed region to smaller values of $s_{e2}^2$ but not $s_{e1}^2$. The red region, on the other hand, shows the scenario in which $\Delta m_{N}$ becomes negative (when $m_{N_{2}}<m_{N_{1}}$). The allowed region instead moves up to larger $s_{e2}^2$ for the same $s_{e1}^2$. We will investigate this behaviour more quantitatively below. Fig.~\ref{fig:mixingangles} (right) shows the same regions but with the axes given by the ratio and sum of the mixing strengths, $s_{e2}^2/s_{e1}^2$ and $s_{e1}^2 + s_{e2}^2$, respectively. It especially illustrates that there exists a lower limit on the total active-sterile mixing strength $s_{e1}^2 + s_{e2}^2$.

\subsection{CP phases}

The limiting behaviours for small and large mixing strengths can be related to the CP-conserving cases when the phases adopt values such that $e^{i\phi_1} = \pm 1$, $ e^{i\phi_2'}= \pm 1$, corresponding to the relative CP parity of the sterile fields. The CP parity of the $m_\nu$ state is defined by convention as $+1$. Three possibilities emerge for the CP parities of the other states:  
\begin{enumerate}
	\item[(A)] $e^{i\phi_1} = e^{i\phi_2'} = +1$: The condition $(\mathcal{M}_\nu)_{11} = 0$ in Eq.~\eqref{eq:cond11} cannot be satisfied (unless trivially when $m_\nu = m_{N_1} = m_{N_2} = 0$) as all three contributions add up constructively, $L_\nu + L_1 + L_2 > 0$. 
	
	\item[(B)] $e^{i\phi_1} = e^{i\phi_2'} = -1$: The contributions of the states $N_1$ and $N_2$ are negative and cancel the active neutrino contribution, $L_\nu - (L_1 + L_2) = 0$. Eq.~\eqref{eq:cond11} can be solved for one of the active-sterile mixing angles as 
	\begin{align}
	\label{eq:caseB}
		s_{e2}^2 
		& \ = \ \frac{m_\nu - (m_{N_1} + m_\nu)s_{e1}^2}{m_{N_1} +\Delta m_N + m_\nu - (m_{N_1} + m_\nu)s_{e1}^2} \nonumber\\
		& \ \approx \ \frac{m_\nu/m_{N_1} - s_{e1}^2}{1+\Delta m_N/m_{N_1}}
		\quad \text{for}\quad m_\nu\ll m_{N_1} \, {\rm and}~ s_{e1}^2 \ll 1\, .
	\end{align}
	Because $s_{e2}^2 \geq 0$ this can only be satisfied if $s_{e1}^2 \leq m_\nu/(m_{N_1} + m_\nu) \lesssim m_\nu/m_{N_1}$, i.e. for $s_{e1}^2$ up to the ordinary single heavy-state seesaw mixing $s_{e1}^2 = m_\nu/m_{N_1}$. Consequently, $s_{e2}^2$ can range from $s_{e2}^2 = 0$ (when $s_{e1}^2 = m_\nu/m_{N_1}$) to $s_{e2}^2 \approx m_\nu/m_{N_2}$ (when $s_{e1}^2 = 0$). This scenario corresponds to the \textit{canonical seesaw} with two heavy Majorana states; the active state can mix with either of them with adjustable strength. In Fig.~\ref{fig:mixingangles}~(left), this particular limit corresponds to the line-like extensions towards vanishing $s_{e2}^2$ at the bottom ($N_2$ decouples, $s_{e1}^2 \to m_\nu / m_{N_1}$) and vanishing $s_{e1}^2$ to the left ($N_1$ decouples, $s_{e2}^2 \to m_\nu / m_{N_2}$). Intermediate solutions lie on the lower left edge of the allowed region in Fig.~\ref{fig:mixingangles} (left). Rearranging Eq.~\eqref{eq:caseB} for small $\Delta m_{N}$ gives $s_{e1}^2+s_{e2}^2\approx m_{\nu}/m_{N_1}$, a behaviour that can clearly be seen in Fig.~\ref{fig:mixingangles} (right).
	
	\item[(C)] $e^{i\phi_1} = + 1$, $e^{i\phi_2'} = - 1$: The contributions of the heavy states can (partially) cancel among each other, $L_\nu + (L_1 - L_2) = 0$. Again, we can solve for the mixing angle $s_{e2}^2$, 
	\begin{align}
		s_{e2}^2 
		& \ = \ \frac{m_\nu+(m_{N_1} - m_\nu)s_{e1}^2}{m_{N_1} + \Delta m_N + m_\nu + (m_{N_1}-m_\nu)s_{e1}^2} \nonumber\\
		& \ \approx \ \frac{s_{e1}^2}{1+\Delta m_N/m_{N_1}}
		\quad \text{for}\quad m _\nu\ll m_{N_1}\, {\rm and}~s_{e1}^2 \ll 1\,.
	\end{align}
	Here, no upper bound on $s_{e1}^2$ exists and it can in principle take values between $0 \leq s_{e1}^2 \leq 1$. For a small mass splitting $\Delta m_N \ll m_{N_1}$ this case corresponds to the \textit{inverse seesaw} scenario where the two heavy Majorana states form a pseudo-Dirac neutrino pair. In Fig.~\ref{fig:mixingangles} (left), this limit corresponds to the thin extension of the allowed region to large mixing strengths. It should be noted that this phenomenological parametrisation does not enforce a small mass splitting and $\Delta m_N$ can be arbitrarily large for a given light neutrino mass $m_\nu$. As we will discuss below, however, this will induce large loop corrections to $m_{\nu}$.  
\end{enumerate}

For arbitrary values of the phases $\phi_1$ and $\phi'_2$ the interior of the shaded regions in Fig.~\ref{fig:mixingangles} is covered. In order to simplify the following discussion we make use of the dimensionless ratios $r_{\nu} = m_\nu/m_{N_1}$ and $r_{\Delta} = \Delta m_N/m_{N_1}$ as already introduced in Fig.~\ref{fig:mixingangles}. For arbitrary phases, Eq.~\eqref{eq:cond11} in fact represents two conditions; $\text{Re}\{(\mathcal{M}_{\nu})_{11}\}=0$ and $\text{Im}\{(\mathcal{M}_{\nu})_{11}\}=0$.  These relations can be rearranged to find two equivalent expressions for $s_{e2}^2$,
\begin{align}
\label{eq:realandimagpart}
1-\frac{1}{s_{e2}^2} 
\ = \ \frac{(1+r_{\Delta})\cos\phi'_2}{r_{\nu}+(\cos\phi_1-r_{\nu})s_{e1}^2}
\ = \ \frac{(1+r_{\Delta})\sin\phi'_2}{\sin\phi_1 s_{e1}^2}\,,
\end{align}
where the first and second equalities are derived from the real and imaginary conditions, respectively. We can also rearrange Eq.~\eqref{eq:realandimagpart} to solve for the tangent of $\phi_2'$,
\begin{align}
\label{eq:tanphi2}
	\tan\phi'_2
	\ = \ \frac{\sin\phi_1 s_{e1}^2}{r_{\nu}+(\cos\phi_1-r_{\nu})s_{e1}^2} 
	\ \approx \ \begin{cases}
		\sin\phi_1 \frac{s_{e1}^2}{r_{\nu}} + \mathcal{O}(s_{e1}^4) 
		&\text{for}~s_{e1}^2\ll r_{\nu} \\
		\tan(\phi_1/2) &\text{for}~s_{e1}^{2} = r_{\nu} \\
		\tan{\phi_1} + \mathcal{O}(r_{\nu}) &\text{for}~s_{e1}^2\gg r_{\nu} \, \\
	\end{cases},
\end{align}
where we also indicate approximate solutions for the different limits of $s_{e1}^2$. In effect, the condition in Eq.~\eqref{eq:cond11} has allowed us to eliminate two parameters, $s_{e2}^2$ and $\phi_2'$, by expressing them in terms of a subset of the remaining free parameters, $r_{\nu}$, $r_{\Delta}$, $s_{e1}^2$ and $\phi_1$. The freedom to divide Eq.~\eqref{eq:cond11} by $m_{N_{1}}$ and using instead the ratios $r_{\nu}$ and $r_{\Delta}$ also effectively removes a mass degree of freedom. This can be seen from the behaviour of the allowed regions in Fig.~\ref{fig:mixingangles}; a shift in the $s_{e1}^2$ -- $s_{e2}^2$ plane only occurs when $r_{\nu}$ and $r_{\Delta}$ are changed. It must however be remembered that the other elements of $\mathcal{M}_{\nu}$ (e.g. $m_{D}$, $m_{S}$, $\mu_R$, $\mu_{S}$) have been divided by $m_{N_{1}}$, so this factor must be taken into account when calculating these flavour-basis parameters as functions of the phenomenological mass-basis parameters.

Alternatively a more physical choice would be to solve for $\cos\phi_1$ and $\cos\phi'_2$ using the cosine rule for the $(\mathcal{M}_{\nu})_{11}=0$ constraint triangle in Fig.~\ref{fig:triangle_diagram},
\begin{align}
\label{eq:cosphi1}
	\cos\phi_1 
	&=  \frac{(1+r_{\Delta})^2 s_{e2}^4 - r_{\nu}^2 c_{e1}^4 c_{e2}^4
		     - s_{e1}^4 c_{e2}^4}{2r_{\nu} s_{e1}^2 c_{e1}^2 c_{e2}^4 }
	\approx \frac{(1+r_{\Delta})^2s_{e2}^4 - r_{\nu}^2 - s_{e1}^4}{2r_{\nu} s_{e1}^2}, \\
\label{eq:cosphi2}
	\cos\phi_2'
	&= \frac{s_{e1}^4 c_{e2}^4 - r_{\nu}^2 c_{e1}^4 c_{e2}^4 
		     - (1+r_{\Delta})^2 s_{e2}^4}{2r_{\nu}(1+r_{\Delta})c_{e1}^2 s_{e2}^2 c_{e2}^2}
	\approx \frac{s_{e1}^4 - r_{\nu}^2 - (1+r_{\Delta})^2 s_{e2}^4}{2r_{\nu}(1+r_{\Delta})s_{e2}^2},
\end{align}
where the approximate expressions hold for small mixing $s_{e1}^2, \,s^2_{e2} \ll 1$. In this way the phases $\phi_1$ and $\phi'_2$ are determined (up to a pair of solutions in the range [0, $2\pi$], modulo $\pi$) by the neutrino masses through the ratios $r_{\nu}$ and $r_{\Delta}$ and the mixing strengths $s_{e1}^2$ and $s_{e2}^2$, all of which are in principle experimentally measurable. If the solution for $\phi_1$ lies in the first or second quadrant (i.e. $\phi_1 \in [0,\pi] $), in order to close the triangle in Fig.~\ref{fig:triangle_diagram} it is necessary for $\phi'_2$ to be in the third or fourth quadrants ($\phi'_2 \in [\pi,2\pi] $) and vice versa.

\begin{figure}[t!]
	\centering
	\includegraphics[width=0.49\textwidth]{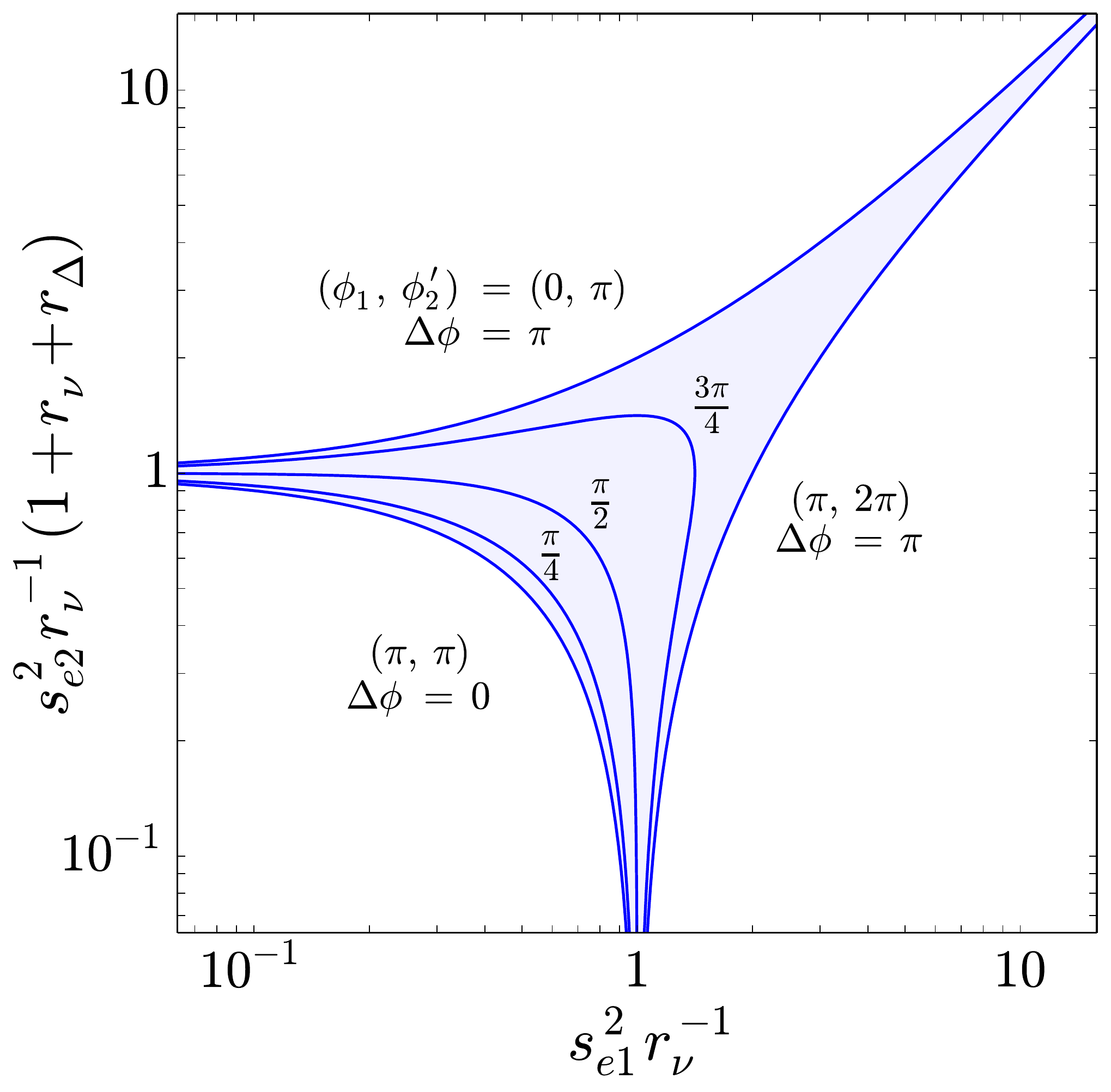}
	\includegraphics[width=0.49\textwidth]{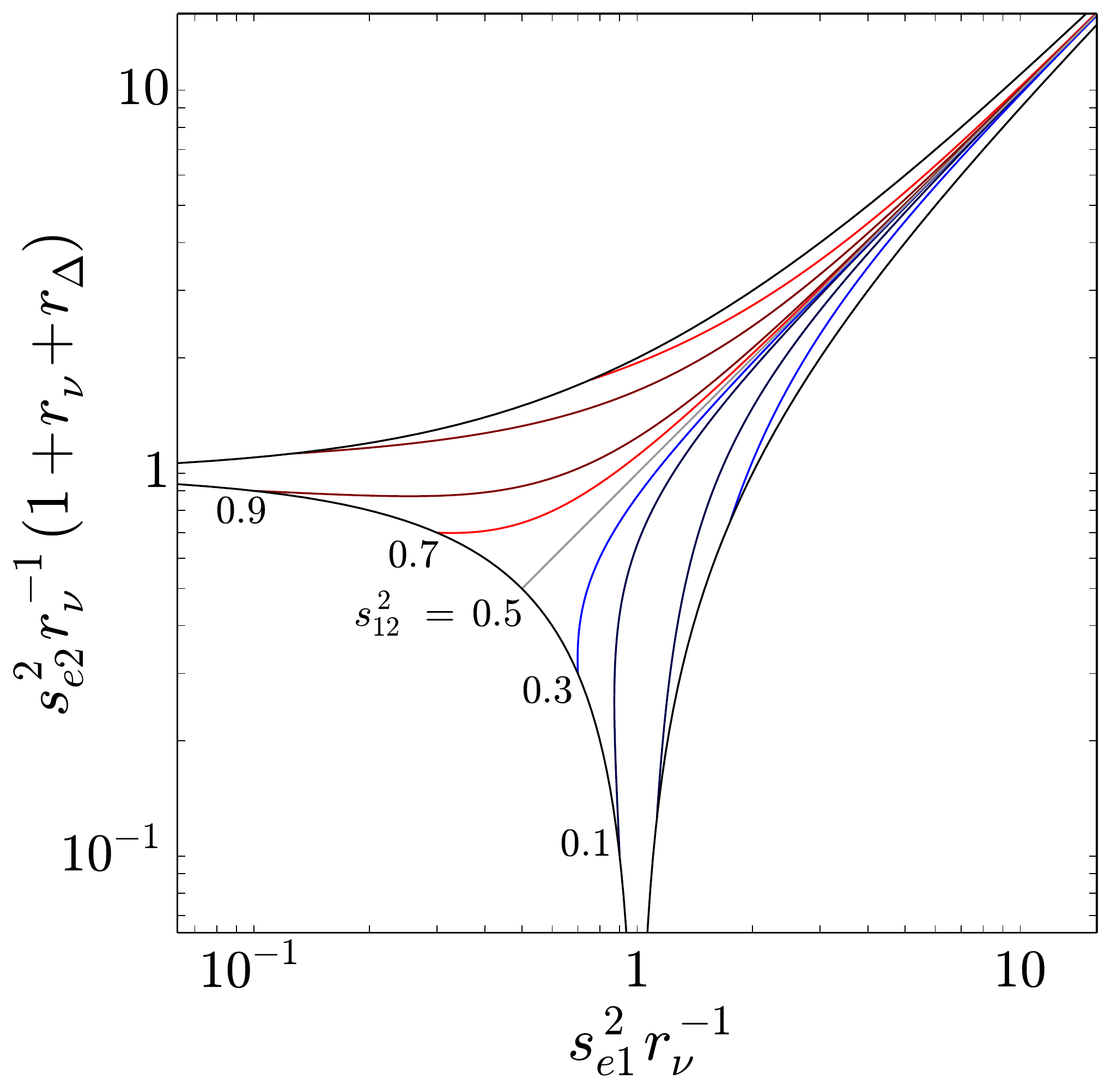}
	\caption{{\it Left:} Heavy state CP phase difference $\Delta\phi =  \phi_1 - \phi_2'$ as a function of the active-sterile mixing strengths $s_{e1}^2$ and $s_{e2}^2$. The axes are normalised by $r_{\nu} = m_\nu/m_{N_1}$ and $r_{\nu}/(1+r_{\nu}+r_{\Delta})$ ($r_{\Delta} = \Delta m_N/m_{N_1}$), respectively, so that allowed region applies for an arbitrary choice of the neutrino mass eigenvalues. The edges of the allowed region are determined by the limiting values for $(\phi_1, \phi'_2)$ as indicated. {\it Right:} Sterile-sterile neutrino mixing strength $s_{12}^2$ as a function of $s_{e1}^2$ and $s_{e2}^2$, setting $\delta = 0$.}
	\label{fig:s12s13}
\end{figure}

An important parameter in determining the nature of the two heavy states is the phase difference $\Delta\phi = \phi_1 - \phi'_2 =\phi_1 - \phi_2 + 2\delta$ between $N_1$ and $N_2$. If $\Delta\phi \approx 0$ we expect the heavy states to behave like Majorana fermions, whereas for $\Delta\phi \approx \pm \pi$ they should form a pseudo-Dirac pair with an associated suppression of LNV effects. Using the solutions Eqs.~\eqref{eq:cosphi1} and \eqref{eq:cosphi2}, or alternatively using the cosine rule for the third angle of the triangle in Fig.~\ref{fig:triangle_diagram}, $\Delta \phi$ is given in terms of the other parameters by
\begin{align}
	\cos{\Delta \phi} = \frac{r_{\nu}^2c_{e1}^4c_{e2}^4-s_{e1}^4c_{e2}^4-(1+r_{\Delta})^2s_{e2}^4}
	     {2(1+r_{\Delta})s_{e1}^2c_{e2}^2s_{e2}^2} 
	\approx \frac{r_{\nu}^2-s_{e1}^4-(1+r_{\Delta})^2s_{e2}^4}{2(1+r_{\Delta})s_{e1}^2s_{e2}^2}\,.
\end{align}
This phase difference is plotted in Fig.~\ref{fig:s12s13} (left) as a function of the mixing strengths $s_{e1}^2$ and $s_{e2}^2$ within the region allowed by the $(1, 1)$ element constraint Eq.~\eqref{eq:cond11}. Note that the active-sterile mixing strengths $s_{e1}^2$ and $s_{e2}^2$ are normalised by $r_{\nu}$ and $r_{\nu}/(1+r_{\nu}+r_{\Delta})$ respectively, making the plot generically applicable for an arbitrary choice of the light and heavy neutrino masses. The edges of the allowed region correspond to the CP-conserving combinations of phases: (i) $\phi_1 = \phi'_2= \pi$ to the lower left corresponding to the \emph{canonical seesaw} with two Majorana heavy states and (ii) $\phi_1 = 0\, (\pi)$, $\phi'_2 = \pi\,(2\pi)$ on the top (lower right) edge, corresponding to an \emph{inverse seesaw}-like scenario. Intermediate scenarios interpolating between these limiting cases are characterised by the phase difference $|\Delta\phi|$ increasing from 0 to $\pi$ as shown.

We have so far seen that is it possible to eliminate the two phases $\phi_1$ and $\phi_2'$ from the nine initial phenomenological parameters. The next question is whether additional relationships can be found between these parameters. While Eq.~\eqref{eq:cond11} is a parametrisation-independent condition, we can make convenient choices for the remaining parameters in $\mathcal{M}_{\nu}$ which assist in this effort. For example, as discussed briefly in Sec.~\ref{sec:model-setup}, without lack of generality we can assume that the $(1,3)$ element of the neutrino mass matrix $\mathcal{M}_\nu$ in Eq.~\eqref{eq:massmatrix} vanishes, $(\mathcal{M}_\nu)_{13} = 0$. This can always be achieved by rotating the heavy states appropriately, even in the linear seesaw scenario~\cite{Ma:2009du}. Using our phenomenological parametrisation this corresponds to
\begin{align} 
\label{eq:const13}
	(\mathcal{M}_\nu)_{13}
	& \ = \ r_{\nu} c_{e1} c_{e2}\left(s_{e1}s_{12} - e^{i\delta}c_{e1}s_{e2}c_{12}\right)\nonumber\\ 
	&~~~\,- e^{i\phi_1} s_{e1} c_{e2}\left(c_{e1}s_{12} 
	   + e^{i\delta}s_{e1}s_{e2}c_{12}\right) + e^{i(\phi_2-\delta)}(1+r_{\Delta})c_{e2}s_{e2}c_{12}  \ = \ 0\,.
\end{align}
Note that in this condition the linear combination $\phi'_2=\phi_2-2\delta$ does not appear explicitly. As we would like to continue using the relations for $\cos\phi_1$ and $\cos\phi'_2$ in Eqs.~\eqref{eq:cosphi1} and \eqref{eq:cosphi2} we introduce the linear combination $\delta'=2\phi_2+\delta$ orthogonal to $\phi_2'$. The phases $\phi_1$, $\phi_2$ and $\delta$ can consequently be written as linear combinations of $\phi_1$, $\phi'_2$ and $\delta'$. As the $(1, 1)$ element constraint we can take both the real and imaginary part of Eq.~\eqref{eq:const13}, rearranging for $s_{12}^2$ as a function of $s_{e1}^2$, $s_{e2}^2$ and the phases,
\begin{align} 
\label{eq:s12squared}
	\frac{1}{s_{12}^2} 
	& \ = \ 1 +
	C_R \frac{s_{e1}^2c_{e1}^2}{s_{e2}^2} \ = \ 1 +
	C_I \frac{s_{e1}^2c_{e1}^2}{s_{e2}^2}\,,
\end{align}
where
\begin{align} 
\label{eq:CR}
C_R
& \ = \
 \frac{(r_{\nu}-\cos\phi _1)^2}{((1+r_{\Delta})\cos(\phi_2-\delta) - r_{\nu}\cos\delta 
	+ \left(r_{\nu}\cos\delta - \cos(\phi_1+\delta)\right)s_{e1}^2)^2}\,,\nonumber\\
C_I
& \ = \ 
\frac{\sin^2\phi _1}{((1+r_{\Delta})\sin(\phi_2-\delta) - r_{\nu}\sin\delta 
	+ \left(r_{\nu}\sin\delta - \cos(\phi_1+\delta)\right)s_{e1}^2)^2}\,.
\end{align}
The sterile-sterile mixing strength $s_{12}^2$ is shown in Fig. \ref{fig:s12s13} (right) as a function of $s_{e1}^2$ and $s_{e2}^2$ for $\delta = 0$. Further, proceeding as before, we can equate the real and imaginary solutions of $s_{12}^2$ in Eq.~\eqref{eq:s12squared}, i.e. $C_{R}=C_{I}$. Rewriting in terms of the phases $\phi_1$, $\phi'_2$ and $\delta'$ and making use of the solutions for $\cos\phi_{1}$ and $\cos\phi_{2}'$ in theory allows to solve for the final phase $\delta'$ in terms of $r_{\nu}$, $r_{\Delta}$, $s_{e1}^2$ and $s_{e2}^2$. In practice it is difficult to do this analytically, but numerically $\delta'$ can be found by finding the intersecting points of the curves $C_{R}(\delta')$ and $C_{I}(\delta')$. 

We have therefore seen that, given values of the parameters $r_{\nu}$, $r_{\Delta}$, $s_{e1}^2$ and $s_{e2}^2$ and assuming a particular parametrisation of the neutrino mass matrix $\mathcal{M}_{\nu}$, the remaining parameters $s_{12}^2$, $\phi_{1}$, $\phi_{2}'$ and $\delta'$ are uniquely determined. Thus, if the absolute neutrino mass scale $m_{\nu}$ were known and an experiment were to observe two sterile states with mass splitting $\Delta m_{N}$ and mixing strengths $s_{e1}^2$ and $s_{e2}^2$, in the generalised inverse seesaw the sterile-sterile mixing strength $s_{12}^2$ and CP phases $\phi_1$, $\phi'_2$ and $\delta'$ are predicted quantities. As we will see in Sec.~\ref{sec:searches}, direct searches for the production and decay of heavy states can probe, if not sensitive to the lepton numbers of the final states, the mixing matrix elements $|V_{eN_1}|^2\approx s^2_{e1}$ and $|V_{eN_2}|^2 \approx s^2_{e2}$ for particular values of $m_{N_{1}}$ or $m_{N_{2}}$. If the splitting $\Delta m_{N}$ is large enough for the two states to be resolved, $|V_{eN_1}|^2$ and $|V_{eN_2}|^2$ can be measured independently, constraining the values of the other parameters. If the splitting is below the energy resolution of an experiment it will instead be sensitive to the sum $|V_{eN_1}|^2+|V_{eN_2}|^2$. As seen in Fig.~\ref{fig:s12s13} (left), this can only put a lower bound on $\Delta \phi$ while $s_{12}^2$ and $\delta'$ are left unconstrained. Most current and future direct searches are still probing the regime $|V_{eN_1}|^2\approx |V_{eN_2}|^2 \gg r_{\nu}$, where the generalised inverse seesaw predicts the phase difference $\Delta \phi = \pm\pi$. Some experiments like the KATRIN upgrade TRISTAN~\cite{Mertens:2018vuu} and the future long-baseline neutrino oscillation experiment DUNE~\cite{Adams:2013qkq} may however reach mixing strengths $|V_{eN_1}|^2\lesssim r_{\nu}$, thus being able to pin down any phase difference in the range $|\Delta \phi| \in [0,\pi]$, cf. Fig. \ref{fig:future_constraints}.

The next question to ask is whether the parameters $s_{12}^2$, $\phi_1$, $\phi_2'$ and $\delta'$ can be measured in order to confirm the predictions of the generalised inverse seesaw. The Majorana and pseudo-Dirac limits (governed by $\phi_1$ and $\phi_2'$) are primarily distinguished by the magnitude of LNV. In the case where the sterile mass splitting is not too small, LNV searches are currently probing mixing strengths in the pseudo-Dirac limit. It is unlikely for future LNV searches to be able to reach the mixing strengths $|V_{eN_1}|^2\lesssim r_{\nu}$ required for the Majorana limit. Put differently, if an experiment sees two sterile states with mixings  $|V_{eN_1}|^2\approx|V_{eN_2}|^2\gg r_{\nu}$, but also a large LNV signal (e.g. from a large asymmetry in the pseudorapidity distribution at the ILC~\cite{Hernandez:2018cgc}), it would strongly imply some other source of LNV~\cite{Ibarra:2011xn}. For example, the states $N_{1}$ and $N_{2}$ could possess additional strong couplings to SM particles from a TeV-scale type-III seesaw mechanism, or the light neutrino masses are not generated by the seesaw (e.g. instead, radiatively)~\cite{Ibarra:2010xw}.

We next consider $s_{12}^2$. In the small mixing limit $s_{e1}^2$, $s_{e2}^2 \ll 1$, the matrix
\begin{align}
V_{N} \ \approx \ \begin{pmatrix}
\phantom{-}c_{12} & s_{12} \\
-s_{12} & c_{12}
\end{pmatrix}\cdot
\begin{pmatrix}
e^{i\phi_1/2} & 0 \\
0 & e^{i\phi_2/2} 
\end{pmatrix}
\end{align}
diagonalises the $2\times 2$  sub-matrix $\mathcal{M}_{S}$ of $\mathcal{M}_{\nu}$ in Eq.~\eqref{eq:massmatrix} as $V_{N}^{T}\mathcal{M}_{S}V_{N}$ in the basis that the charged lepton Yukawa coupling is diagonal. In Ref.~\cite{Dev:2019rxh} it was noted that the Dirac sub-matrix $\mathcal{M}_{D}$ can always be redefined as $\mathcal{M}'_{D} = \mathcal{M}_{D}V_{N}^{\dagger}$ so that it is impossible to measure the angle $\vartheta_{12}$, making it unphysical (see also Ref.~\cite{Chao:2009ef}). If right-handed currents are introduced, for example in left-right symmetric models,  $s^2_{12}$ in theory becomes observable because the lower two sub-matrices of $V$ in Eq.~\eqref{eq:Vmatrix} rotate the $W_{R}$ gauge boson interaction. In the pseudo-Dirac case it becomes possible to observe heavy neutrino mixing via the ratio of same-sign to opposite-sign charged lepton production rates in colliders~\cite{Dev:2015pga, Anamiati:2016uxp, Das:2017hmg},
\begin{align}
\label{eq:Rll}
R_{\ell\ell}=\frac{\Delta m^2_{N}}{2\Gamma_{N}^2+\Delta m_{N}^2}\,,
\end{align}
where $\Gamma_{N}$ is the average decay width of the sterile neutrinos. The distinguishing signal here is that $R_{\ell\ell}$ can take an intermediate value between 0 (Dirac limit) and 1 (Majorana limit). While the elements of the mixing matrix $V_{N}$ containing $s_{12}^2$ appear in the same-sign and opposite-sign rates, they cancel in the numerator and denomator for $\Delta \phi = \pm\pi$. This is generally not true if $|\Delta \phi| < \pi$.

\subsection{Including loop corrections}

The mixing strength $s_{12}^2$ is nonetheless important for evaluating the radiatively generated neutrino mass at one-loop in Eq.~\eqref{eq:oneloopexact} (exact expression) and Eq.~\eqref{eq:oneloopapprox} (in the limit $\mu_{R,S} \ll m_{S}$). When written in terms of the masses, mixing angles and CP phases (in the particular parametrisation setting the $(1, 3)$ element of $\mathcal{M}_{\nu}$ to zero), the flavour-space parameters $m_{D}$, $m_{S}$, $\mu_{S}$ and $\mu_{R}$ are functions of $s_{12}^2$. In evaluating these parameters for the purpose of evaluating $\delta m_{\nu}^{\text{1-loop}}$, we will for simplicity assume $\delta = 0$ from the start instead of numerically solving $C_{R}=C_{I}$ for $s_{12}^2$ and $\delta'$ for given values of $m_{\nu}$, $m_{N_{1}}$, $r_{\Delta}$, $s_{e1}^2$ and $s_{e2}^2$. We reiterate that $m_{\nu}$ and $m_{N_{1}}$ must be chosen independently (instead of just the ratio $r_{\nu}$) because an overall factor $m_{N_{1}}$ cannot be eliminated from $m_{D}$, $m_{S}$, $\mu_{S}$ and $\mu_{R}$ as for the $(\mathcal{M}_{\nu})_{11}=0$ and $(\mathcal{M}_{\nu})_{13}=0$ constraints.

In this scenario we can investigate the value of $\vartheta_{12}$ for the limiting cases of $\phi_1$ and $\phi_2'=\phi_2$ along the edges of the allowed region in Fig.~\ref{fig:s12s13}. Applying the limits $s_{e1}^2, s_{e2}^2 \ll 1$ and $m_\nu\ll m_{N_1}$ to the expression for $s_{12}^2$ in Eq.~\eqref{eq:s12squared}, the cases resolve to:
\begin{enumerate}
	\item[(A)] $e^{i\phi_1} = e^{i\phi_2} = +1$: No solution. 
	
	\item[(B)] $e^{i\phi_1} = e^{i\phi_2} = -1$: In this case we have
	\begin{align}
	\label{eq:th12-cs}
		\tan\vartheta_{12} 
		\ = \ \sqrt{(1+r_{\Delta})(r_{\nu}/s_{e1}^2-1)}~,
	\end{align}
	where $s_{e1}^2\leq r_{\nu}$ as discussed before in this case, making the root well defined.
	
	\item[(C)] $e^{i\phi_1} = \pm 1$, $e^{i\phi_2} = \mp 1$: Now the sterile-sterile mixing angle is determined as
	\begin{align}
	\label{eq:th12-is}
		\tan\vartheta_{12} \ = \ \sqrt{(1+r_{\Delta})(1\pm r_{\nu}/s_{e1}^2)}~.
	\end{align}
	which is only valid for $s_{e1}^2\geq r_{\nu}$ in the $e^{i\phi_1} = - 1$, $e^{i\phi_2} = + 1$ case.
\end{enumerate}
The general behaviour for $s_{12}^2$ is shown in Fig.~\ref{fig:s12s13} (right) as a function of the active-sterile mixing strengths $s_{e1}^2$ and $s_{e2}^2$. At each point in the allowed region the phases $\phi_1$ and $\phi_2$ are calculated according to Eqs.~\eqref{eq:cosphi1} and \eqref{eq:cosphi2} as shown in Fig.~\ref{fig:s12s13} (left), while $\delta$ is set to zero. We see that the sterile-sterile mixing is $\vartheta_{12} = \pi/2$ when $s_{e1}^2\ll r_{\nu}$. As $s_{e1}^2$ approaches $r_{\nu}$ along the canonical seesaw side of the allowed region the mixing angle falls to $\vartheta_{12} = 0$. These two values are physically equivalent, signifying an exchange in the role of the two heavy states as one state becomes decoupled while the other state's mixing strength increases to $r_{\nu}$ or $r_{\nu}/(1+r_{\nu}+r_{\Delta})$. In the inverse seesaw limit the sterile-sterile mixing angle approaches $\vartheta_{12} = \pi/4$, i.e. maximal mixing. 

With the sterile-sterile mixing strength $s_{12}^2$ taken care of, we now return to the neutrino mass generated at one-loop. So far in this section we have worked at tree-level. From gauge invariance of the SM Lagrangian under $\text{SU(2)}_{L}$, it is not possible to write a Majorana mass term $\bar{\nu}_{L}^C\nu_{L}$ for the left-handed neutrino field, and thus the $(1,1)$ element of the neutrino mass matrix is zero. The inclusion of loop corrections will to first loop-order however lead to the appearance of a finite value for the $(1,1)$ element in Eq.~\eqref{eq:generalized},
\begin{align}
\label{eq:generalizedloop}
\mathcal{M}_\nu \ = \  
\begin{pmatrix}
\delta m_{\nu}^{\text{1-loop}} & m_D & 0     \\
m_D\phantom{-} & \mu_R & m_S   \\
0\phantom{-} & m_S  & \mu_S 
\end{pmatrix},
\end{align}
where $\delta m_{\nu}^{\text{1-loop}}$ is given by Eq.~\eqref{eq:oneloopexact}. This will contribute to the mass eigenvalue of the lightest state as
\begin{align}
\label{eq:treeplusloop}
	m_\nu \ = \ m_\nu^\text{tree} + \delta m_\nu^\text{1-loop},
\end{align}
where $m_\nu^\text{tree} = -\mu_{S}m_{D}^2/m_{S}^2$ is the tree-level mass from the diagonalisation of the mass matrix Eq.~\eqref{eq:generalized} as discussed in Sec.~\ref{sec:model-setup}. When using $m_{\nu}$ from now on we assume that it is the physical mass as measured by an experiment, including both the tree-level and one-loop contributions.

\begin{figure}[t!]
	\centering
	\includegraphics[width=0.49\textwidth]{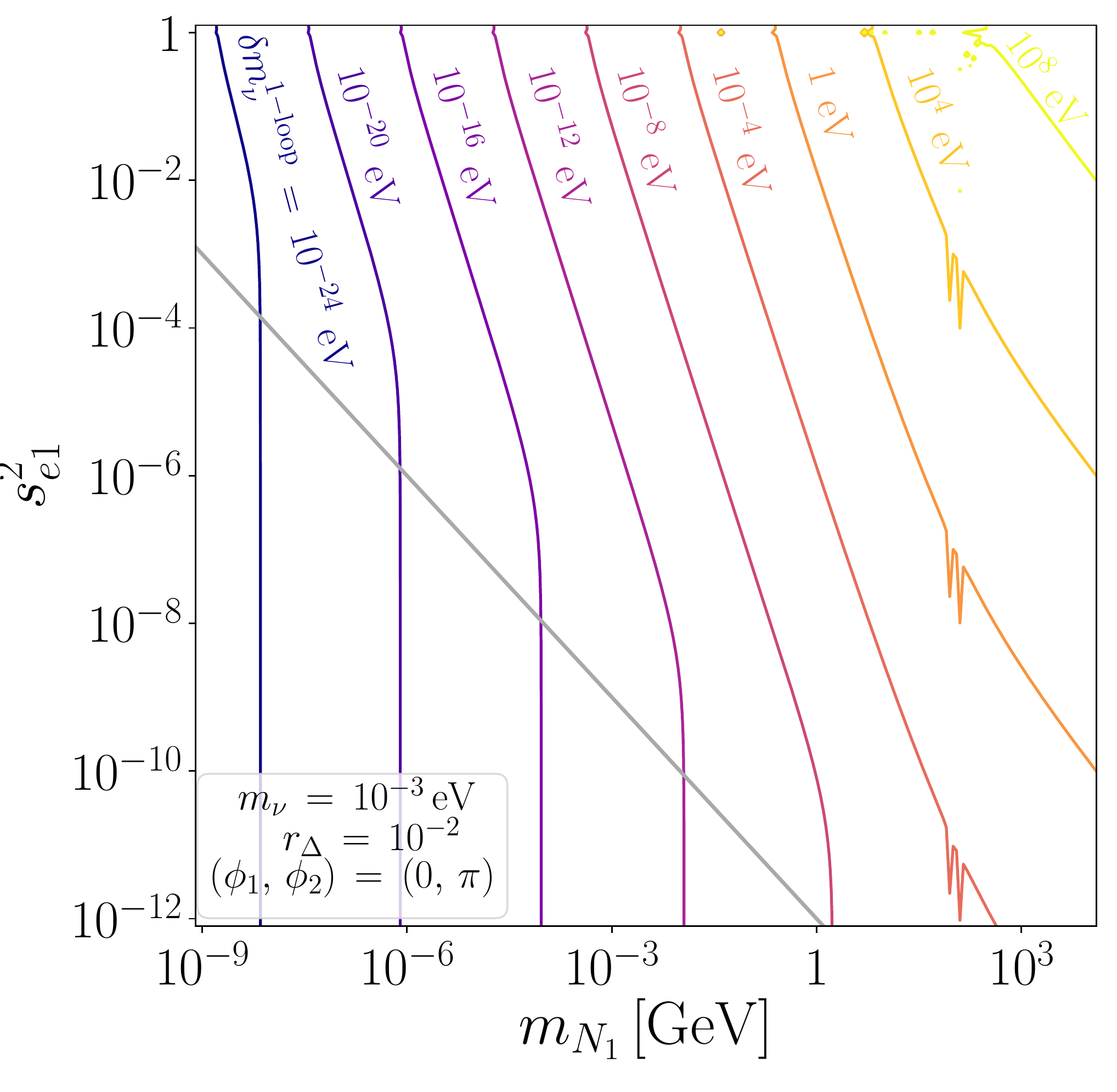}
	\includegraphics[width=0.49\textwidth]{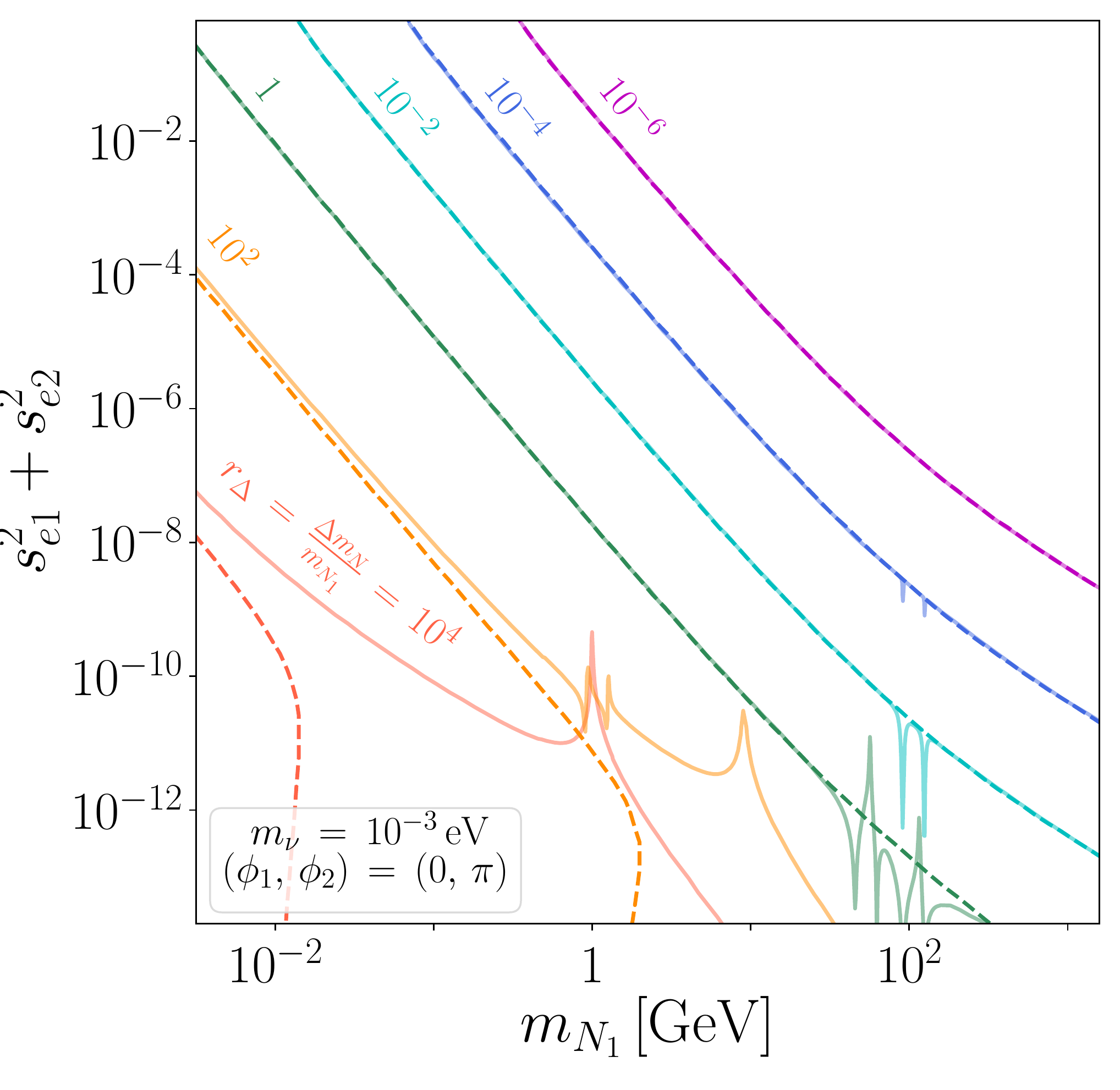}
	\caption{{\it Left:} Absolute magnitude of the one-loop neutrino mass contribution $|\delta m^{\mathrm{1-loop}}_{\nu}|$ as a function of the lighter sterile mass $m_{N_{1}}$ and mixing strength $s_{e1}^2$ for indicated values of the other parameters. The canonical seesaw case with $s^2_{e1}=r_{\nu}$ is indicated by the diagonal grey line. {\it Right:} Maximally allowed value of $s_{e1}^2+s_{e2}^2$ from the condition $|\delta m^{\mathrm{1-loop}}_{\nu}| < 0.1 m_\nu$, as a function of $m_{N_1}$  for different values of the heavy neutrino splitting ratio $r_{\Delta} = \frac{\Delta m_N}{m_{N_1}}$. Solid lines are found by using the exact formula Eq.~\eqref{eq:oneloopexact}, while the dashed lines use this same formula but in the limit $\mu_{R,S}\ll m_{S}$, given by Ref.~\cite{Dev:2012sg}.}
	\label{fig:numass-loop}
\end{figure}

In Fig.~\ref{fig:numass-loop} (left), we plot the exact formula for $\delta m_\nu^\text{1-loop}$ in Eq.~\eqref{eq:oneloopexact} as a function of the heavy neutrino mass $m_{N_1}$ and the mixing strength $s_{e1}^2$. The parameters $m_{\nu}$, $r_{\Delta}$, $\phi_1$ and $\phi_2$ (for $\delta = 0$) are fixed as indicated in the figure, while $s_{e2}^2$ and $s_{12}^2$ are calculated according to Eqs.~\eqref{eq:realandimagpart} and \eqref{eq:s12squared}, respectively. Specifically the tree-level mass and the relative heavy neutrino splitting are given for the benchmark values $m_\nu = 10^{-3}$~eV and $r_{\Delta}=10^{-2}$, while the Majorana phases are chosen such that the scenario is located on the right edge of the allowed parameter space in Fig.~\ref{fig:s12s13} (left). We also plot the \textquoteleft seesaw' line $s_{e1}^2=r_{\nu}=m_{\nu}/m_{N_{1}}$ in grey. Below this line $s_{e2}^2$ will tend to the constant value $r_{\nu}/(1+r_{\nu}+r_{\Delta})\approx r_{\nu}$, while $s_{12}^2$ tends towards $\pi/2$. Above this line is the inverse seesaw limit with $s_{e2}^2=s_{e1}^2/(1+r_{\Delta})\approx s_{e1}^2$ and $s_{12}^2=\pi/4$. This plot demonstrates the strong dependence of $|\delta m_{\nu}^{\text{1-loop}}|$ on the model parameters. For large $m_{N_{1}}$, we can already see that the one-loop corrections are dangerously large as a consequence of the comparatively large splitting between the heavy states $\Delta m_{N} = r_{\Delta} m_{N_{1}}$. Looking at the approximate loop formula in Eq.~\eqref{eq:oneloopapprox} and recalling that $m_{D}$, $m_{S}$, $\mu_{R}$ contain terms proportional to $m_{N_{1}}$ (when written in terms of the mass-basis parameters and mixing angles), the strong dependence on $m_{N_1}$ is not surprising because $\delta m_{\nu}^{\text{1-loop}}$ naively scales as $m_{N_1}^3\ln(m_{N_{1}})$ for  $m_{N_{1}}<m_{Z,H}$ and as $m_{N_1}\ln(m_{N_{1}})$ for $m_{N_{1}}>m_{Z,H}$. The two discontinuities in Fig.~\ref{fig:numass-loop} occur at $m_{N_1} = m_{Z}$ and $m_{N_{1}}=m_{H}$, i.e. when the one-loop contributions are enhanced.

As stated before, in this work we will maintain the assumption that the loop corrections to the light neutrino mass are sub-dominant, i.e. we assume that the neutrinos largely acquire their masses via the tree-level seesaw mechanism. A reasonable requirement that $|\delta m_\nu^\text{1-loop}| < 0.1m_{\nu}$ [cf.~Eq.~\eqref{eq:loop-tree}] can subsequently be used to set an upper limit on the active-sterile mixing strengths. This is shown in Fig.~\ref{fig:numass-loop} (right) as a function of the heavy neutrino mass $m_{N_1}$ for different values of $r_{\Delta} = \Delta m_N/m_{N_1}$. It can be seen that as the relative splitting $r_{\Delta}$ becomes smaller, the associated upper limit on the mixing strength becomes weaker. The solid and dashed lines correspond to the upper limit derived from the exact formula Eq.~\eqref{eq:oneloopexact} and the approximation Eq.~\eqref{eq:oneloopapprox}, respectively. It can be seen that the exact and approximate upper limits diverge for small $m_{N_{1}}$ and $s_{e1}^2$ -- this is because $\mu_{R,S}\ll m_{S}$ no longer holds in this particular region of the parameter space. 

\begin{figure}[t!]
	\centering
	\includegraphics[width=0.55\textwidth]{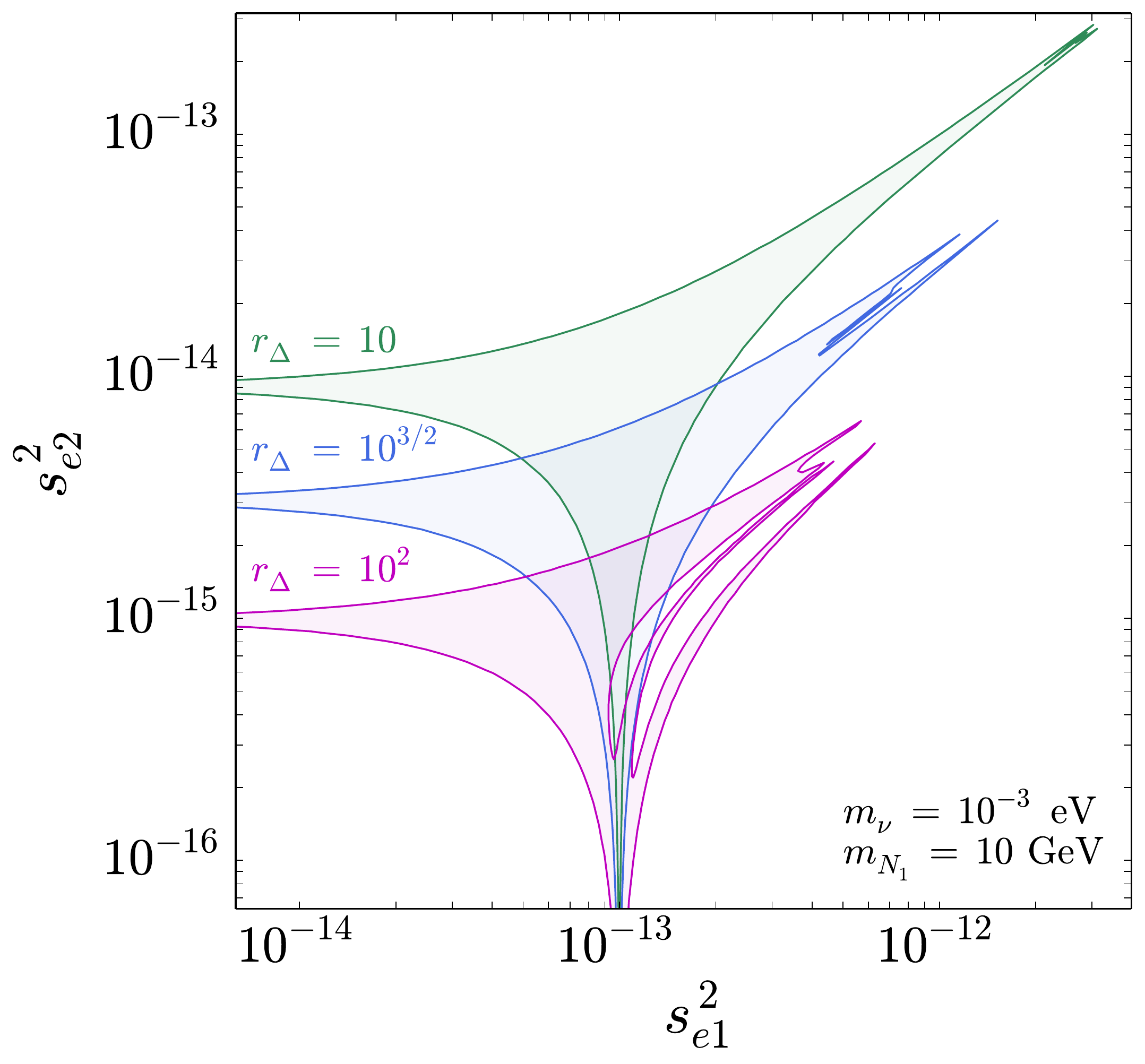}
	\caption{Modified allowed regions for the active-sterile mixing strengths satisfying the tree-level constraint $(\mathcal{M}_{\nu})_{11}=0$ and the condition for the one-loop contribution to be small, $|\delta m_{\nu}^{\mathrm{1-loop}}|<0.1 m_{\nu}$.}
	\label{fig:s12s13-loop}
\end{figure}

In Fig.~\ref{fig:s12s13-loop} we again plot the region satisfying the tree-level constraint $(\mathcal{M}_{\nu})_{11} = 0 $ in Eq.~\eqref{eq:cond11}, but now also exclude the region not satisfying the $|\delta m_\nu^\text{1-loop}| < 0.1m_{\nu}$ loop requirement for different values of the relative splitting $r_{\Delta}$. It can be seen that as $r_{\Delta}$ increases the allowed region is reduced, excluding much of the inverse seesaw region. It is worth mentioning that in order to see this effect around $s_{e1}^2\sim r_{\nu}$ requires large relative splittings, otherwise the loop requirement only excludes much larger mixings strengths $s_{e1}^2\approx s_{e2}^2$ in the inverse seesaw limit. While combining the constraints $(\mathcal{M}_{\nu})_{11} = 0$ and $|\delta m_\nu^\text{1-loop}| < 0.1m_{\nu}$ is true to first order, it breaks down when $|\delta m_\nu^\text{1-loop}| $ becomes large. An exact treatment would of course need to combine the conditions $(\mathcal{M}_{\nu})_{11} = \delta m^{\text{1-loop}}_{\nu}$ and $|\delta m_\nu^\text{1-loop}| < 0.1m_{\nu}$. Consequently, 
\begin{align}
\label{eq:cond11loop}
(\mathcal{M}_\nu)_{11} \ = \ \delta m^{\text{1-loop}}_{\nu} \quad\Rightarrow\quad
c_{e1}^2 c_{e2}^2 \,\frac{m^{\text{tree}}_\nu}{m_{N_1}} 
+ s_{e1}^2 c_{e2}^2 \, e^{i\phi_1}
+ s_{e2}^2 \,\frac{m_{N_2}}{m_{N_1}} \,e^{i\phi'_2} \ = \ \delta m^{\text{1-loop}}_{\nu}\,,
\end{align} 
where we take the neutrino mass on the LHS to be the tree-level mass to first approximation. Writing $m_{\nu}^{\text{tree}}=m_{\nu}-\delta   m^{\text{1-loop}}_{\nu}$ via Eq.~\eqref{eq:treeplusloop}, Eq.~\eqref{eq:cond11loop} can be rearranged as before to solve for $s_{e2}^2$ and $\cos\phi'_2$, but now as a function of the loop mass. Paradoxically, $s_{e2}^2$ and $\phi'_2$ are themselves required to evaluate the loop mass in Eq.~\eqref{eq:oneloopexact} as a function of $m_{N_{1}}$ and $s_{e1}^2$. Inserting the new expressions for $s_{e2}^2$ and $\phi'_2$, the loop mass can be evaluated iteratively by first setting $(\mathcal{M}_{\nu})_{11} = 0$ and then re-inserting each new value back into the one-loop formula. We find that the difference between the initial (setting $(\mathcal{M}_{\nu})_{11} = 0$) and iterated loop mass is negligibly small when the initial loop mass satisfies $|\delta m_\nu^\text{1-loop}| < 0.1m_{\nu}$. When the initial loop mass is larger this iterative approach is strictly no longer valid, but we assume that is viable up to $|\delta m_\nu^\text{1-loop}| \sim 0.1m_{\nu}$. This should then not significantly affect the upper bounds on $s_{e1}^2$ derived from the loop condition. In other words, we keep the constraints derived using $(\mathcal{M}_{\nu})_{11} \approx 0$ and $|\delta m_\nu^\text{1-loop}| < 0.1m_{\nu}$. 

%

%%%%%%%%%%%%%%%%%%%%%%%%%%%%%%%%%%%%%%%%%%%%%%%%%%%%%%%%%%%%%%%%%%%%%%%%%%%%%%%%%
\section{Constraints on heavy sterile neutrinos}
\label{sec:searches}
%%%%%%%%%%%%%%%%%%%%%%%%%%%%%%%%%%%%%%%%%%%%%%%%%%%%%%%%%%%%%%%%%%%%%%%%%%%%%%%%%

In this section we will summarise the results of experimental searches for sterile neutrinos and hence constraints on the active-sterile mixing $|V_{\ell N}|^2$ over the sterile neutrino mass range $m_N\in [1 ~\mathrm{eV},\: 10 ~\mathrm{TeV}]$. For lighter masses it becomes possible for one of the sterile states to form a \textit{quasi-Dirac} state with the active state. A large portion of this parameter space is constrained by solar neutrino oscillations~\cite{Cirelli:2004cz,Donini:2011jh}. For heavier masses $m_{N}\gtrsim 10$ TeV, sterile neutrinos can generate the light active neutrino masses via the conventional seesaw mechanism. These neutrinos, however, are not kinematically accessible to direct searches. The constraints from existing searches and observations in the $m_N-|V_{e N}|^2$ parameter space are shown in Fig.~\ref{fig:current_constraints} by various shaded regions, whereas Fig.~\ref{fig:future_constraints} illustrates the sensitivity of expected future experiments and observations. As our ultimate focus is on a comparison with constraints from $0\nu\beta\beta$ decay in Sec.~\ref{sec:0vbb}, we focus on the first generation mixing element $|V_{e N}|^2$. However, for the sake of completion and future reference, we also compile and update the constraints on $|V_{\mu N}|^2$ and $|V_{\tau N}|^2$ in Appendix~\ref{sec:app}. 
For earlier summary plots showing a partial list of these constraints, see e.g. Refs.~\cite{Atre:2009rg, Deppisch:2015qwa, deGouvea:2015euy, Chrzaszcz:2019inj}. Most limits shown in the plots were derived assuming a single heavy neutrino. For small splitting, the limit is applicable on $|V_{eN_{1}}|^2+|V_{eN_{2}}|^2$; for large splitting the limits apply separately for each species.

\subsection{High-energy collider searches}
Heavy states are produced in charged-current (CC) and neutral-current (NC) processes through their admixture with the active states, and thus their decay products can be searched for at high-energy colliders which copiously produce $W$ and $Z$ bosons. For sufficiently small mixing angles, the macroscopic decay- ength of the heavy neutrinos can result in displaced vertices with distinct detector signatures. We consider the following searches (keywords in bold match the corresponding regions in Figs. \ref{fig:current_constraints} and \ref{fig:future_constraints}):
\begin{figure}[t!]
	\centering
	\includegraphics[width=\textwidth]{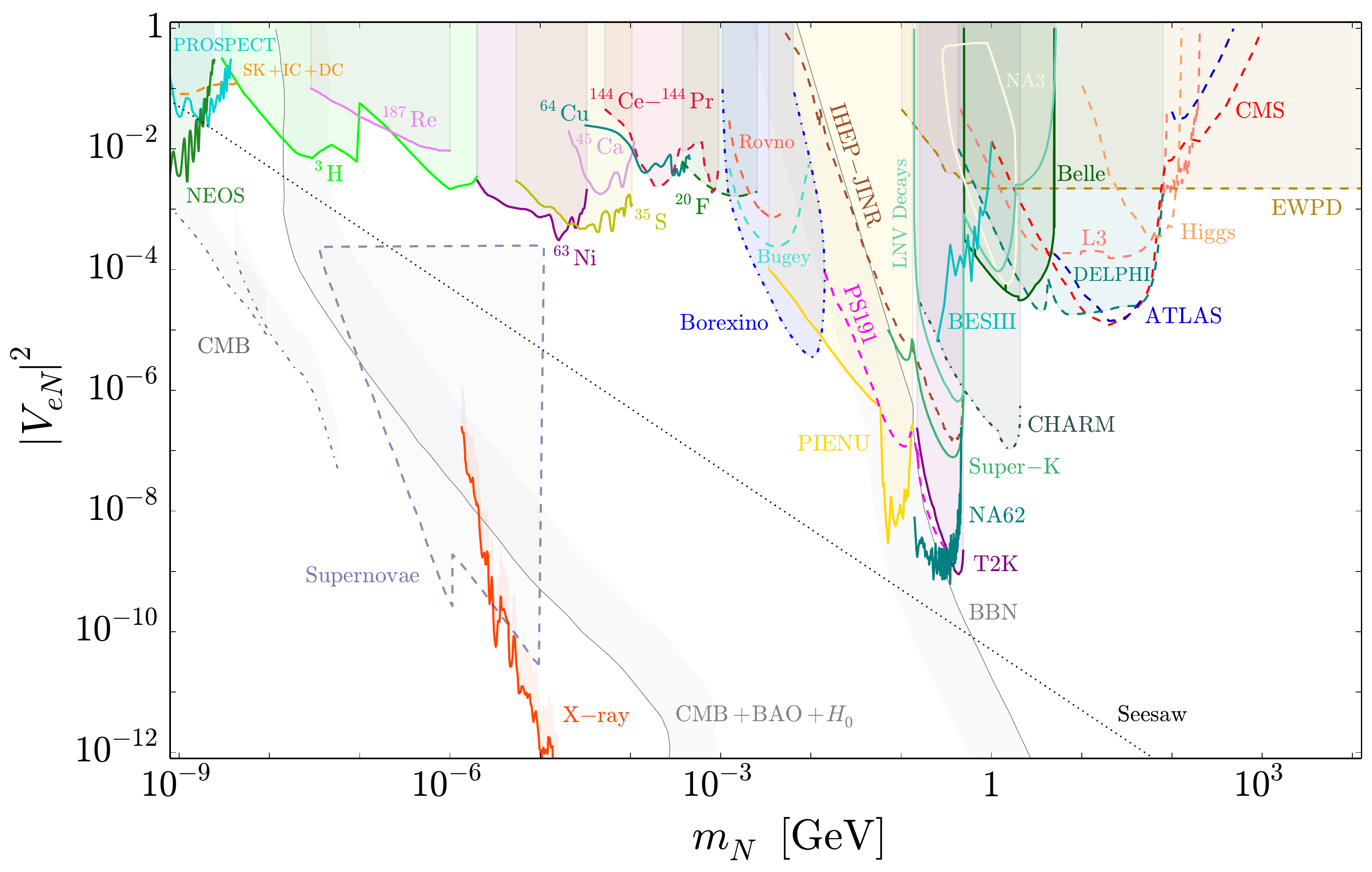}
	\caption{Constraints on the mass $m_N$ of the sterile neutrino and its squared mixing $|V_{e N}|^2$ with the electron neutrino. The shaded regions are excluded by the searches and observations indicated and discussed in Sec.~\ref{sec:searches}. The diagonal line labelled \textquoteleft Seesaw' indicates the canonical seesaw relation $|V_{eN}|^2=m_{\nu}/m_{N}$ with $m_{\nu} = 0.05$ eV.}
	\label{fig:current_constraints}
\end{figure} 
\begin{itemize}	
	\item The LHC collaborations \search{ATLAS} and \search{CMS} have searched for $N$ production and decay through various channels. Both have recently searched for decays of $W$-produced $N$ to three charged leptons, $W^{\pm}\rightarrow\ell^{\pm}N, ~N\rightarrow\ell^{\pm}\ell^{\mp}\nu_{\ell}$ ($\ell = e, \mu$), either in the LNC or LNV mode. ATLAS used the prompt final state of three isolated leptons and no opposite-charge same-flavour lepton pairs (LNV channel) to reject Drell-Yan, $W$ + jets and $t\bar{t}$ backgrounds. CMS broadened the search to the LNC channel with a sensitivity to displaced decays. The analyses impose the limits $|V_{e N}|^2,\,|V_{\mu N}|^2 < 10^{-5}-10^{-4}$ over the mass range $5 ~\mathrm{GeV} < m_{N} < 50~\mathrm{GeV}$~\cite{Aad:2019kiz,Sirunyan:2018mtv}. ATLAS and CMS have also conducted searches for the LNV same-sign dilepton + jets channel, $W^{\pm}\rightarrow\ell^{\pm}N, ~N\rightarrow\ell^{\pm}jj$~\cite{Aad:2015xaa,Sirunyan:2018xiv}. Above the $Z$ boson mass limits can be improved in future by ATLAS and CMS during the high luminosity ($\mathcal{L} = 3~\mathrm{ab}^{-1}$) LHC phase (\search{HL-LHC}) and by a future $\sqrt{s} = 27$ or $100$ TeV Future Circular Collider (\search{FCC-hh})~\cite{Benedikt:2018csr,Pascoli:2018heg}. Around the \search{Higgs} mass, limits can also be set from the SM Higgs decay to sterile neutrinos~\cite{Das:2017zjc}.
	
	\item In the future, \search{ATLAS}, \search{CMS} and \search{LHCb} are expected to probe smaller $|V_{\ell N}|^2$ through displaced vertex searches. For a given mixing, $m_{N}$ must lie in a specific range in order to avoid $N$ decaying promptly or outside the detector. The best projected limit is $|V_{eN}|^2,\,|V_{\mu N}|^2 \lesssim 10^{-9} $ for $m_{N} \approx 30$ GeV~\cite{Aad:2019kiz}.
	
	\item At the LEP collider, the collaborations \search{L3}~\cite{Adriani:1992pq, Achard:2001qv} and \search{DELPHI}~\cite{Abreu:1996pa} searched for $N$ produced through on-shell $Z$ production, $e^{+}e^{-} \rightarrow Z \rightarrow N\nu_{\ell}$, followed by the decays $N \rightarrow \ell^{\mp} W^{\pm}$, $N \rightarrow \nu_{\ell}Z$ and $N \rightarrow \nu_{\ell}H$. Using $N \rightarrow e^{\mp}W^{\pm}$ and $W^{\pm} \rightarrow jj$, L3 enforced a limit of $|V_{e N}|^2 < 10^{-4}$ in the range $5~\mathrm{GeV}<m_{N}<80~\mathrm{GeV}$. This was reduced to $|V_{e N}|^2 < 10^{-5}$ by an improved DELPHI analysis. At a future linear electron-electron collider such as the \search{ILC}~\cite{Banerjee:2015gca}, for a benchmark $\sqrt{s} = 500$ GeV and $\mathcal{L} = 100~ \mathrm{fb}^{-1}$ limits may be improved to $|V_{e N}|^2 < 10^{-4}$ above the $Z$ mass. At a proposed Compact Linear Collider (\search{CLIC}), for $\sqrt{s} = 3$ TeV and $\mathcal{L} = 1~ \mathrm{ab}^{-1}$ limits are $|V_{e N}|^2 \lesssim 10^{-5}-10^{-4}$ for $600~\mathrm{GeV} < m_{N} < 2.3~\mathrm{TeV}$~\cite{Chakraborty:2018khw,Das:2018usr}. Furthermore, a future \search{FCC-ee} collider, acting as a powerful $e^{+}e^{-}\rightarrow Z$ factory and exploiting low backgrounds in displaced vertex searches, can improve the sensitivity drastically; down to $|V_{e N}|^2 \lesssim 10^{-11}$ for $m_{N}\approx 50$ GeV~\cite{Blondel:2014bra}. At the ILC it may also be possible to distinguish LNC and LNV $W^{\pm}$ exchange channels between the $e^{+}e^{-}$ pair by measuring the asymmetry of the outgoing lepton pseudorapidity distribution~\cite{Hernandez:2018cgc}. Finally, the proposed Large Hadron-Electron Collider (\search{LHeC}) LHC upgrade may also provide competitive constraints for $m_{N}>m_{Z}$~\cite{Mondal:2016kof,Das:2018usr,Antusch:2019eiz}. An overview of proposed collider sensitivities is given in Ref.~\cite{Antusch:2016ejd}.
	
	\item Proposed detectors placed near existing LHC interaction points have been designed specifically to search for displaced vertex signatures. These include \search{AL3X}~\cite{Dercks:2018wum}, \search{CODEX-b}~\cite{Gligorov:2017nwh}, \search{FASER2}~\cite{Feng:2017uoz}, \search{MATHUSLA}~\cite{Chou:2016lxi} and the  MoEDAL experiment's MAPP detector~\cite{Frank:2019pgk}. In Fig.~\ref{fig:future_constraints}, we show the expected sensitivity of AL3X, FASER--2~\cite{Kling:2018wct} and MATHUSLA~\cite{Curtin:2018mvb} for illustration. The best projected limits of MATHUSLA are $|V_{eN}|^2,\,|V_{\mu N}|^2 \lesssim 10^{-9}$ for $1~\mathrm{GeV}<m_{N}<2~\mathrm{GeV}$, while AL3X and FASER--2 are slightly less stringent but extend to higher $m_{N}$.
	
\end{itemize}

\begin{figure}[t!]
	\centering
	\includegraphics[width=\textwidth]{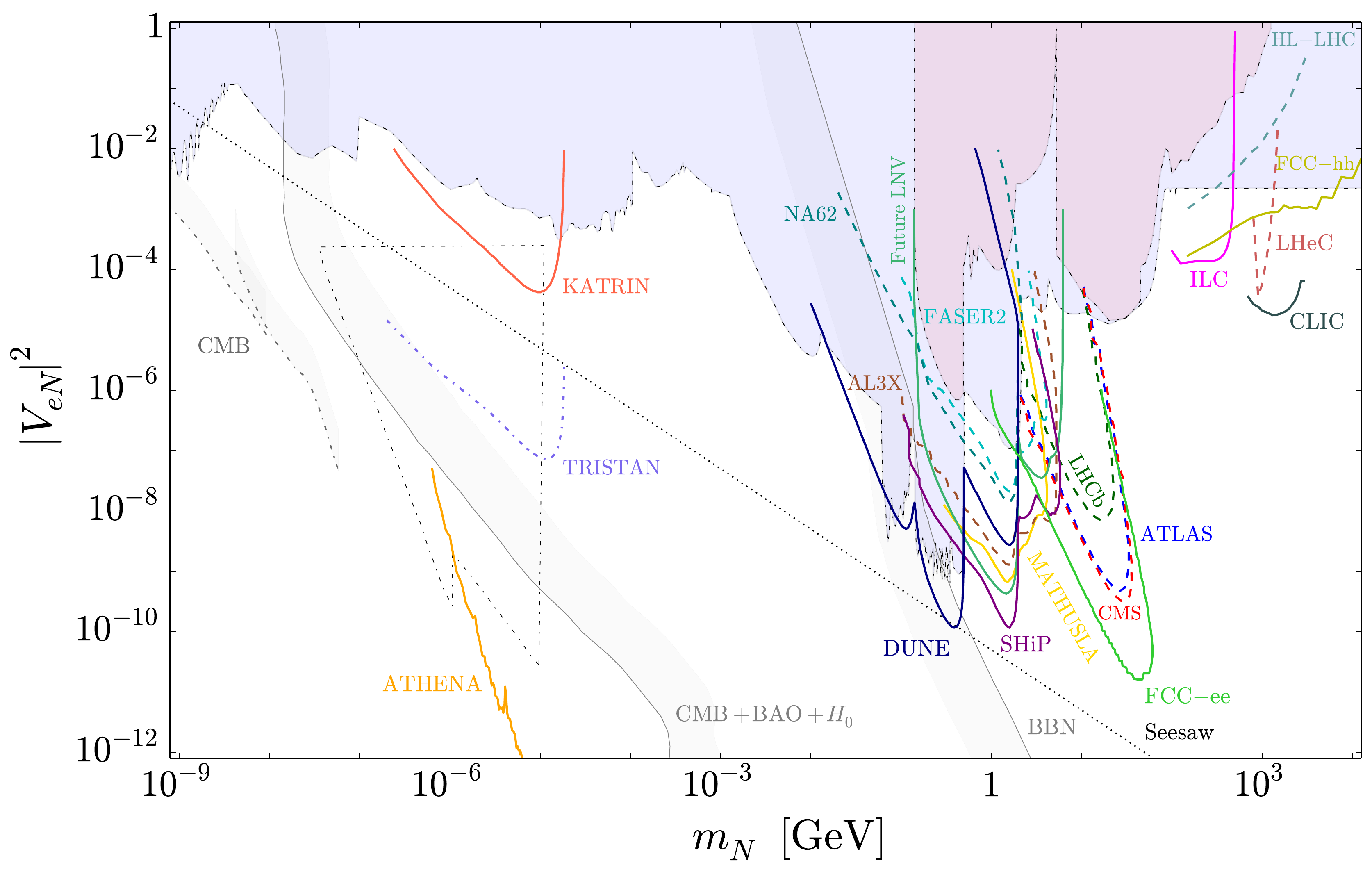}
	\caption{As Fig.~\ref{fig:current_constraints}, but showing the expected sensitivity of future searches and observations. The shaded blue region indicates the parameter space already excluded from current searches as detailed in Fig.~\ref{fig:current_constraints}. The shaded red region contained therein further details current limits from searches for LNV signals.}
	\label{fig:future_constraints}
\end{figure} 

\subsection{On the LNV signal at colliders}
\label{sec:LNVcolliders}

As for the LNV signature at colliders, in a natural seesaw scenario with approximate lepton number conservation, the LNV amplitude for the on-shell production of heavy neutrinos at average four-momentum squared $\bar{s}=(m_{N_1}^2+m_{N_2}^2)/2$ can be written as~\cite{Bray:2007ru, Dev:2013wba} 
\begin{align}
\label{ampLNV}
\mathcal{A}_\text{LNV}(\bar s) 
\ = \ V_{\ell N}^2\frac{2\Delta m_N}{\Delta m_N^2 + \Gamma_N^2} 
+ \mathcal{O}\left(\frac{\Delta m_N}{m_N}\right),
\end{align}
for $\Delta m_N\lesssim \Gamma_N$, i.e. for a small mass splitting $|\Delta m_N| = |m_{N_2} - m_{N_1}|$ between the heavy neutrinos compared to their average decay width $\Gamma_N\equiv (\Gamma_{N_1} + \Gamma_{N_2})/2$. Thus, the LNV amplitude in Eq.~\eqref{ampLNV} will be suppressed by the small mass splitting, except for the case $\Delta m_N\simeq \Gamma_N$ when it can be resonantly enhanced~\cite{Pilaftsis:1997dr, Bray:2007ru}.

For the $5 - 50 ~ \mathrm{GeV}$ range of sterile neutrino masses probed by the ATLAS and CMS same-sign trilepton and dilepton + jets analyses, the total sterile neutrino decay width, if decays only takes to place to SM leptonic and hadronic degrees of freedom, is given by
\begin{align}
\Gamma_{N} & \ = \ \sum_{\ell }a_{\ell}(m_{N})\,|V_{\ell N}|^2\,,
\end{align}
where the complete expressions for the factors $a_{\ell}(m_{N})$ are given in Refs.~\cite{Atre:2009rg,Helo:2010cw}. The factors $a_{\ell}(m_N)$ include the contributions from two-body semi-leptonic and three-body leptonic decays, and are approximately given by
\begin{align}
a_{\ell}(m_N) \ \approx \ N^{\mathrm{2-body}}\cdot\Gamma^{\mathrm{2-body}}+N^{\mathrm{3-body}}\cdot\Gamma^{\mathrm{3-body}}\,,
\end{align}
where $N^{\mathrm{2-body}}$ and $N^{\mathrm{3-body}}$ are the number of decay channels open for each decay topology. $\Gamma^{\mathrm{2-body}}$ and $\Gamma^{\mathrm{3-body}}$ are given roughly by
\begin{align}
\Gamma^{\mathrm{2-body}} \ \sim \ \frac{G_{F}^2f_{M}^2m_{N}^3}{5\pi}\,, \qquad \Gamma^{\mathrm{3-body}}\ \sim \ \frac{G_{F}^2m_{N}^5}{200\pi^3}\,,
\end{align}
where $f_{M}$ is the order of magnitude of the meson decay constants \cite{Atre:2009rg}. For $m_{N} \approx 50$ GeV all three-body leptonic decays and two-body semi-leptonic decays to pseudoscalar mesons ($\pi^0$, $\eta$, $\eta'$, $\eta_c$, $\eta_b$, $\pi^\pm$, $K^{\pm}$, $D^{\pm}$, $D_{s}^{\pm}$, $B^{\pm}$, $B_{c}^{\pm}$) and vector mesons ($\rho^0$, $\omega$, $\phi$, $J/\psi$, $\Upsilon(4S)$, $K^{*0}$, $D^{*0}$, $B^{*0}$, $B^{*0}_s$, $\rho^{\pm}$, $K^{*\pm}$, $D^{*\pm}$, $D^{*\pm}_s$, $B^{*\pm}$, $B_{c}^{*\pm}$) are open, and so the total decay width (for $|V_{\mu N}|^2 = |V_{\tau N}|^2 = 0 $) is approximately
\begin{align} \label{eq:approxwidth}
\Gamma_{N} & \ \sim \ \left(30 \cdot  \Gamma^{\mathrm{2-body}}+10\cdot\Gamma^{\mathrm{3-body}}\right)|V_{eN}|^2 
%\nonumber\\& \ \sim \  (30\cdot10^{-8}+10\cdot 10^{-5})~|V_{e N}|^2 ~\mathrm{GeV} 
\ \sim \ 10^{-4}~|V_{e N}|^2 ~\mathrm{GeV}\,.
\end{align}
For small splittings, e.g.~$r_{\Delta} = 10^{-4}$ and hence $\Delta m_{N} \approx 5 ~\mathrm{MeV}$ for $m_{N} \approx 50 ~\mathrm{GeV}$, and the $|V_{eN}|^2 \sim 10^{-5}$ mixing probed by the LNV analyses, Eq.~\eqref{eq:approxwidth} implies that $\Gamma_N/\Delta m_{N} \sim 10^{-6}$. Collider searches specifically looking for an LNV signal in Fig.~\ref{fig:future_constraints} are therefore still valid for this splitting and splittings down to $r_{\Delta} \sim 10^{-10}$. As will be discussed later, this is important for the comparison with $0\nu\beta\beta$ decay in this mass range. We finally note that the analysis of Ref. \cite{Drewes:2019byd} gives an estimate for the regions of the $m_{N}-|V_{\ell N}|^2$ parameter space where the ratio $R_{\ell\ell}$ in Eq.~\eqref{eq:Rll} is less than or greater than a third. Comparing with Fig. 1 of that work, we again confirm that LNV signals searched for by colliders below the electroweak scale remain unsuppressed, particularly for $\Delta m_{N}$ of order the light neutrino mass splittings (motivated by naturalness).

\subsection{Meson decays and beam-dump experiments}

At the intensity frontier $N$ can be produced abundantly in beam-dump experiments and through various meson decays. We consider the following limits:

\begin{itemize}
	
	\item The TRIUMF \textbf{PIENU} experiment~\cite{PhysRevD.84.052002} conducted a search for $N$ produced in pion decays at rest. Utilising the helicity suppression of the $\pi \rightarrow e \nu$ decay channel in comparison to $\pi \rightarrow \mu \nu$ channel, the presence of $N$ induces extra peaks in the lower positron energy region. Improving on previous results limited by the background $\mu^{+}\rightarrow e^{+}\nu_{e}\bar{\nu}_{\mu}$, the collaboration set limits at the level of $|V_{\ell N}|^2 \lesssim 10^{-8}$ in the range $60~\mathrm{MeV} < m_{N} < 129 ~\mathrm{MeV}$~\cite{Aguilar-Arevalo:2017vlf,Bryman:2019ssi,Bryman:2019bjg}.
	
	\item The \search{NA62} experiment~\cite{CortinaGil:2017mqf} used a secondary $75 ~\mathrm{GeV}$ hadron beam containing a fraction of kaons, and has been able to probe the decays $K^{+}\rightarrow  \ell^{+}N$ $(\ell = e, \mu)$. For small $|V_{\ell N}|^2$ the $N$ decay length is much longer than the 156 m detector volume and the process is characterised by a single detected track -- a positive signal is a peak in the missing mass distribution. Limits $|V_{eN}|^2,~|V_{\mu N}|^2< 10^{-8}-10^{-9} $ in the range $170~\mathrm{MeV} < m_{N}< 450~\mathrm{MeV}$ (up to the kaon mass) have been made. In future NA62 will be converted to a beam-dump configuration and will be able to probe hadronic decays to $N$, followed by $N$ decays, up to the $D$ meson mass. The projected sensitivity is $|V_{eN}|^2,~|V_{\mu N}|^2< 10^{-8} $ for $1~\mathrm{GeV} < m_{N} < 2 ~\mathrm{GeV}$~\cite{Drewes:2018gkc}. A recent recalculation of the impact of sterile neutrinos on kaon decays was conducted in Ref.~\cite{Abada:2016plb}.
	
	\item The \search{Belle} experiment~\cite{Abashian:2000cg} was a $B$ factory that extended the peak search method to higher energies -- using $B\overline{B}$ pairs collected at the $\Upsilon(4S)$ resonance, the decay mode $B \rightarrow (X) \ell N $, with $X$ a charmed meson $D^{(*)}$ or light meson, could be followed by $N\rightarrow \ell \pi $ ($\ell = e, \mu$). Constraints were made between the $K$ and $B$ meson masses and at best were $|V_{\ell N}|^2 \lesssim 3\times 10^{-5}$ for $m_{N}\approx 2$ GeV~\cite{Liventsev:2013zz}.

	\item The \search{NA3} experiment~\cite{Badier:1986xz} collided  a secondary $300 ~\mathrm{GeV}$ $\pi^{-}$ beam with an iron absorber, producing hadronic states which subsequently decayed to leptonic, semi-leptonic or fully hadronic final states. $N$ decays producing leptonic or semi-leptonic final states could be produced from the decays of $\pi$, $K$, $D$ and $B$ mesons. NA3 was most sensitive up to the $D$ meson mass, setting limits of $|V_{eN}|^2,~|V_{\mu N}|^2< 10^{-4} $ for $1~\mathrm{GeV} < m_{N} < 2 ~\mathrm{GeV}$.
	
	\item Accelerated neutrino beam experiments have conducted a variety of parallel searches. The \search{CHARM}~\cite{Bergsma:1985is,Vilain:1994vg} and \search{PS191}~\cite{Bernardi:1987ek} experiments and the \search{IHEP-JINR} neutrino detector~\cite{Baranov:1992vq,Aguilar-Arevalo:2017vlf} searched for a small fraction of $N$ in a predominantly $\nu_{\mu}$ beam. The beams were produced by colliding a primary beam of protons with an iron or copper fixed target, with the hadronic products decaying as $\pi/K/D \rightarrow \ell \nu (N)$ ($\ell = e,\mu$). If sufficiently massive, $N$ may decay before reaching the detector via the channel $N\rightarrow \ell^{+}\ell^{-}\nu_{\ell}$. CHARM also used a wide-band neutrino beam to constrain the NC process $\nu_{\mu}n(p)\rightarrow N X$ followed by $N\rightarrow \mu X$ within the detector. IHEP-JINR and PS191 provide constraints (down to $|V_{eN}|^2 \lesssim 10^{-7}$ and $|V_{eN}|^2 \lesssim 10^{-9}$, respectively) up to the kaon mass. CHARM provides constraints up to the $D$ meson mass, at best $|V_{eN}|^2, |V_{\mu N}|^2 \lesssim 10^{-7}$ for $1~\mathrm{GeV}< m_{N}<2~\mathrm{GeV}$.
	
	\item The long-baseline neutrino oscillation experiment \search{T2K}~\cite{Abe:2019kgx} searched for an admixture of $N$ in its initial neutrino beam flux, produced by colliding 30 GeV protons with a graphite target at J-PARC. Daughter $K^{\pm}$ of a given charge are focused and decay via $K \rightarrow \ell \nu (N)$. The off-axis near-detector at a baseline of 280 m searched for $N$ decays via the channel $N\rightarrow \ell\pi$, improving on the constraints made by PS191. In future, the near detector of the oscillation experiment \search{DUNE} will be highly sensitive for $m_{N}$ up to the $D_{s}$ meson mass~\cite{Krasnov:2019kdc,Ballett:2019bgd}.
	
	\item The future beam-dump experiment \search{SHiP}~\cite{Alekhin:2015byh} is purposely designed to look for exotic long-lived particles. Utilising a 400 GeV proton beam from the CERN Super Proton Synchrotron, it is expected to be sensitive to sterile neutrinos with $m_{N}$ up to the $B_{c}$ meson mass ($\sim 6$ GeV). In a benchmark scenario where the electron-sterile coupling dominates, SHiP is expected to be sensitive down to $|V_{eN}|^2 \lesssim 10^{-10}$ for $m_{N}\approx 1.6$ GeV~\cite{SHiP:2018xqw}.
	
	\item In parallel with collider searches it is possible to look for \search{LNV Decays} of tau leptons and pseudoscalar mesons as discussed in Refs.~\cite{Kovalenko:2009td,Atre:2009rg,Helo:2010cw,Abada:2017jjx}. One issue is that if the LNV process is mediated by the light neutrinos the amplitude is proportional to and suppressed by the small $m_{\nu}^2$, while if mediated by heavy neutrinos it is suppressed by $1/m_{N}$ and $|V_{\ell N}|^2$. LNV decay widths however can be strongly enhanced if a sterile state is produced on-shell. The sensitivity of NA62 to three-body LNV light mesons decays ($K^{+}\rightarrow\ell^{+}\ell^{'+}\pi^{-}$), BESIII to charmed meson decays ($D^{+}/D_{s}^{+}\rightarrow\ell^{+}\ell^{'+}\pi^{-}/K^{-}$) and BaBar, Belle and LHCb for $B$ meson decays ($B^{+}\rightarrow\ell^{+}\ell^{'+}\pi^{-}/K^{-}/D^{-}/\rho^{-}/K^{*-}$) for $\ell,\,\ell' = e,\,\mu$ were estimated most recently by Ref.~\cite{Abada:2017jjx}. The \search{BESIII} has also conducted its own analysis on the ($D^{+}\rightarrow\ell^{+}\ell^{'+}\pi^{-}/K^{-}$) decay channel~\cite{Ablikim:2019gvd}. Finally, the \search{Future LNV} decay sensitivities of NA62, LHCb, Belle-II, MATHUSLA, SHiP and FCC-ee have been explored in Ref.~\cite{Chun:2019nwi}
\end{itemize}

\subsection{Beta decays and nuclear processes}
Active neutrinos are produced in the $\beta$-decays of unstable isotopes and in nuclear fission processes. Heavy sterile neutrinos can also be produced via the active-sterile mixing if the sterile mass is smaller than the energy release ($Q$-value) of the relevant nuclear process. The production of a sterile state results in a distortion or \textquoteleft kink' in the $\beta$-decay spectrum and associated Kurie plot. It is also possible for the sterile state to decay before detection. We include the following searches:

\begin{itemize}
	\item Heavy neutrinos produced in $\beta$-decays significantly alter the energy spectrum of the emitted $\beta$ electron. In order to be kinematically accessible the sterile neutrino mass must be smaller than the $Q$-value of the process, $m_{N}< Q_{\beta}$. If this is satisfied and the sterile states are sufficiently more massive than the active states, the $\beta$-decay spectrum becomes the incoherent sum
	\begin{align}
	\frac{d\Gamma}{dE} \ = \ \left(1-\sum_{i}|V_{ eN_i}|^2\right)\,\frac{d\Gamma}{dE}(m^2_{\beta})+\sum_{i}|V_{e N}|^2\,\frac{d\Gamma}{dE}(m^2_{N_i})\,\Theta(Q_{\beta}-m_{N_i})\,,
	\end{align}
	where $m^2_{\beta} = \sum_{k}|U_{e k}|^2m^2_k$ is the usual scale probed by $\beta$-decay \cite{Shrock:1980vy}. This expression can give rise to multiple kinks in the spectrum of relative size $|V_{e N_i}|^2$ and at energies $E_{\mathrm{kink}}=Q_{\beta} - m_{N_i}$.
	Such an effect for a single sterile neutrino has been probed for a variety of isotopes with a range of different $Q$-values, and therefore sensitive to different $m_{N}$. Isotopes include \search{\boldmath $^3$H}~\cite{Hiddemann:1995ce,Kraus:2012he,Belesev:2013cba,Abdurashitov:2017kka}, \search{\boldmath $^{20}$F}~\cite{PhysRevC.27.1175}, \search{\boldmath $^{35}$S}~\cite{Holzschuh:2000nj}, \search{\boldmath $^{45}$Ca}~\cite{Derbin:1997ut}, \search{\boldmath $^{63}$Ni}~\cite{Holzschuh:1999vy}, \search{\boldmath $^{64}$Cu}~\cite{Schreckenbach:1983cg}, \search{\boldmath $^{144}$Ce--$^{144}$Pr}~\cite{Derbin2018} and \search{\boldmath $^{187}$Re}~\cite{PhysRevLett.86.1978}.
	In the future, strongly improved limits by the operating tritium $\beta$-decay experiment \search{KATRIN} and the proposed \search{TRISTAN} upgrade are expected~\cite{Mertens:2018vuu}. The capability of the PROJECT 8 experiment, which uses the alternative method of cyclotron radiation emission spectroscopy, has also been briefly explored~\cite{Adhikari:2016bei}. 
	
	\item Reactor neutrino experiments are sensitive to sterile neutrinos with masses in the range $1~\mathrm{MeV}< m_N < 10~\mathrm{MeV}$. At these masses it is possible for $N$ to decay within the detector via the channel $N\rightarrow e^{+}e^{-}\nu$. Limits have been set by searches at the \search{Rovno}~\cite{Derbin:1993wy} and \search{Bugey}~\cite{PhysRevD.52.1343} reactors. This effect was also searched for by the \search{Borexino} experiment~\cite{PhysRevD.88.072010}, which detected neutrinos produced by the fission processes in the Sun -- heavy neutrinos with masses up to $14$~MeV can be produced in the decay of $^8$B. Borexino has set the best limits; $|V_{eN}|^2\lesssim 10^{-6} - 10^{-5}$ for $m_{N}\sim 10 ~\mathrm{MeV}$.
	
\end{itemize}

\subsection{Active-sterile neutrino oscillations}

Persistent anomalies in neutrino oscillation experiments are still providing intriguing hints for the existence of an additional mass squared splitting $\Delta m^2 \sim 1 ~\mathrm{eV}^2$ to the well-established solar and atmospheric mass squared splittings $\Delta m_{\mathrm{sol}}^2 = 7.55\times 10^{-5} ~\mathrm{eV}^2$ and $|\Delta m_{\mathrm{atm}}^2| = 2.5 \times 10^{-3}~\mathrm{eV}^2$, respectively~\cite{deSalas:2017kay,Diaz:2019fwt}. This apparent splitting has been established in the measurement of multiple oscillation processes, including $\nu_{\mu}\rightarrow\nu_e$ accelerator neutrino appearance (LSND anomaly), $\bar{\nu}_{e}\rightarrow\bar{\nu}_e$ reactor neutrino disappearance (reactor anomaly) and the $\nu_{e}\rightarrow\nu_e$ disappearance of $^{37}$Ar and $^{51}$Cr electron capture decay neutrinos (gallium anomaly). Attempts have been made to fit the data to models with additional eV-scale neutrinos, e.g. (3+1) and (3+2) phenomenological models. While recent reactor experiments such as \search{DANSS}~\cite{Alekseev:2018efk} and \search{NEOS}~\cite{Ko:2016owz} have improved the statistical significance of an additional eV-scale sterile state, when combined with the $\nu_e$ appearance data of MiniBooNE they are in strong tension with the observed $\nu_{\mu}\rightarrow\nu_{\mu}$ accelerator neutrino disappearance of the MINOS, NO$\nu$A and IceCube experiments.

In the context of the single-generation simplification of this work we interpret the mass squared splitting to be $\Delta m_{41}^2 = m^2_{N}-m_{\nu}^2$. As we are focused on the electron-sterile coupling it is thus only the $\nu_e\rightarrow\nu_e$ and $\bar{\nu}_e\rightarrow\bar{\nu}_e$ experiments sensitive to $\sin^2 2\theta_{ee} \approx 4|V_{eN}|^2$ that are relevant. For sub-eV sterile neutrino masses the \search{Daya Bay}~\cite{An:2016luf}, \search{KamLAND}~\cite{Cirelli:2004cz} and upcoming \search{JUNO}~\cite{Berryman:2019nvr} experiments can probe the mixing down to $|V_{eN}|^2 \lesssim 10^{-3}$. However it should be noted that if one wants to fit the solar and atmospheric mass splittings in a minimal (3+1) or (3+2) extension, solar data excludes the region $10^{-9} ~\mathrm{eV}< m_{N}< 0.6~\mathrm{eV}$~\cite{Donini:2011jh,Donini:2012tt}. Below this region is the pseudo-Dirac scenario and above the mini-seesaw extending to the conventional high-scale seesaw. Light sterile neutrinos can be implemented in the context of an inverse seesaw as considered in Ref.~\cite{Barry:2011wb, Dev:2012bd, Abada:2017ieq}.

In Figs.~\ref{fig:current_constraints} and \ref{fig:future_constraints} we therefore start $m_{N}$ at the eV-scale. Above this the DANSS and NEOS experiments provide limits down to $|V_{eN}|^2 \lesssim 10^{-2}$ (as both exclusions are similar, Fig.~\ref{fig:current_constraints} shows NEOS only) while the operating \search{PROSPECT}~\cite{Ashenfelter:2018iov} experiment provides constraints up to $m_{N} = \sqrt{\Delta m_{41}^2+m^2_{\nu}}\sim 5$~eV. Over the same mass range Super-Kamiokande, IceCube and DeepCore (\search{SK+IC+DC}) provide complementary limits~\cite{Dentler:2018sju}. We note that the above limits are from oscillations conserving total lepton number. While it is in principle possible to observe LNV in oscillations, this requires new physics beyond sterile neutrinos such as right-handed currents \cite{Bolton:2019wta}.

\subsection{Electroweak precision data and other indirect laboratory constraints}
Any mixing between active and sterile neutrinos necessarily induces non-unitarity effects among the active neutrinos visible in CC and NC processes~\cite{Abada:2007ux,Fernandez-Martinez:2016lgt, Blennow:2016jkn}. This is most easily parametrised by a non-unitary light neutrino mixing matrix
\begin{align}
	U_\nu = (1 - \eta)\cdot U_\text{PMNS},
\end{align}
where the matrix $\eta$ measures deviations from unitarity. The elements of $\eta$ are given in a generic seesaw model by $\sqrt{2|\eta_{\ell\ell'}|} = \sum_{i}\sqrt{V_{\ell N_{i}}V^*_{\ell' N_{i}}}$ and alter electroweak precision data (\search{EWPD}) observables. These include leptonic and hadronic measurements of the weak mixing angle $s^2_W$,
the $W$ boson mass $m_{W}$, ratios of fermionic $Z$ boson decay rates $R_{l}$, $R_{c}$, $R_{b}$ and $\sigma_{\mathrm{had}}^{0}$, the $Z$ invisible decay width $\Gamma^{\mathrm{inv}}_{Z}$ and ratios of leptonic weak decays testing EW universality $R^{\pi}_{\ell\ell'}$, $R^{W}_{\ell\ell'}$, $R^{K}_{\ell\ell'}$ and $R^{l}_{\ell\ell'}$. Furthermore, by modifying $G_{F}$, the non-unitarity of $U_{\nu}$ impacts the values of CKM mixing matrix elements extracted from experiments. Numerous weak decays have been used to pin down the CKM elements $V_{ud}$, $V_{us}$, $V_{ub}$ and the unitarity condition $|V_{ud}|^2 + |V_{us}|^2 + |V_{ub}|^2 = 1$. Assuming a single sterile state coupling to just the first generation, all of these measurements enforce a constant bound of $\sqrt{2|\eta_{ee}|} = |V_{eN}| < 0.050$ for $m_N \gtrsim 1$ GeV~\cite{delAguila:2008pw, Akhmedov:2013hec, deBlas:2013gla, Antusch:2014woa, Blennow:2016jkn, Flieger:2019eor}.

Another indirect measurement of $\eta_{\ell\ell'}$ and hence different combinations of the active-sterile mixings comes from the non-observation of lepton flavour violating (LFV) processes $\ell\rightarrow \ell'\gamma$ and $\mu-e$ conversion in nuclei \cite{Deppisch:2012vj}. 
%Rates and branching ratios for these processes are given in Appendix A. 
Due to the different flavours of charged leptons involved in these processes, active-sterile mixings to at least two active generations are required. For the purpose of our single active generation picture we may convert the constraint on $|V_{eN} V^*_{\mu N}|$ obtained from the limits $\mathrm{Br}(\mu\rightarrow e\gamma) < 4.2 \times 10^{-13}$ \cite{Tanabashi:2018oca} and $R^{\mathrm{Ti}}_{\mu\rightarrow e}< 4.3\times 10^{-12}$~\cite{Dohmen:1993mp} to a constraint in the $m_{N} - |V_{eN}|^2$ parameter space by assuming $|V_{\mu N}| = |V_{eN}|$. We find $|V_{eN}|^2 \lesssim 10^{-3}$ for $m_{N}\approx 10$ GeV, improving to $|V_{eN}|^2 \lesssim 10^{-5}$ for $100~\mathrm{GeV} \lesssim m_{N}\lesssim  10~\mathrm{TeV}$. In making the assumption $|V_{\mu N}| = |V_{eN}|$ however, the constraints in the $m_{N} - |V_{\mu N}|^2$ parameter space equally apply for $|V_{e N}|^2$. For clarity and consistency we therefore do not show the LFV constraints in Fig.~\ref{fig:current_constraints}.

\subsection{Cosmological and astrophysical constraints}

The presence of sterile states with masses $m_{N}$ and mixings $|V_{\ell N}|^2$ (and therefore predicted production rates, decay lengths and active-sterile oscillations) can have drastic consequences on early-universe observables, and have been explored extensively in the literature~\cite{Abazajian:2012ys}. These include the abundances of light nuclei formed during Big Bang Nucleosynthesis (BBN), temperature anisotropies in the Cosmic Microwave Background (CMB) radiation and the large-scale clustering of galaxies. Deviations from the standard smooth, isotropic background evolution and perturbations around this background impose severe constraints, especially for sterile states with masses $m_{N}\lesssim 100$~MeV. The limits are however highly sensitive to the production and decay mechanism of the sterile state and can be relaxed in certain models. For the purpose of comparison we consider the following scenarios:

\begin{itemize}
	\item Sterile neutrinos with masses $m_{N}\lesssim 1 ~\mathrm{GeV}$ can be sufficiently long-lived to disrupt the standard formation of light nuclei $^{4}$He, D, $^{3}$He and $^{7}$Li during \search{BBN}~\cite{Boyarsky:2009ix,Ruchayskiy:2012si}. For larger masses the decay products from the accessible two-body and three-body decays have enough time to thermalise with the plasma. 
	For decay times $\tau\gtrsim 1$ s occuring below $T \lesssim 1$ MeV, i.e. roughly after $\nu/N$ -- $e^{\pm}$ decoupling and the onset of BBN, both the modified background expansion due to the presence of non-relativistic $N$ and the altered weak processes $n+\nu\leftrightarrow p+e^{-}$ and $p+\bar{\nu}\leftrightarrow n+e^{+}$ involving non-thermal decay product neutrinos lead to modified nuclei abundances. The condition $\tau = \Gamma_{N}^{-1} \gtrsim 1$ s naively translates to a lower limit of $|V_{eN}|^2\gtrsim 10^{-11}  \,(\mathrm{GeV}/m_{N})^5$ for $N\rightarrow 3\nu$, $N\rightarrow\nu e^{+}e^{-}$ and the sub-dominant radiative decay $N\rightarrow \nu\gamma$. Above the pion mass threshold the already considerably less stringent constraints are made even weaker by including the decays $N\rightarrow \nu\pi^{0}$ and $N\rightarrow e^{\pm}\pi^{\mp}$.
	
	\item Sterile neutrinos decaying at later times (with $\tau \lesssim t_{\mathrm{rec}} \approx 1.2\times 10^{13}$ s) to non-thermally distributed active neutrinos can modify the amount of dark radiation measured (beyond the usual value including active neutrino oscillations, $N_{\mathrm{eff}} \simeq 3.046$) at recombination, $\Delta N_{\mathrm{eff}}$.  Decays after recombination but before the current epoch ($t_{\mathrm{rec}} \lesssim \tau \lesssim t_0 \approx 4.3\times 10^{17}$ s) can also be important. Useful probes of these effects on the smooth, isotropic expansion history include the CMB shift parameter $R_{\mathrm{CMB}}$ (related to the position of the first acoustic peak in the CMB temperature power spectrum), the first peak of Baryon Acoustic Oscillation (BAO) sound waves imprinted on the large-scale distribution of galaxies and finally the value of the Hubble parameter $H(z)$ inferred from Type Ia supernova, BAO and Lyman-$\alpha$ survey data. These exclude values of $m_{N}$ and $|V_{eN}|^2$ corresponding to lifetimes up to $t_0$, where the condition that $N$ does not make up more than the observed matter density $\Omega_{\mathrm{sterile}}<\Omega_{\mathrm{DM}} \approx 0.12 \,h^{-2}$ and thus overcloses the Universe also applies. This constraint can naturally be evaded in exotic models~\cite{Bezrukov:2009th,Nemevsek:2012cd,El-Zant:2013nta,Biswas:2018iny}, for example those that inject additional entropy and dilute the dark matter (DM) energy density. We indicate the combined constraints from Ref.~\cite{Vincent:2014rja} in Fig.~\ref{fig:current_constraints} as \search{\boldmath CMB+BAO+$H_0$}.
	
	\item Sterile neutrinos with masses $1 ~\mathrm{keV} \lesssim m_{N} \lesssim 100 ~\mathrm{keV} $ can avoid the global constraints above if the active-sterile mixing is sufficiently small, i.e $|V_{eN}|^2\lesssim 10^{-10} - 10^{-8}$. With lifetimes longer than the current age of the Universe these sterile states are viable DM candidates if efficiently produced~\cite{Abada:2014zra,Adhikari:2016bei,Abazajian:2017tcc}. Depending on the size of the lepton-antilepton asymmetry $\eta_{L} \equiv n_{L}/n_{\gamma}$, population can occur either through resonant ($\eta_{L} > 10^{6}\,\eta_{b}$) or non-resonant ($\eta_{L} \approx 0 $) active-sterile oscillations. The former (Shi-Fuller mechanism~\cite{Shi:1998km}) is independent of $|V_{\ell N}|^2$ while the latter (Dodelson-Widrow mechanism~\cite{Dodelson:1993je}) requires values of $|V_{\ell N}|^2$ now excluded by the global constraints. If DM is composed entirely of keV sterile neutrinos their fermionic nature limits the phase space density of DM-rich dwarf galaxies and imposes the Tremaine-Gunn bound, $m_{N}\gtrsim 0.4$ keV. It is also possible to search for anomalous \search{X-ray} lines from the radiative decays $N\rightarrow\nu\gamma$ in the diffuse X-ray background and from DM-rich astrophysical objects. An intriguing signal at $E \simeq 3.55 $ keV implying a sterile neutrino with a mass of 7.1 keV has continued to persist in observations of stacked galaxy clusters~\cite{Bulbul:2014sua}, the Perseus galaxy cluster and Andromeda M31 galaxy~\cite{Boyarsky:2014jta} and the centre bulge of the Milky Way~\cite{Boyarsky:2014ska}. In Fig.~\ref{fig:current_constraints} we include the most recent observations of M31 and the Milky Way by NuSTAR~\cite{Ng:2019gch,Roach:2019ctw}. In Fig.~\ref{fig:future_constraints} we show the slightly improved future sensitivity of \search{ATHENA}~\cite{Neronov:2015kca}. These constraints assume $\Omega_{\mathrm{DM}}=\Omega_{\mathrm{sterile}}$, but can be multiplied by $\Omega_{\mathrm{sterile}}/\Omega_{\mathrm{DM}}$ to account for other DM species~\cite{Vincent:2014rja}.
	
	\item Active-sterile mixings can be excluded for sterile neutrinos in the mass range $10~\mathrm{eV} \lesssim m_{N}\lesssim 10~\mathrm{keV}$ by examining their impact on Type II \search{Supernovae}. Active-sterile neutrino oscillations hinder the standard neutrino reheating of the reflected shock wave which becomes stalled in the first fraction of a second after the core bounce. For the explosion to proceed and additionally produce the observed SN1987A $\bar{\nu}_e$ signal of terrestrial detectors such as Kamioka~\cite{hirata:1987hu} and IMB~\cite{bionta:1987qt}, a certain region of the $m_{N} - |V_{\ell N}|^2$ parameter space must be excluded. Refs.~\cite{Kainulainen:1990bn,Shi:1993ee,Nunokawa:1997ct,Hidaka:2006sg,Hidaka:2007se,Tamborra:2011is,Warren:2014qza} have studied in detail the resonant conversion $\nu_{e}\rightarrow N$ in the dense medium of collapsing stars and the necessary conditions to prevent impeding the supernova explosion. Refs.~\cite{Fuller:2009zz,Raffelt:2011nc,Arguelles:2016uwb} have similarly investigated $\nu_{\mu,\tau}\rightarrow N$ conversions for which the Mikheyev–Smirnov–Wolfenstein resonance conditions are different. An open question is whether the conditions for $r$-process nucleosynthesis to produce heavy elements in the supernova outflows are met in these cases~\cite{Nunokawa:1997ct,Tamborra:2011is}. Lastly, sterile neutrinos that escape supernovae can subsequently decay radiatively via $N\to \nu_e\gamma$ and $N\to \nu_e e^+ e^- \gamma$, producing an excess of gamma rays arriving soon after the detection of the $\nu_e$. The non-observation of such an excess for SN1987A provides a stringent limit in the mass range $1~\mathrm{MeV} \lesssim m_{N} \lesssim 30~\mathrm{MeV}$~\cite{Oberauer:1993yr}. Given the various assumptions and calculational differences of the constraints discussed we show for illustration in Fig.~\ref{fig:current_constraints} the excluded region from Ref.~\cite{Shi:1993ee}.
	
	\item Sufficiently stable and light sterile neutrinos with masses $m_{N}\lesssim 50$ eV can be produced with quasi-thermal temperatures before the decoupling of active neutrinos via active-sterile oscillations~\cite{Adhikari:2016bei,Gariazzo:2015rra,Asaka:2006nq}. While relativistic they contribute themselves towards the extra effective number of light fermionic degrees of freedom $\Delta N_{\mathrm{eff}}$. Once becoming non-relativistic they contribute towards the matter density as $\Omega_{\mathrm{sterile}}\,h^2 = (m_{\mathrm{eff}}^{\mathrm{sterile}}/94.1\,\mathrm{eV})$ while also damping density perturbations below a mass-dependent free-streaming scale. The most simple case of a single sterile neutrino thermalising through oscillations at the active neutrino temperature has $\Delta N_{\mathrm{eff}} = 1$ and $m_{\mathrm{eff}}^{\mathrm{sterile}} \simeq m_{N}$~\cite{Dolgov:2003sg, Cirelli:2004cz,Hannestad:2015tea} which is now likely excluded~\cite{Mirizzi:2013gnd}. The Planck collaboration has made fits of CMB (TT+lowP+lensing+BAO) data to the parameters ($\sum m_{\nu}$, $N_{\mathrm{eff}}$) and ($m_{\mathrm{eff}}^{\mathrm{sterile}}$, $\Delta N_{\mathrm{eff}}$)~\cite{Aghanim:2018eyx}. In Refs.~\cite{Vincent:2014rja} and~\cite{Bridle:2016isd} these constraints are mapped to the ($\Delta m_{41}^2$, $\sin^2{2\theta_{ee}}$) parameter space which we use to plot the grey dot-dashed \search{CMB} constraints in Fig.~\ref{fig:current_constraints}.
\end{itemize}

%%%%%%%%%%%%%%%%%%%%%%%%%%%%%%%%%%%%%%%%%%%%%%%%%%%%%%%%%%%%%%%%%%%%%%%%%%%%%%%%%
\section{Neutrinoless double beta decay}
\label{sec:0vbb}
%%%%%%%%%%%%%%%%%%%%%%%%%%%%%%%%%%%%%%%%%%%%%%%%%%%%%%%%%%%%%%%%%%%%%%%%%%%%%%%%%

In this section we will first review the treatment of $0\nu\beta\beta$ decay in the presence of sterile neutrinos, having previously been covered in detail in the literature in the context of the type-I seesaw~\cite{Mitra:2011qr, LopezPavon:2012zg}, inverse and extended seesaws~\cite{Awasthi:2013ff, Abada:2018qok} and left-right symmetric models~\cite{Tello:2010am, Chakrabortty:2012mh, Barry:2013xxa,Dev:2013vxa, Huang:2013kma,Stefanik:2015twa,Horoi:2015gdv,Deppisch:2017vne,Bolton:2019bou}.

Of particular importance is the dependence of the $0\nu\beta\beta$ decay rate on the sterile neutrino mass $m_{N}$ and the average momentum exchange squared of the process $\langle p^2\rangle$. We will see that if $m^2_{N} > \langle p^2\rangle$ the contribution from a \textquoteleft heavy' sterile neutrino is suppressed by $1/m_{N}$ and $|V_{eN}|^2$. In the limit $m_{N}^2\gg \langle p^2\rangle$ the heavy states are integrated out and $0\nu\beta\beta$ decay becomes a probe of generic short and long-range exchange mechanisms with dimension-7 and above effective operators (depending on the model of interest) at the interaction vertices~\cite{P_s_1999}. If $m_{N}^2 \ll \langle p^2\rangle$ the \textquoteleft light' sterile neutrino contributes much like a light active neutrino. In this case the condition $(\mathcal{M}_{\nu})_{11} =\sum_{i}V_{ei}^2m_{i}= 0$ suppresses the total $0\nu\beta\beta$ decay rate $[T^{0\nu}_{1/2}]^{-1}\propto (\mathcal{M}_{\nu})_{11}$. Multiple sterile states, some with masses above and some below $\langle p^2\rangle$ is an intriguing intermediate scenario. It was observed in Ref.~\cite{LopezPavon:2012zg} that here the \textquoteleft light' sterile neutrino contribution may even dominate over the light active contribution; the necessary and contradictory prerequisites are a large source of LNV and a small loop contribution to the light neutrino masses. This was found to be possible either in an extended seesaw or by having fine-tuned cancellations between generations.

We will also give a broad comparison between the discussed $0\nu\beta\beta$ decay constraints and those from the numerous searches discussed in Sec.~\ref{sec:searches}, particularly where the $0\nu\beta\beta$ decay constraints become relevant ($m_{N} \gtrsim 100 $ keV). One of the most interesting aspects of this comparison is the change of the $0\nu\beta\beta$ decay constraints  as a function of the mass splitting between the heavy states $\Delta m_{N}$. Because $0\nu\beta\beta$ decay is an LNV process we know specifically in the inverse seesaw that it must vanish in the LNC limit $\mu_{R,S}\rightarrow\mathbf{0}$. The LNV matrices $\mu_{R,S}$ also control the splitting between the heavy states, so in the limit $\Delta m_{N}\rightarrow 0$ (the heavy states form a pseudo-Dirac fermion) the $0\nu\beta\beta$ decay limits vanish. Following Sec.~\ref{sec:LNVcolliders}, we will compare this with the suppression of LNV collider and meson decay constraints. No such suppression occurs for the LNC search constraints discussed generally in Sec.~\ref{sec:searches}.

It is also crucial to consider how the sterile neutrino mass splitting $\Delta m_{N}$ affects the interpretation of the direct searches. For example, the analyses of $\beta$-decay kink searches and meson decay peak searches assume a single sterile state and constrain the associated mixing $|V_{eN}|^2$ and mass $m_{N}$. On the other hand, it could be the case that there are two sterile neutrinos with a splitting $\Delta m_{N}$ below the energy resolution of the experiment -- the searches are then sensitive to the sum of mixings $|V_{\ell N_1}|^2+|V_{\ell N_2}|^2$. It is easy to see that, again in the single-generation case, there is a lower limit on this sum from the $(\mathcal{M}_{\nu})_{11} = 0$ condition (or the requirement to produce the observed light neutrino mass $m_{\nu}$),
\begin{align}
|V_{\ell N_1}|^2+|V_{\ell N_2}|^2 \ \approx \ s_{e1}^2+s_{e2}^2 \ = \ s_{e1}^2+\frac{1}{1-\frac{(1+r_{\Delta})\cos\phi_2}{r_{\nu}+(\cos\phi_1-r_{\nu})s_{e1}^2}}\gtrsim \frac{m_{\nu}}{m_{N_1}}\,,
\end{align}
where we assume $r_\Delta \ll 1$. This is qualitatively identical to the discussion of Ref.~\cite{Drewes:2019mhg}, where it is made clear that for any individual mixing $V_{\ell N_{i}}$ it is not possible to impose a lower limit from the seesaw relation because we are free to set $|V_{\ell N_{1}}|^2 = 0$ and $|V_{\ell N_{2}}|^2 = \frac{m_{\nu}}{m_{N_1}}$. The equivalent freedom in the three-generation picture can be for example the choice of orthogonal matrix $\mathcal{R}$ entering the generalised Casas-Ibarra parametrisation~\cite{Casas:2001sr,Donini:2012tt}. If $\Delta m_{N}$ is instead larger than the energy resolution of direct searches, the non-observation of a sterile state excludes regions in both the $m_{N_{1}}-|V_{\ell N_{1}}|^2$ and $m_{N_{2}}-|V_{\ell N_{2}}|^2$ parameter spaces. As direct searches have so far only probed mixing strengths viable in the inverse seesaw region of the parameter space, $|V_{\ell N_{1}}|^2\approx |V_{\ell N_{2}}|^2(1+r_{\Delta})$, the excluded region in $m_{N_{2}}-|V_{\ell N_{2}}|^2$ excludes additional portions of $m_{N_{1}}-|V_{\ell N_{1}}|^2$. In our subsequent Fig.~\ref{fig:largeychange} this is simply represented in the excluded region shifted to smaller $m_{N_{1}}$ and larger $|V_{\ell N_{1}}|^2$ by the factor $(1+r_{\Delta})$.

\subsection{Coherent contribution of light and heavy neutrinos}
\label{sec:0vbbcontribs}
The $0\nu\beta\beta$ decay rate or inverse half-life, taking into account the exchange of both three active and $n_{S}$ sterile neutrinos, can be written as
\begin{align}
\label{eq:halflife1}
\frac{\Gamma^{0\nu\beta\beta}}{\ln{2}}=\frac{1}{T^{0\nu}_{1/2}} 
= G^{0 \nu}  g_A^{4}\,m_p^2\left|\sum_{i=1}^{3} U_{ei}^2 m_{i} \mathcal{M}^{0\nu}(m_{i})+\sum_{i=1}^{n_{S}} V_{eN_{i}}^2 m_{N_i} \mathcal{M}^{0\nu}(m_{N_i})\right|^2,
\end{align}
where $G^{0\nu}$ is a kinematic phase space factor for the outgoing electron pair, $g_A$ the axial coupling strength, $m_{p}$ the proton mass and $\mathcal{M}^{0\nu}(m_i)$ the nuclear matrix element (NME) of the process for an exchanged Majorana neutrino of mass $m_{i}$ \cite{Benes:2005hn}. 

The most recent calculations of $G^{0\nu}$ for relevant $0\nu\beta\beta$ decay isotopes have included effects such as the Coulomb distortion of the electron wave functions due to the finite size of the daughter nucleus and electron screening~\cite{Kotila:2012zza, Stoica:2013lka, Mirea:2014dza}. The NMEs are in principle far more difficult to compute as they encode the non-trivial transition between the initial and final state nuclei in the process. The NMEs entering Eq.~\eqref{eq:halflife1} take the form
\begin{align}
\label{eq:NME}
\mathcal{M}^{0\nu}(m_i) = \frac{1}{m_{p}m_{e}}\frac{R}{g_{A}(0)^2}\int d^3\mathbf{x} \int d^3\mathbf{y}\int \frac{d\mathbf{p}}{2\pi^2}e^{i\mathbf{p}\cdot(\mathbf{x}-\mathbf{y})}\sum_{n}\frac{\braket{F|J^{\mu\dagger}(\mathbf{x})|n}\braket{n|J^{\dagger}_{\mu}(\mathbf{y})|I}}{\omega_{i}(\omega_{i}+\mu)}\,,
\end{align}
where $J^{\mu}$ is the hadronic current, $R$ the nuclear radius and $\omega_{i} = \sqrt{\mathbf{p}^{2}+m_{i}^2} $ the energy of the exchanged neutrino. It is necessary to sum over all possible intermediate nuclear states $n$ between the initial and final states $I$ and $F$ respectively, and $\mu = E_{n}-\frac{1}{2}(E_{I}+E_{F})$ is the relative energy of these virtual states with respect to the average energy of the process. This sum, along with the non-perturbative nature of the hadronic currents, has made the calculation of Eq.~\eqref{eq:NME} extremely difficult, and at present there are still large theoretical uncertainties in computed values. Four common simplifying assumptions are (i) the closure approximation, (ii) the impulse approximation, (iii) $J^{P}=0^{+}$ final nuclear states and (iv) electrons emitted in $s$-wave. (i) assumes that only exchanged neutrino momenta $|\mathbf{p}|$ of similar size to the nucleon-nucleon spacing contribute to the amplitude -- this allows the denominator in Eq.~\eqref{eq:NME} to be pulled out of the sum and removes the contribution of intermediate odd-odd nuclei. (ii) allows the expression of the hadronic current matrix elements in terms of the nucleon-level current form factors associated with the vector ($g_V$), axial-vector ($g_A$), induced weak-magnetic ($g_M$) and induced pseudo-scalar ($g_P$) couplings. As $0\nu\beta\beta$ decay parent and daughter isotopes have even numbers of protons and neutrons, their ground state is always $J^{P}=0^{+}$, while decays to excited states are suppressed, thus justifying the assumption (iii). Finally, $p$-wave emitted electrons are also suppressed and the computation of $G^{0\nu}$ is greatly simplified in the $s$-wave case, as assumed in (iv). 
\begin{table}[t]
	\begin{tabular}{|l|l|l|l|l|}
		\hline
		\multirow{2}{*}{~~~~NME Calculation} & \multicolumn{2}{l|}{~~~~~~~$|\mathcal{M}^{0\nu}_{\nu}|~~(\delta|\mathcal{M}^{0\nu}_{\nu}|)$} & \multicolumn{2}{l|}{~~~~~~~~~$|\mathcal{M}^{0\nu}_{N}|~(\delta|\mathcal{M}^{0\nu}_{N}|)$}  \\ \cline{2-5} 
		&    ~~~~$^{76}\mathrm{Ge}$       &    ~~~$^{136}\mathrm{Xe}$       &    ~~~~$^{76}\mathrm{Ge}$       &    ~~~$^{136}\mathrm{Xe}$    \\ \hline
		~~QRPA T\"{u}bingen~~\cite{Faessler:2014kka} &    ~4.73~(0.18)~    &    ~2.05~(0.20)~\hspace{0.1em}       &    ~318.5~(0.36)~       &  ~168.0~(0.36)~ \\ \hline
		~~QRPA Jyv\"{a}skyl\"{a}~~\cite{Hybvarinen:2015} &     ~5.90~(0.11) ~     &    ~3.21~(0.09)\hspace{0.1em}        &     ~437.5~(0.08)~      &     ~202.3~(0.08)~    \\ \hline
		~~~~~~~~IBM-2~~\cite{Barea:2015zfa}  &     ~4.68~(0.32)~      &    ~3.05~(0.32)~\hspace{0.1em}        &    ~104.0~(0.54)~      &     ~73.0~(0.54)~   \\ \hline
		~~~~~~~~~ISM~~\cite{Blennow:2010th} &    ~2.79~(0.30)~       &    ~2.15~(0.30)~\hspace{0.1em}        &     ~132.7~(0.38)~      &     ~114.9~(0.38)~     \\ \hline
		
	\end{tabular}
	\caption{Light $|\mathcal{M}^{0\nu}_{\nu}|$ and heavy $|\mathcal{M}^{0\nu}_{N}|$ NMEs and associated fractional uncertainties $\delta|\mathcal{M}^{0\nu}_{\nu}|$ and $\delta|\mathcal{M}^{0\nu}_{N}|$ for $^{76}\mathrm{Ge}$ and $^{136}\mathrm{Xe}$ used in this work, taken from QRPA, IBM and ISM calculations in the literature, which are the only available ones that quote both light and heavy neutrino NMEs. When not explicitly given in the reference we estimate the uncertainties from the variation of NMEs with $g_{A}$ and the choice of short-range correlations. }
	\label{tab:NMEs}
\end{table}

A useful interpolating formula for the NMEs can be derived examining the limits of Eq.~\eqref{eq:NME} for the neutrino mass much smaller and much larger than the average momentum exchange,
\begin{align} 
\mathcal{M}^{0\nu}(m_{i}\ll |\mathbf{p}|) 
\ = \ \frac{\mathcal{M}_{\nu}^{0\nu}}{m_p m_e},\qquad \qquad 
\mathcal{M}^{0\nu}(m_{i}\gg |\mathbf{p}|) 
\ = \ \frac{\mathcal{M}_N^{0\nu}}{m_{i}^2}\,,
\end{align}
where $\mathcal{M}_{\nu}^{0\nu}$ and $\mathcal{M}_{N}^{0\nu}$ are dimensionless \textquoteleft light' and \textquoteleft heavy' NMEs respectively. It is possible to write an approximate interpolating formula that includes both of these scaling behaviours,
\begin{align}
\label{eq:interpolate}
\mathcal{M}^{0\nu}(m_{i}) 
\ \approx \ \frac{|\mathcal{M}_N^{0\nu}|}{\langle \mathbf{p}^2\rangle + m_{i}^2},\qquad
\langle  \mathbf{p}^2 \rangle \ = \ åm_p m_e
\left|\frac{\mathcal{M}_N^{0\nu}}{\mathcal{M}_\nu^{0\nu}}\right|,
\end{align}
so that the half-life formula \eqref{eq:halflife1} including sterile states becomes~\cite{Kovalenko:2009td,Faessler:2014kka}
\begin{align}
\label{eq:halflife2}
\frac{1}{T^{0\nu}_{1/2}} 
\ = \ G^{0 \nu}  g_A^{4}m_p^2|\mathcal{M}_{N}^{0\nu}|^2\left|\sum_{i=1}^{3} \frac{U_{ei}^2 m_{i}}{\langle  \mathbf{p}^2\rangle}+\sum_{i=1}^{n_S} \frac{V_{eN_i}^2 m_{N_i}}{\langle  \mathbf{p}^2\rangle + m_{N_i}^2}\right|^2,
\end{align}

Using the above-discussed approximations in Eq.~\eqref{eq:NME} the values of $|\mathcal{M}^{0\nu}_\nu|$ and $|\mathcal{M}^{0\nu}_N|$ have been calculated in a variety of different frameworks. These include the quasiparticle random phase approximation (QRPA) \cite{Faessler:2014kka,Hybvarinen:2015}, interacting boson model (IBM-2) \cite{Barea:2013bz,Barea:2015zfa,Graf:2018ozy} and interacting shell model (ISM) \cite{Blennow:2010th}. A review of these methods as well as their respective strengths and weaknesses is given in Ref.~\cite{Engel:2016xgb}. In Tab. \ref{tab:NMEs} we show the light and heavy NMEs and their associated fractional uncertainties for the $0\nu\beta\beta$ decay isotopes $^{76}$Ge and $^{136}$Xe. The QRPA calculations of the T\"{u}bingen and Jyv\"{a}skyl\"{a} groups and the IBM-2 calculations of the Yale group give NME values for quenched ($g_{A} = 1$) and non-quenched ($g_{A} = 1.269$) values of the axial coupling and also for phenomenological Argonne~\cite{Wiringa:1994wb} and CD-Bonn~\cite{Machleidt:2000ge} forms of the Jastrow potential describing two-nucleon short-range correlations. We use the average of these NME values and take the uncertainty to be half the maximum spread. It was noted in Ref.~\cite{Hybvarinen:2015} that the QRPA Jyv\"{a}skyl\"{a} and IBM-2 Yale heavy NMEs change by a common factor when changing potentials, while for an unknown reason the changes for the QRPA T\"{u}bingen heavy NMEs are significantly different. There are now numerous other computational tools being used for \textit{ab initio} calculations of light NMEs for both light and heavy nuclei, including improved chiral effective field theory~\cite{Machleidt:2011zz}, renormalisation group~\cite{Hergert:2015awm,Cirigliano:2019vdj} and lattice QCD techniques~\cite{Detmold:2018zan}.

\begin{figure}[t!]
	\centering
	\includegraphics[width=0.49\textwidth]{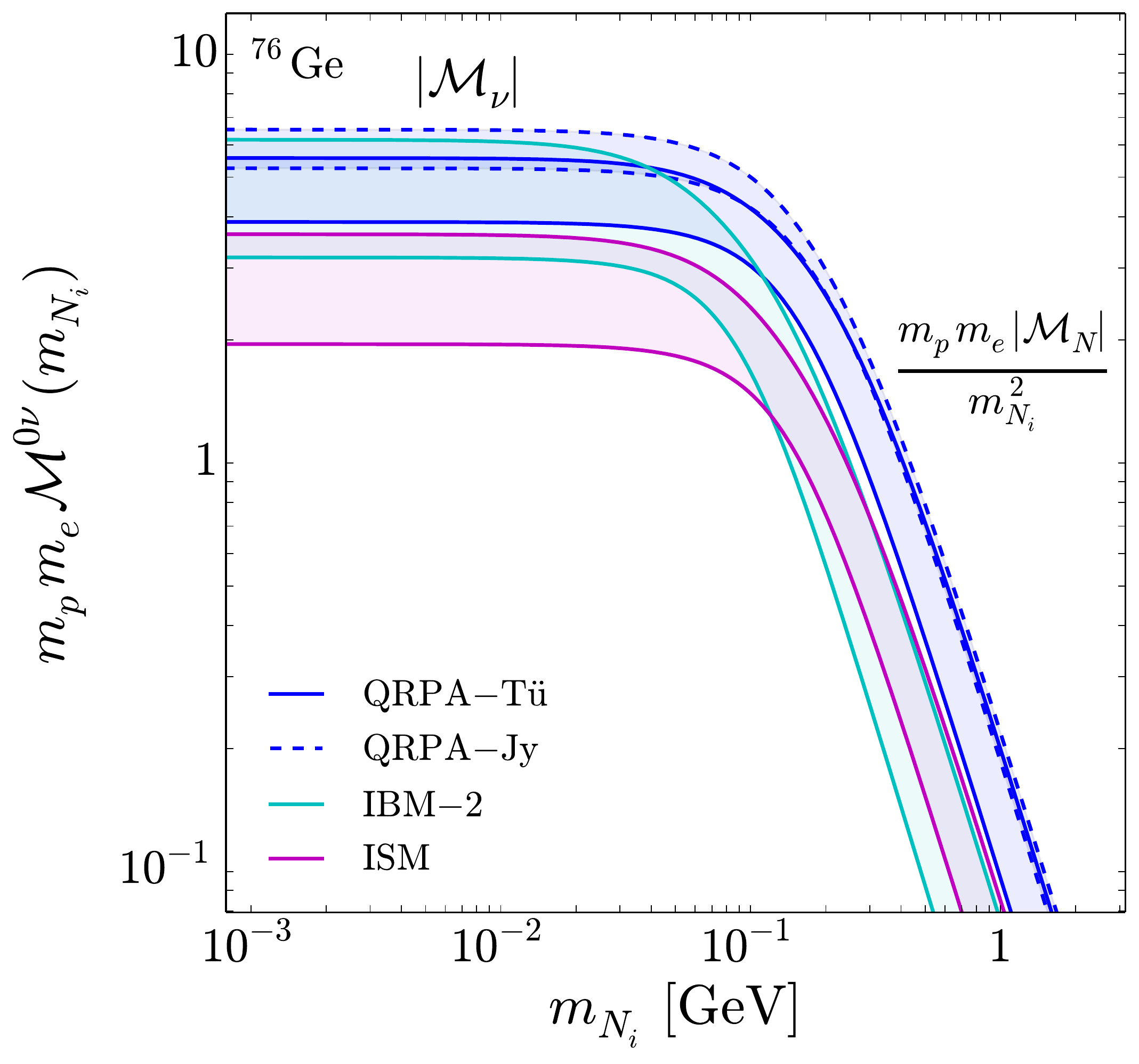}
	\includegraphics[width=0.49\textwidth]{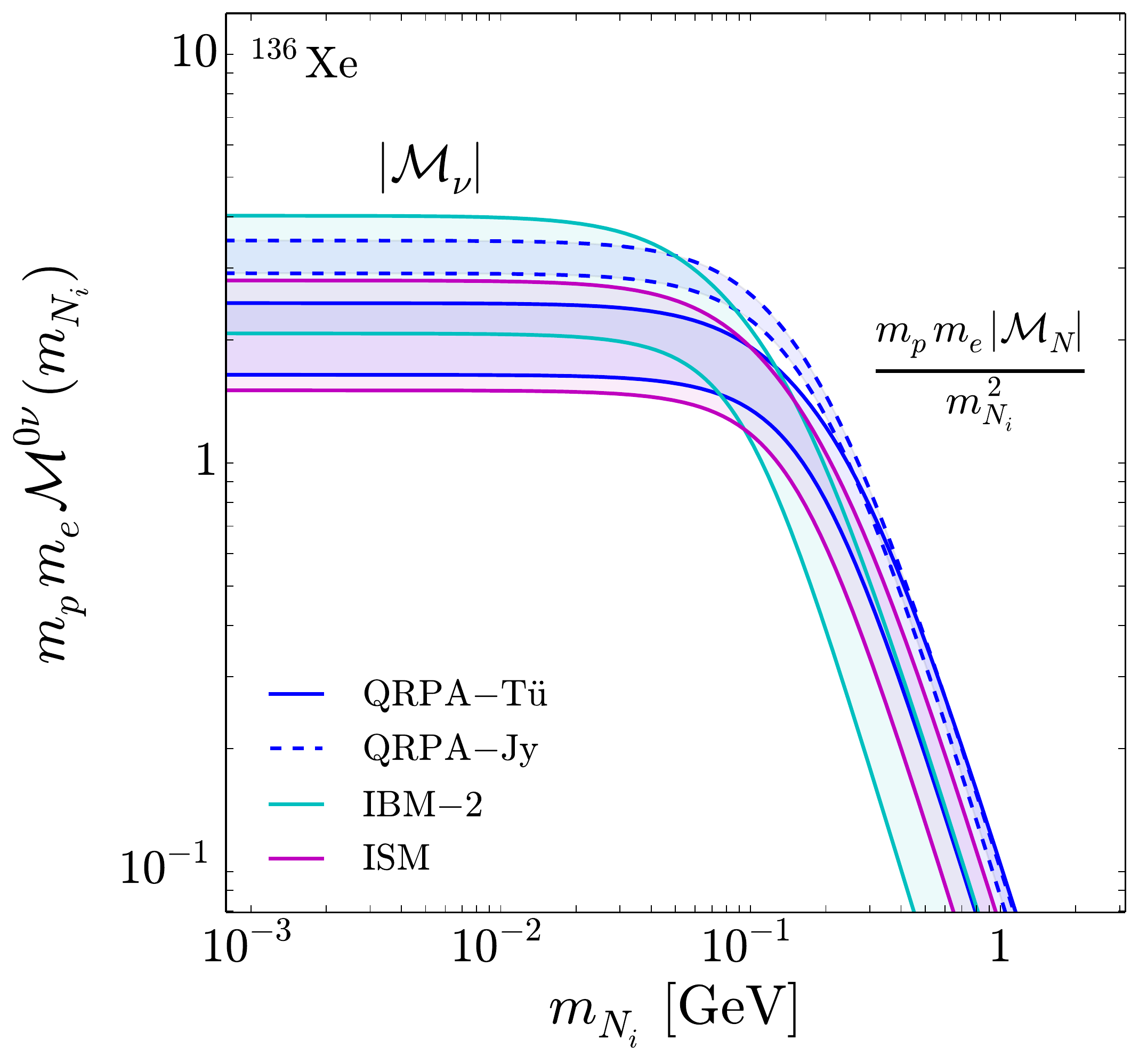}
	\caption{Normalised $0\nu\beta\beta$ decay NMEs for $^{76}$Ge (left) and $^{136}$Xe (right) as a function of the exchanged sterile neutrino mass $m_{N_i}$ using the interpolating formula Eq.~\eqref{eq:interpolate}. We make use of the light and heavy NMEs shown in Table \ref{tab:NMEs}. The bands indicate the NME uncertainties arising from the choice of quenched $g_A$ and short-range correlations. }
	\label{fig:nmeuncertainties} 
\end{figure}

In Fig.~\ref{fig:nmeuncertainties} we plot the $^{76}$Ge and $^{136}$Xe NMEs as a function of the exchanged neutrino mass $m_{N_{i}}$ using the interpolating formula of Eq.~\eqref{eq:interpolate} and the different light and heavy NMEs given in Table \ref{tab:NMEs}. It can clearly be seen that the NMEs are constant below $\langle \mathbf{p}^2\rangle \sim 100 ~\mathrm{MeV}^2$ and suppressed by $1/m_{N_{i}}^2$ above. If all masses are below $\langle p^2\rangle$ we will see that it is instead the seesaw relation suppressing the $0\nu\beta\beta$ decay rate. To plot the uncertainty bands in Fig.~\ref{fig:nmeuncertainties} we propagate the uncertainties of $|\mathcal{M}^{0\nu}_{\nu}|$ and $|\mathcal{M}^{0\nu}_{N}|$ through Eq.~\eqref{eq:NME} as
\begin{align}
\delta{\mathcal{M}^{0\nu}} 
\ = \ \sqrt{\left(\frac{\partial \mathcal{M}^{0\nu}}{\partial 
		|\mathcal{M}_N^{0\nu}|}\right)^2\delta{|\mathcal{M}_N^{0\nu}|}^2 + 
	\left(\frac{\partial \mathcal{M}^{0\nu}}{\partial |\mathcal{M}_\nu^{0\nu}|}\right)^2
	\delta{|\mathcal{M}_\nu^{0\nu}|}^2}\,.
\end{align}
It can be seen that the largest uncertainties are in the IBM-2 NMEs -- for illustrative purposes and to give conservative estimates we use these NMEs in the following discussion.

In our single-generation simplification the summation appearing in the interpolating formula is approximately
\begin{align}
\label{eq:onubb1}
\frac{m_\nu}{\langle \mathbf{p}^2\rangle} 
+ \frac{e^{i\phi_1} m_{N_1} s_{e1}^2}{\langle  \mathbf{p}^2\rangle + m_{N_1}^2}
+ \frac{e^{i\phi_2} m_{N_1}(1+r_{\Delta})s_{e2}^2}{\langle  \mathbf{p}^2\rangle + m_{N_1}^2 (1+r_{\Delta})^2}
\ = \ \alpha + \beta s_{e1}^2 e^{i\phi_1}\,,
\end{align}
where we have used the approximate seesaw relation $m_\nu + e^{i\phi_1}m_{N_1} s_{e1}^2 = - e^{i\phi_2}m_{N_1}(1+r_{\Delta}) s_{e2}^2$ to eliminate $s_{e2}^2$ and rewrite the summation using the factors
\begin{align}
\label{eq:alphabeta}
\alpha
& \ \equiv \ m_\nu\left(\frac{1}{\langle  \mathbf{p}^2\rangle} 
- \frac{1}{\langle  \mathbf{p}^2\rangle + m_{N_1}^2(1+r_{\Delta})^2}\right),\nonumber\\
\beta
&\ \equiv \ m_{N_1} \left(\frac{1}{\langle  \mathbf{p}^2\rangle + m_{N_1}^2} 
- \frac{1}{\langle  \mathbf{p}^2\rangle + m_{N_1}^2(1+r_{\Delta})^2}\right).
\end{align}
Alternatively, one could eliminate $s_{e2}^2$ and $\phi_2$ using the exact seesaw relations Eqs.~\eqref{eq:realandimagpart} and \eqref{eq:tanphi2}. However, taking the small mixing approximation $c_{e1}^2\approx c_{e2}^2\approx 1$ as done above makes a very small difference to the following results.
It is easy to see that these two substitutions are equivalent -- if we set $s_{e1}^2 = 0$ in Eq.~\eqref{eq:onubb1} we would be left with the contributions from the light and second heavy state. There is a relative minus sign between terms because in this limit $\phi_2 = \pi $ in both the canonical seesaw $\phi_1 = \pi$ and inverse seesaw  $\phi_1 = 0$ cases (and any intermediate $\phi_1$ value), as can be seen in Fig.~\ref{fig:s12s13}. Taking the square of the summation in Eq.~\eqref{eq:onubb1} and inserting into Eq.~\eqref{eq:halflife2} now gives
\begin{align}
\chi^2 \ = \ \alpha^2 + \beta^2 s_{e1}^4 + 2\alpha\beta s_{e1}^2\cos{\phi_1} \,;\quad
\chi \ \equiv \ \sqrt{\frac{1}{T^{0\nu}_{1/2} G^{0\nu} g_A^4|\mathcal{M}_N^{0\nu}|^2m_p^2}}\,\,.
\end{align}
Experimental lower bounds on the $0\nu\beta\beta$ decay half-life $T_{1/2}^{0\nu}>(T_{1/2}^{0\nu})_{\mathrm{exp}}$ (or $\chi^2 < \chi_{\mathrm{exp}}^2$) can therefore be used to put an \textit{upper} bound on $s_{e1}^2$ as a function of $m_{N_1}$, $m_\nu$, $r_{\Delta} =\frac{\Delta m_{N}}{m_{N_1}}$, $\phi_1$ and (through dependence on $\langle\mathbf{p}^2\rangle$ and $\chi_{\mathrm{exp}}$) the light and  heavy NMEs $|\mathcal{M}^{0\nu}_\nu|$ and $|\mathcal{M}^{0\nu}_N|$,
\begin{align}
\label{eq:se1upperbound2}
s_{e1}^2 \ < \  -\frac{\alpha}{\beta}\cos{\phi_1} + \frac{1}{\beta}\sqrt{\chi_{\mathrm{exp}}^2-\alpha^2\sin^2{\phi_1}}\,.
\end{align}
Of course there is another limit derived from the quadratic inequality $\chi^2<\chi_{\mathrm{exp}}^2$, that is a \textit{lower} bound on $s_{e1}^2$
\begin{align}
\label{eq:se1lowerbound}
s_{e1}^2 \ > \  -\frac{\alpha}{\beta}\cos{\phi_1} - \frac{1}{\beta}\sqrt{\chi_{\mathrm{exp}}^2-\alpha^2\sin^2{\phi_1}}\,.
\end{align}
We will see that for most choices of parameters this is negative and unphysical. It will be important when $\cos{\phi_1}<0$ and $\alpha>\chi_{\mathrm{exp}}$.

As detailed earlier, we work in a one-generation framework with one light state $\nu_{L,e}$ which we identify with the electron neutrino. As such, the effective $0\nu\beta\beta$ mass is not a coherent sum as usually defined,
\begin{align}
\label{eq:effmass}
	m_{\beta\beta} = \left|\sum_{i=1}^3 U^2_{ei} m_{\nu_i}\right|,
\end{align}
but is simply given by the electron neutrino mass,
\begin{align}
	m_{\beta\beta} = m_\nu.
\end{align}
In our parametrization, $m_\nu$ is always real and positive. We calculate $0\nu\beta\beta$ consistently in this framework; specifically, we include a coherent summation in the contributions of the light neutrino state and the two heavy neutrino states including the relative phases as detailed above. In this sense, $m_\nu$ is a surrogate for the general effective $0\nu\beta\beta$ mass $m_{\beta\beta}$ in Eq.~\eqref{eq:effmass}, but we cannot include the potential destructive interference therein due to the Majorana phases within the light PMNS matrix. This effect has been discussed extensively in the literature (for a review, see e.g.~\cite{DellOro:2016tmg}) whereas our focus is on the constraints on the heavy neutrino parameters. The precise value of the light neutrino contribution may only be important if it saturates the limit from $0\nu\beta\beta$ decay searches. This is discussed in Fig.~\ref{fig:phiandmchange}~(right) and the accompanying text where the choice $m_\nu = 6\times 10^{-2}$~eV is near the excluded $m_{\beta\beta}$ limit and thus the constraints on the extra contributions of the heavy neutrinos become overly restrictive. These may instead be relaxed if there is a sizeable cancellation among the light neutrino contributions reducing $m_{\beta\beta}$. A full analytic discussion of $0\nu\beta\beta$ in presence of three active neutrinos mixing with sterile neutrinos is beyond the scope of the present work and will be pursued in a follow-up work.

\subsection{Sensitivity to sterile neutrino parameters}

\begin{figure}[t!]
	\centering
	\includegraphics[width=\textwidth]{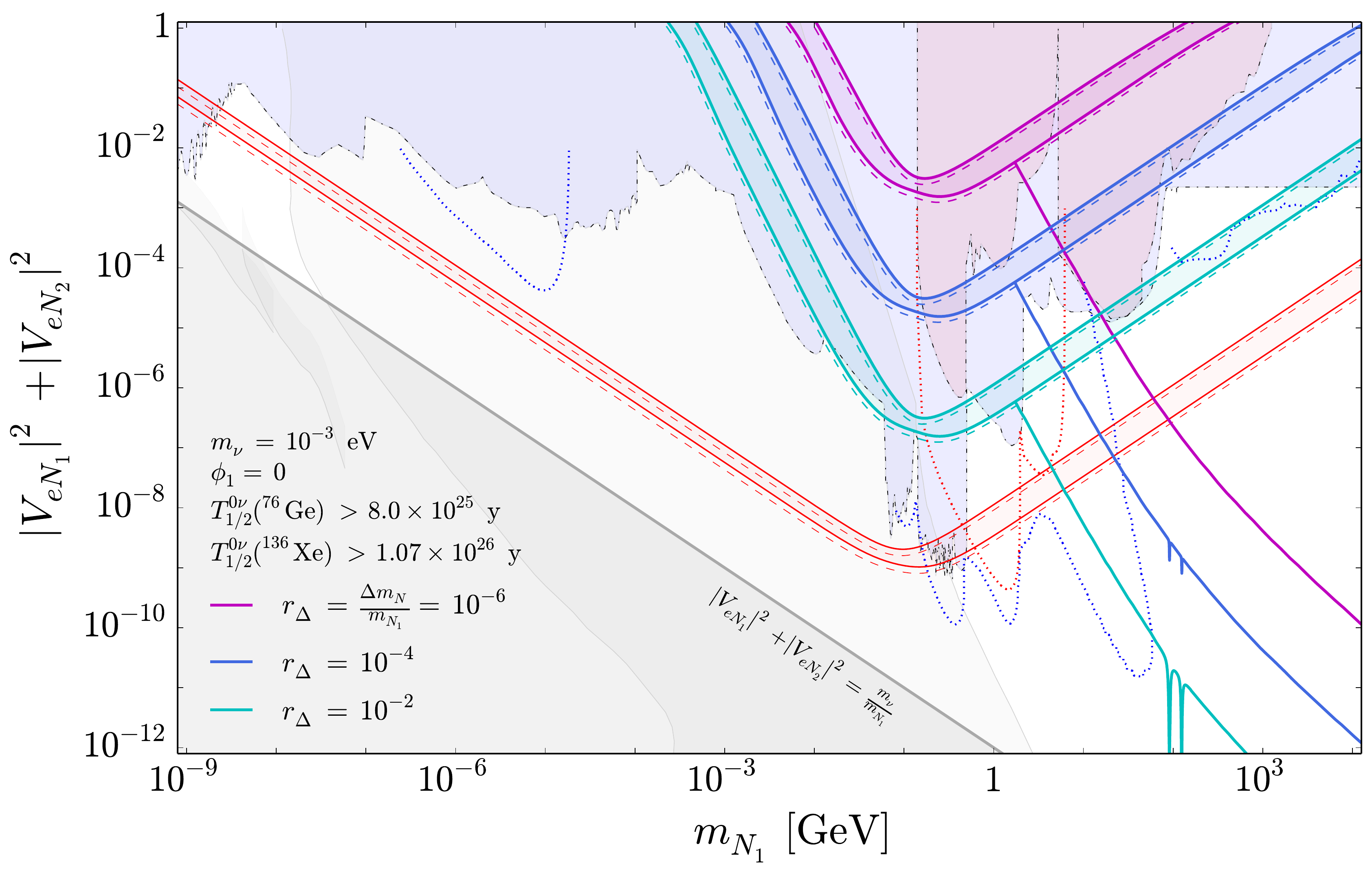}
	\caption{Upper limits on the sum of squared active-sterile mixings for three values of the sterile neutrino mass splitting ratio $r_{\Delta} =\frac{\Delta m_{N}}{m_{N_1}}\ll 1$. We show the limits from $^{136}$Xe (solid) and $^{76}$Ge (dashed) experiments with the bands indicating the respective uncertainties. The red curves highlight the limit in which $0\nu\beta\beta$ decay is driven by a single sterile neutrino. The curves sloping down to the lower right indicate the upper bounds by enforcing $|\delta m_{\nu}^{\mathrm{1-loop}}|<0.1 m_{\nu}$. These constraints are compared with the current and future sensitivities of LNC (blue shaded/dotted) and LNV (red shaded/dotted) searches, cf. Figs.~\ref{fig:current_constraints} and \ref{fig:future_constraints}.}
	\label{fig:ychange}
\end{figure}

In Fig.~\ref{fig:ychange} we display the upper bounds on the sum of squared active-sterile mixings $|V_{eN_{1}}|^2+|V_{eN_{2}}|^2\approx s_{e1}^2+s_{e2}^2$ as a function of the first sterile neutrino mass $m_{N_1}$ for three \textit{small} values of the sterile neutrino mass splitting ratio $r_{\Delta} \ll 1$ and for benchmark values of the light neutrino mass $m_{\nu} = 10^{-3}$ eV and Majorana phase $\phi_1 = 0$. The sum is used in the assumption that for small splitting the energy resolutions of direct searches are larger than $\Delta m_{N}$ and consequently constrain $|V_{eN_{1}}|^2+|V_{eN_{2}}|^2$ as a function of the mass $m_{N_{1}}\approx m_{N_{2}}$. Making use of the $s_{e1}^2$ inequality in Eq.~\eqref{eq:se1upperbound2} we take the most recent lower limits on $T_{1/2}^{0\nu}$ from the $^{136}$Xe KamLAND-Zen~\cite{KamLAND-Zen:2016pfg} and  $^{76}$Ge GERDA-II~\cite{Agostini:2018tnm} experiments and the IBM-2 light and heavy NMEs in Table~\ref{tab:NMEs} to plot the solid (and dashed) curves in the upper right portion of Fig.~\ref{fig:ychange}. The bands illustrate the uncertainty on $|V_{eN_{1}}|^2+|V_{eN_{2}}|^2$ as a function of $m_{N_1}$ found by propagating the conservative IBM-2 uncertainties through Eq.~\eqref{eq:se1upperbound2}. The red curves in Fig.~\ref{fig:ychange} depict the upper limits on $|V_{eN_{1}}|^2+|V_{eN_{2}}|^2$ when including only the contribution of a single sterile state (neglecting light active exchange) towards $0\nu\beta\beta$ decay. Finally, we show for these choices of $r_{\Delta}$ the upper limits on $|V_{eN_{1}}|^2+|V_{eN_{2}}|^2$ from the requirement that $|\delta m_{\nu}^{\mathrm{1-loop}}|<0.1m_{\nu}$, taken directly from Fig.~\ref{fig:numass-loop} (right).

We compare these $0\nu\beta\beta$ decay bounds to the direct search limits discussed in Sec.~\ref{sec:searches}. These include the current (blue-shaded) and future (blue dot-dashed line) sensitivities of LNC probes including $\beta$-decay kink searches, meson decay peak searches, beam dump experiments and collider constraints. We also display separately the current (red-shaded) and future (red dot-dashed line) sensitivities of LNV meson decay and collider probes. Faint grey regions correspond to the cosmological excluded regions. Finally, the dark grey shaded region below the seesaw line $|V_{eN_{1}}|^2+|V_{eN_{2}}|^2 = \frac{m_{\nu}}{m_{N_1}}$ is excluded as explored at the start of Sec.~\ref{sec:0vbb}.
\begin{figure}[t!]
	\centering
	\includegraphics[width=\textwidth]{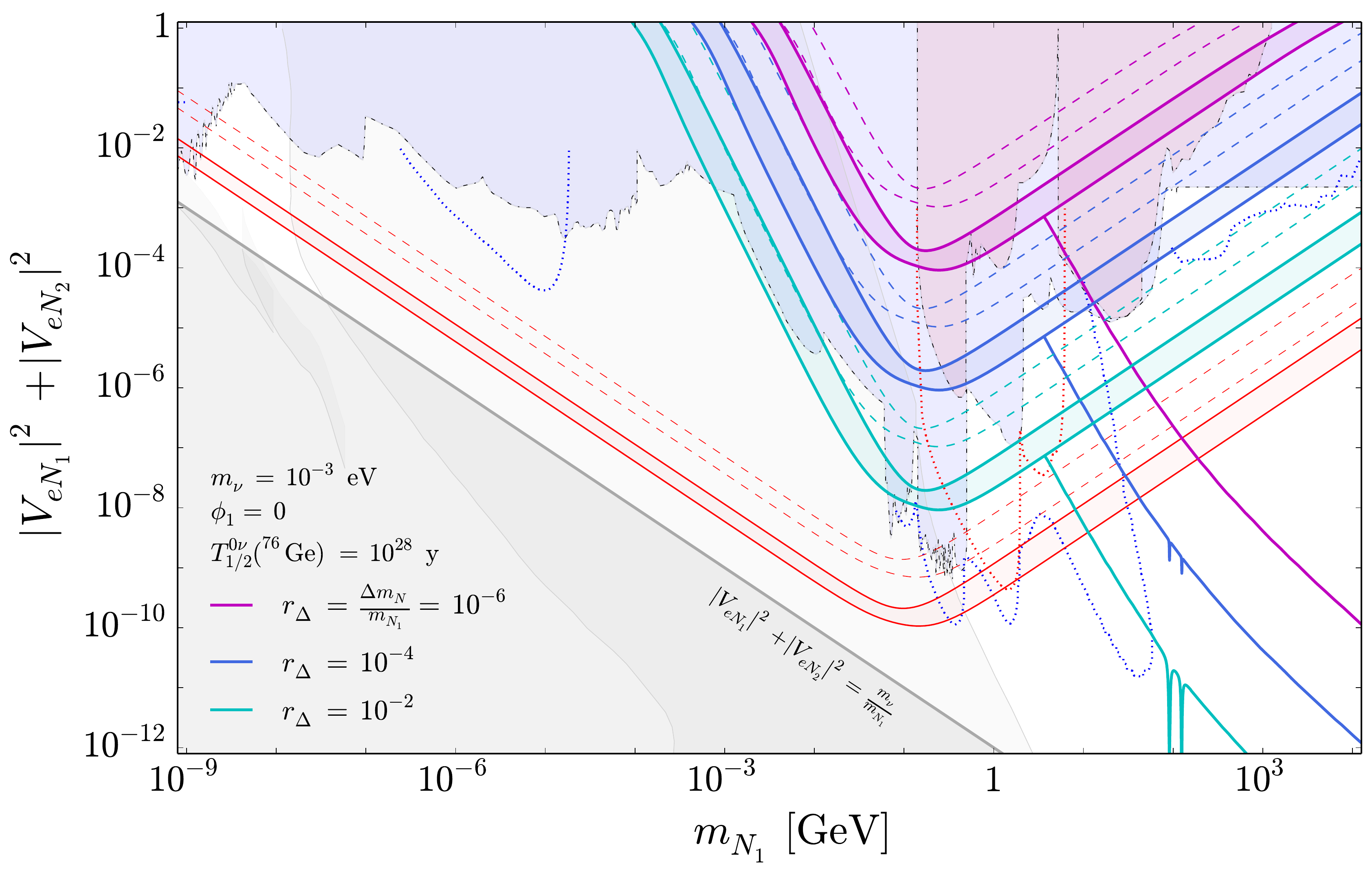}
	\caption{As Fig.~\ref{fig:ychange}, but showing the future $0\nu\beta\beta$ decay sensitivity for $^{76}$Ge at $T_{1/2}^{0\nu}=10^{28}$~y.}
	\label{fig:ychangefuture}
\end{figure} 

In Fig.~\ref{fig:ychangefuture} we similarly show the upper bounds from $0\nu\beta\beta$ decay and loop considerations for the same (small) values of the sterile mass splitting ratio but instead use the predicted sensitivity of future experiments, $T_{1/2}^{0\nu}\gtrsim 10^{28}$ y. This reach may be achievable at the proposed PandaX-III~\cite{Chen:2016qcd} and nEXO~\cite{Albert:2017hjq} $^{136}$Xe experiments, the LEGEND~\cite{Abgrall:2017syy} $^{76}$Ge experiment and the CUPID~\cite{Wang:2015raa} $^{130}$Te, $^{100}$Mo, $^{82}$Se and $^{112}$Cd experiment.

We first observe that the upper bounds are most stringent for $m_{N_1}\sim \sqrt{\langle\mathbf{p}^2\rangle}\sim 200$ MeV, reaching $|V_e|^2\lesssim 10^{-7} $. Towards lower $m_{N_1}$ both sterile states are \textquoteleft light' and the $0\nu\beta\beta$ decay rate is suppressed by the seesaw relation, eventually erasing the upper bounds for $m_{N_1}\lesssim 1$ MeV. For higher $m_{N_1}$ both sterile states are \textquoteleft heavy' and the limits become weaker as $m_{N_1}$ increases due to the growing NME suppression by $1/m_{N_1}^2$. We also see a strong dependence on the sterile mass splitting ratio; decreasing $r_{\Delta}$ by a factor of $\sim 10^2$ weakens the upper bound by a similar factor both above and below $m_{N_{1}}\sim \sqrt{\langle\mathbf{p}^2\rangle}$. This is to be expected as $r_{\Delta} \rightarrow 0$ corresponds the pseudo-Dirac limit in which lepton number is approximately conserved and the $0\nu\beta\beta$ decay process is forbidden. Comparing the $^{76}$Ge and $^{136}$Xe bounds it is interesting to note that those for the former are slightly more stringent despite the smaller experimental half-life lower bound. As can be seen in Fig.~\ref{fig:nmeuncertainties} this is counteracted by $^{76}$Ge possessing larger NMEs on average compared to $^{136}$Xe. Comparing with direct searches we see that for these small choices of $r_{\Delta}$ the current upper bounds are at best comparable with non-resonant meson decay limits for $1~\mathrm{MeV}<m_{N_1}< 1~\mathrm{GeV}$ and more stringent than collider constraints for $m_{N_1}> 5$ GeV. 

We saw in Sec.~\ref{sec:LNVcolliders} that when the sterile mass splitting ratio $r_{\Delta}$ is decreased the LNV collider constraints shaded in red do not weaken significantly -- this is because the amplitide of LNV is controlled by the ratio $\Gamma_{N}/\Delta m_{N}$. By considering the open sterile neutrino decays to SM particles we found  $\Gamma_{N}/\Delta m_{N} = \Gamma_N/(r_\Delta m_{N_{1}})\ll1$ in the mass range $5~\mathrm{GeV}\lesssim m_{N_1}\lesssim 50~\mathrm{GeV}$ for $r_{\Delta} \gtrsim 10^{-10}$. Thus, when $r_{\Delta} \lesssim 10^{-2}$ the $0\nu\beta\beta$ decay constraints become less stringent than the same-sign dilepton and LNV trilepton collider constraints.

\begin{figure}[t!]
	\centering
	\includegraphics[width=0.49\textwidth]{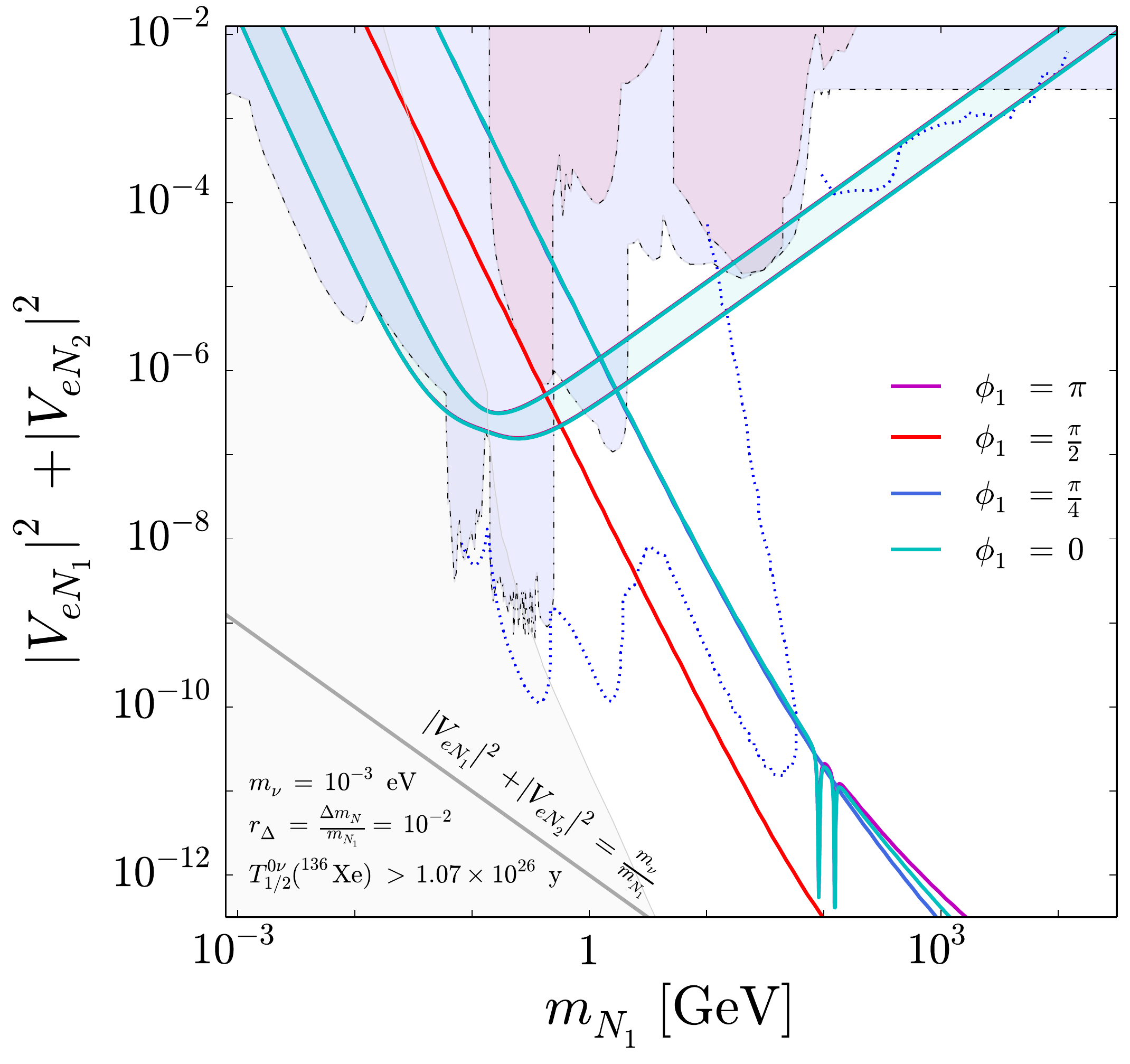}  
	\includegraphics[width=0.49\textwidth]{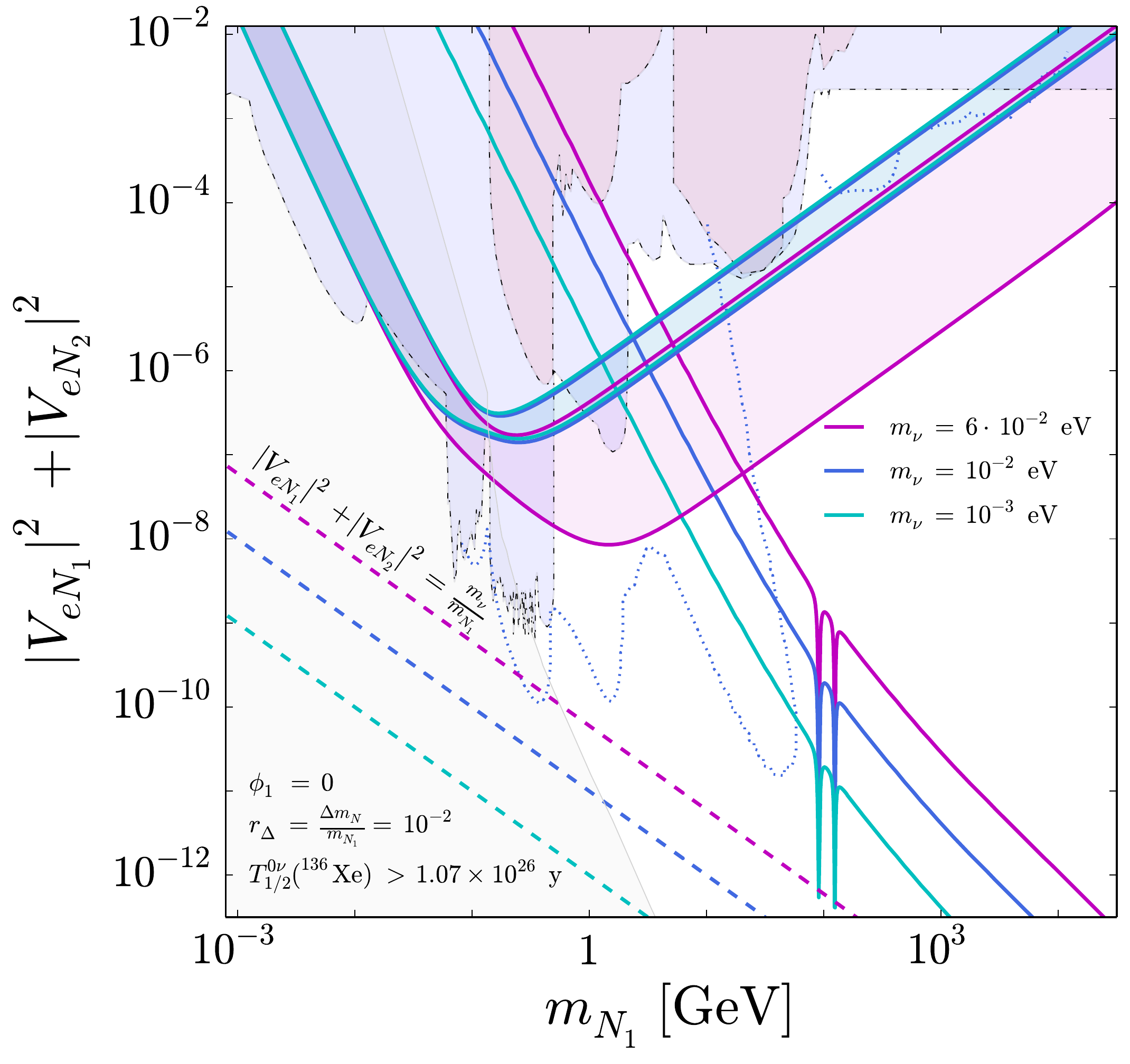}
	\caption{Upper limits on the sum of squared active-sterile mixing for the sterile neutrino mass splitting ratio $r_{\Delta} =\frac{\Delta m_{N}}{m_{N_1}}=10^{-2}$ derived from $0\nu\beta\beta$ decay and loop constraints. We show the limits from $^{136}$Xe for different values of $\phi_1$ (left) and $m_{\nu}$ (right).}
	\label{fig:phiandmchange}
\end{figure}

The behaviour of the $0\nu\beta\beta$ decay upper bound in the \textquoteleft light' and \textquoteleft heavy' regimes can be quantified by taking the Taylor expansion of Eq.~\eqref{eq:se1upperbound2} in the opposing limits $m_{N_1}/\sqrt{\langle \mathbf{p}^2\rangle} \ll 1$ and $m_{N_1}/\sqrt{\langle \mathbf{p}^2\rangle} \gg 1$. In the light regime we derive
\begin{align}
\label{eq:lightreg}
s_{e1}^2 \ \lesssim \ \frac{\langle \mathbf{p}^2\rangle^2 \chi_{\mathrm{exp}}}{m_{N_1}^3r_{\Delta}(2+r_{\Delta})} \,,
\end{align}
while in the heavy regime
\begin{align}
\label{eq:heavyreg}
s_{e1}^2 
\ \lesssim \ \left(-\frac{m_\nu}{\langle \mathbf{p}^2\rangle}\cos\phi_1+\sqrt{\chi^2_{\mathrm{exp}}-\frac{m^2_{\nu}\sin^{2}\phi_1}{\langle\mathbf{p}^2\rangle^2}}\right)
\frac{m_{N_1}(1+r_{\Delta})^2}{r_{\Delta}(2+r_{\Delta})}\,.
\end{align}
The $m_{N_1}$ dependence of these upper bounds agrees qualitatively with Fig. (\ref{fig:ychange}) -- in the light regime the upper bounds scale as $ 1/m_{N_1}^3$ and in the heavy regime as $m_{N_1}$. The dependence on $r_{\Delta}$ is also in agreement -- for $r_\Delta \ll 1$ both Eqs.~\eqref{eq:lightreg} and \eqref{eq:heavyreg} are inversely proportional to $r_{\Delta}$. Thus decreasing or increasing $r_{\Delta}$ shifts the entire upper bound to higher and lower mixings for the whole range of $m_{N_1}$.

In Fig.~\ref{fig:phiandmchange} we study more closely the $|V_{eN_{1}}|^2+|V_{eN_{2}}|^2=s_{e1}^2c_{e2}^2+s_{e2}^2$ upper bound in the $r_{\Delta} =10^{-2}$ case for different values of the Majorana phase $\phi_1$ (left) and the light neutrino mass $m_{\nu}$ (right). To the left it is clear that changing $\phi_1$ has little effect on the $0\nu\beta\beta$ decay constraints for these choices of parameters. As shown in Eq.~\eqref{eq:lightreg}, in the light regime the $s_{e1}^2$ upper bound is independent of $\phi_1$ because the suppression of the $0\nu\beta\beta$ decay rate through the seesaw relation is also independent of $\phi_1$.  From Eq.~\eqref{eq:heavyreg} we see that in the heavy regime changing $\phi_1$ has little effect for these parameter choices because $m_{\nu}/\langle\mathbf{p}^2\rangle \ll \chi_{\mathrm{exp}} $, i.e. the light neutrino contribution is negligible. $0\nu\beta\beta$ decay is therefore driven by the two heavy states. This is the limit $\alpha \ll 1$ and $s_{e1}^2\lesssim\frac{\chi_{\mathrm{exp}}}{\beta}$ in Eq.~\eqref{eq:se1upperbound2}. For $m_{N_{1}}\gg \langle\mathbf{p}^2\rangle^{1/2}$ we have
\begin{align}
\beta \ \approx \ \frac{r_{\Delta}(2+r_{\Delta})}{m_{N_{1}}(1+r_{\Delta})^2}\,,
\end{align}
which gives the expected dependence on $r_{\Delta}$ and $m_{N_{1}}$ in Eq.~\eqref{eq:heavyreg}.

To the right we see that the effect of increasing $m_{\nu}$ for $\phi_1 = 0 $ is to strengthen the upper bound in the heavy regime. This again is described by Eq.~\eqref{eq:heavyreg} -- there is a cancellation between the two terms in the brackets as $m_{\nu}/\langle\mathbf{p}^2\rangle$ approaches $\chi_{\mathrm{exp}}$. In this limit the light active contribution becomes non-negligible compared to the difference between the heavy sterile contributions. For the inverse seesaw region of the parameter space
\begin{align}
\chi^2 \ \approx \ \left|\frac{m_{\nu}}{\langle \mathbf{p}^2\rangle}+\frac{r_{\Delta}(2+r_{\Delta})}{m_{N_{1}}(1+r_{\Delta})^2}s_{e1}^2e^{i\phi_1}\right|^2 \ < \ \chi_{\mathrm{exp}}^2\,.
\end{align}
If for example $(\phi_1,\,\phi_2) = (0,\,\pi)$, the light contribution adds constructively with the difference and the upper bound on $s_{e1}^2$ (multiplying the heavy contributions) must be smaller to account for the observed half-life lower bound. If on the other hand $(\phi_1,\,\phi_2) = (\pi,\,0)$, the light and heavy contributions add destructively and the $s_{e1}^2$ upper bound can be relaxed. 

If $m_{\nu}/\langle\mathbf{p}^2\rangle>\chi_{\mathrm{exp}}$ (which may be the case for a large lower limit on $T_{1/2}^{0\nu}$) no value of $s_{e1}^2$ in the heavy regime is permitted for $\phi_1=0$. In Eq.~\eqref{eq:se1upperbound2} this corresponds more generally to the case $\alpha>\chi_{\mathrm{exp}}$ in which the upper bound on $s_{e1}^2$ becomes negative and unphysical. Constructive interference between the light active contribution and the difference between the heavy sterile contributions, e.g. as for $(\phi_1,\,\phi_2) = (0,\,\pi)$, now gives a $T_{1/2}^{0\nu}$ less than the experimental lower limit, or $\chi>\chi_{\mathrm{exp}}$. Conversely, if the light and heavy contributions interfere destructively, e.g. for $(\phi_1,\,\phi_2) = (\pi,\,0) $ above the seesaw line and $(\phi_1,\,\phi_2) = (\pi,\,\pi) $ below, then $s_{e1}^2$ multiplying the heavy contributions can be made large enough to meet the condition $\chi<\chi_{\mathrm{exp}}$ (but not so large as to dominate over the light contribution). As well as an upper bound, this sets a \textit{lower} bound on $s_{e1}^2$ in the heavy regime. This is the lower bound in Eq.~\eqref{eq:se1lowerbound} becoming non-negative.

It is worth reminding the reader that we are considering a value of $r_{\Delta}$ in the range $[0,\,\infty]$ and so the introduced quantities in Eq.~\ref{eq:alphabeta} satisfy $\alpha>0$ and $\beta>0$. As explained in Sec.~\ref{sec:parametrization}, a value of $r_{\Delta}$ in the range $[-1,\,0]$ is equivalent to swapping the roles of the sterile states, now having $m_{N_{2}}<m_{N_{1}}$. In this equally valid range the introduced quantities satisfy $\alpha>0$ and $\beta<0$. Because of this we see by examining Eqs.~\eqref{eq:se1upperbound2} and \eqref{eq:se1lowerbound} that the behaviours of the active-sterile mixings ($s_{e1}^2,\,s_{e2}^2$) and Majorana phases ($\phi_1,\,\phi_2$) are also swapped, with cancellation between light active and heavy sterile contributions taking place for $(\phi_1,\,\phi_2) = (0,\,\pi)$.

\begin{figure}[t!]
	\centering
	\hspace{-1.5em}\includegraphics[width=14cm]{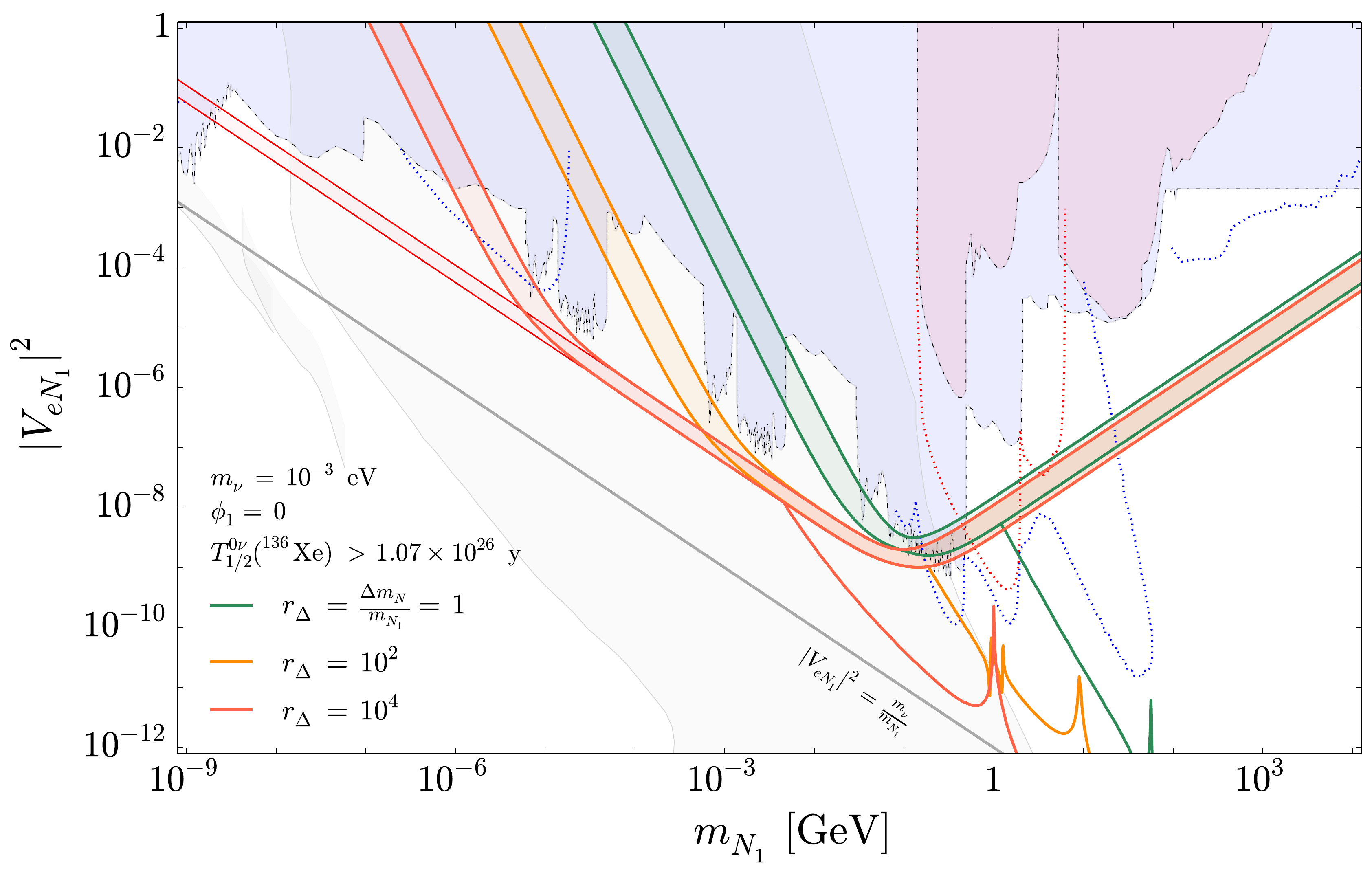}
	\caption{Upper limits on the active-sterile mixing with $N_{1}$ for three values of the sterile neutrino mass splitting ratio $r_{\Delta} =\frac{\Delta m_{N}}{m_{N_1}}\geq 1$. We show the limits for $^{136}$Xe with shaded bands indicating the respective uncertainties. The red curve highlights the limit in which $0\nu\beta\beta$ decay is driven by a single sterile neutrino. The curves sloping down to the lower right indicate the upper bounds by enforcing $|\delta m_{\nu}^{\mathrm{1-loop}}|<0.1 m_{\nu}$. These constraints are compared with the current and future sensitivities of LNC (blue shaded/dotted) and LNV (red shaded/dotted) searches, cf. Figs.~\ref{fig:current_constraints} and \ref{fig:future_constraints}.}
	\label{fig:largeychange}
\end{figure} 

In Fig.~\ref{fig:phiandmchange} we also see how the loop constraints change when varying $\phi_1$ and $m_{\nu}$. For the extreme values $\phi_1 = 0,\,\pi$ and intermediate value $\phi_1 = \frac{\pi}{4}$ the loop constraints are broadly the same. However for $\phi_1=\frac{\pi}{2}$ the upper bound becomes nearly two orders of magnitude more stringent. As $m_{\nu}$ is increased by an order of magnitude (we do not go to $m_{\nu} > \langle \mathbf{p}^2\rangle\chi_{\exp}\approx 0.083$ eV for the reasons discussed previously) we can also see that the loop constraints are correspondingly weakened by an order of magnitude.

In Fig.~\ref{fig:largeychange} we display the active-sterile mixing $|V_{eN_{1}}|^2\approx s_{e1}^2$ as a function of $m_{N_1}$ for three \textit{large} values of the sterile neutrino mass splitting ratio $r_{\Delta}  \geq 1$ and for benchmark values of the light neutrino mass $m_{\nu} = 10^{-3}$ eV and Majorana phase $\phi_1 = 0$. We do not show the sum in this case because it is assumed that the splittings are large enough for the two states to be resolved individually in direct search experiments. We compare these bounds to the direct search limits discussed in Sec.~\ref{sec:searches}. Due to the large splitting, shifted versions of the excluded region depending on the value of $r_{\Delta}$ now apply --  this is a shift to smaller $m_{N_{1}}$ and to larger $|V_{eN_{1}}|^2$ by a factor $(1+r_{\Delta})$. For example, if the T2K experiment excludes a second state of mass $m_{N_{2}}$ and mixing $|V_{eN_{2}}|^2$, it also implies the non-existence of the first state at $m_{N_{1}}\approx m_{N_{2}}/(1+r_{\Delta})$ and $|V_{eN_{1}}|^2\approx|V_{eN_{2}}|^2(1+r_{\Delta})$. These particular relations apply because the T2K bounds are in the inverse seesaw region of the parameter space. For these large splittings we immediately see that the $0\nu\beta\beta$ decay constraints converge towards the upper bound in the limit of single heavy neutrino exchange (commonly used in the literature), shown by the thin red curve in Figs. \ref{fig:ychange}, \ref{fig:ychangefuture} and \ref{fig:largeychange}.

We have so far neglected the one-loop contribution to the neutrino mass $\delta m^{\mathrm{1-loop}}_{\nu}$ in this discussion. Initially one could ask if this has a large impact in the $m_{N_i} \ll \langle \mathbf{p}^2\rangle$ case because
\begin{align}
\label{eq:halflife3}
\frac{1}{T^{0\nu}_{1/2}} 
\ \propto \ \left|\sum_{i=1}^{1+2} V_{ei}^2 \,m_{i}\right|^2 \  \propto \ \left|\delta m_{\nu}^{\mathrm{1-loop}}\right|^2\,,
\end{align}
which would be expected to alter the suppression and $1/m_{N_{1}}^3$ scaling due to the tree-level seesaw relation. However, when we look at Fig.~\ref{fig:numass-loop} we see that $|\delta m_{\nu}^{\mathrm{1-loop}}|\sim 10^{-12}$ eV in the light regime, even after the iterative procedure on $\delta m_{\nu}^{\mathrm{1-loop}}$ is applied. Thus we safely expect this effect on the $0\nu\beta\beta$ decay constraint curves to be negligible.

%%%%%%%%%%%%%%%%%%%%%%%%%%%%%%%%%%%%%%%%%%%%%%%%%%%%%%%%%%%%%%%%%%%%%%%%%%%%%%%%%
\section{Conclusions}
\label{sec:conclusions}
%%%%%%%%%%%%%%%%%%%%%%%%%%%%%%%%%%%%%%%%%%%%%%%%%%%%%%%%%%%%%%%%%%%%%%%%%%%%%%%%%
Heavy sterile neutrinos represent one of the most interesting candidates for particles beyond the Standard Model. They are conspicuously absent from the Standard Model particle content which means SM neutrinos are the only fermions that do not have an electroweak singlet partner field. It is not far-fetched to assume that this has something to do with the fact that sterile neutrinos, as their name implies, are singlets under all the SM gauge groups and a Majorana mass term breaking total lepton number is therefore not protected.

There is a strong ongoing and planned effort to search for sterile neutrinos over a wide range of masses and active-sterile mixing strengths. The main focus of this work is to compare direct searches such as at the LHC and in meson decays with constraints from $0\nu\beta\beta$ decay. The latter is the most important probe of lepton number violation and light Majorana neutrino masses. Heavy neutrinos will generically contribute to $0\nu\beta\beta$ decay as well. They are thus constrained by current searches and can be probed in future $0\nu\beta\beta$ decay experiments.

In this work, we have introduced a phenomenological parametrisation of a one-generation seesaw model in terms of experimentally measurable quantities, such as active-sterile neutrino mixing angles, CP phases, masses and mass splittings. We have identified the regions of parameter space allowed by consistency conditions in the neutrino mass matrix in the single-generation case, and have showed how the type-I and inverse seesaw limits can be recovered (cf.~Fig.~\ref{fig:mixingangles}). Imposing the additional consideration that the loop contribution to the active neutrino mass must be less than 10\% of the tree-level mass further reduces this allowed parameter space, as shown in Fig.~\ref{fig:s12s13-loop}. 

We summarise current and future experimental constraints on the sterile neutrino mass-mixing parameter space over a wide range of interest, including both lepton number conserving and violating processes (cf.~Figs.~\ref{fig:current_constraints} and \ref{fig:future_constraints}), emphasising that the LNV constraints could change depending on the mass splitting between the two sterile states and the relative CP phase between them. This is particularly relevant for $0\nu\beta\beta$ decay searches, which are significantly weakened for quasi-Dirac sterile neutrinos, as shown in Fig.~\ref{fig:ychange}, while for large mass splitting, the $0\nu\beta\beta$ decay constraint remains strong in the electron sector; cf.~Fig.~\ref{fig:largeychange}, and it is especially relevant for heavy neutrino masses in the region $m_N \approx 100$~MeV to a few GeV, where the future $0\nu\beta\beta$ decay sensitivities can reach a level close to the small active-sterile mixing strengths expected in a vanilla seesaw scenario. 
%It is important to keep this in mind while comparing the LNV constraints from $0\nu\beta\beta$ decay with the direct searches from high-energy colliders. 

%%%%%%%%%%%%%%%%%%%%%%%%%%%%%%%%%%%%%%%%%%%%%%%%%%%%%%%%%%%%%%%%%%%%%%%%%%%%%%%%%
\acknowledgments
%%%%%%%%%%%%%%%%%%%%%%%%%%%%%%%%%%%%%%%%%%%%%%%%%%%%%%%%%%%%%%%%%%%%%%%%%%%%%%%%%
PDB and FFD would like to acknowledge support from the UK Science and Technology Facilities Council (STFC) via a Consolidated Grant (Reference ST/P00072X/1). The work of BD is supported in part by the US Department of Energy under Grant No.  DE-SC0017987 and in part by the MCSS. 

\appendix
	\section{Constraints on muon- and tau-sterile neutrino mixings} \label{sec:app}
	
	\begin{figure}[t!]
		\centering
		\includegraphics[width=\textwidth]{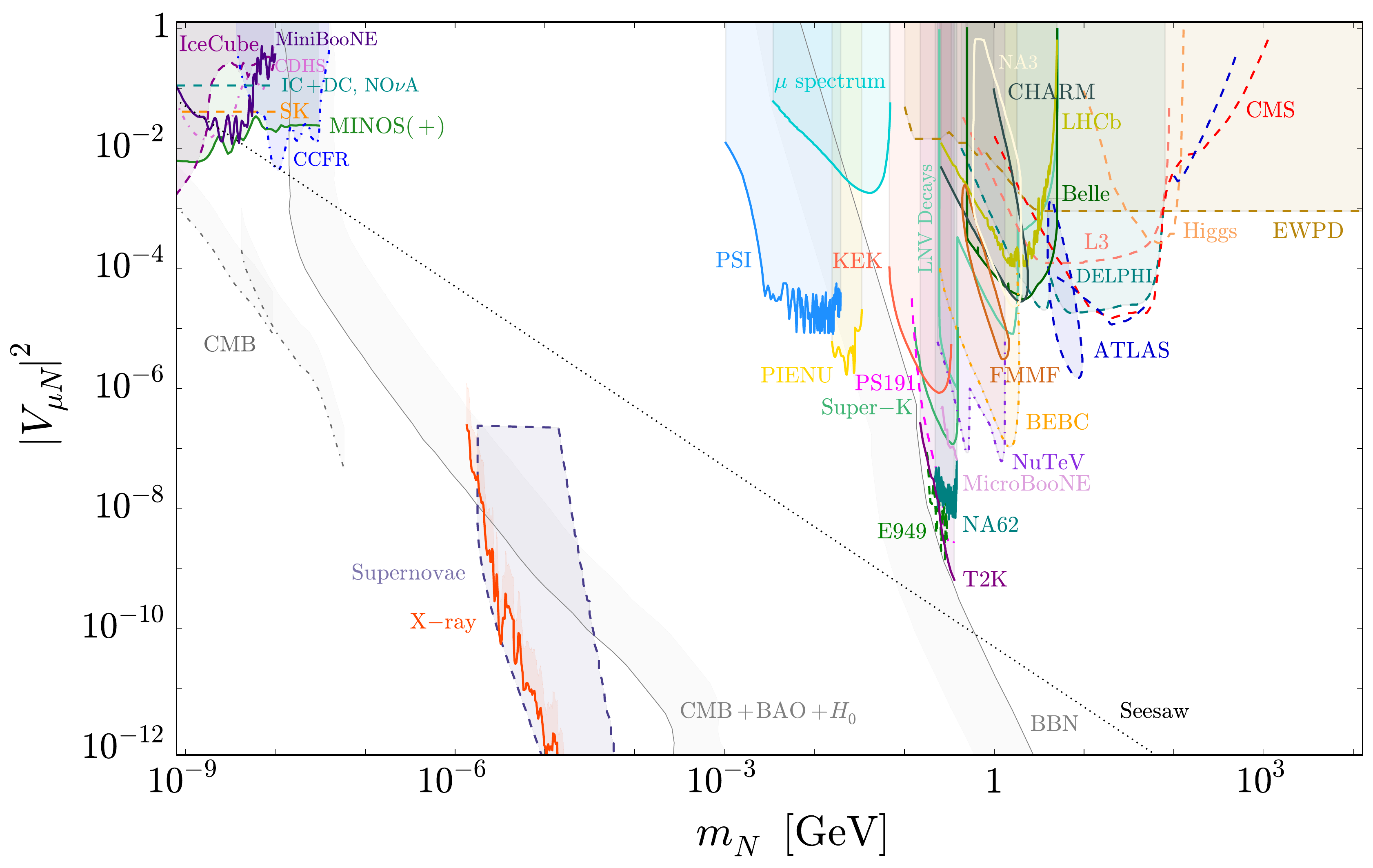}
		\includegraphics[width=\textwidth]{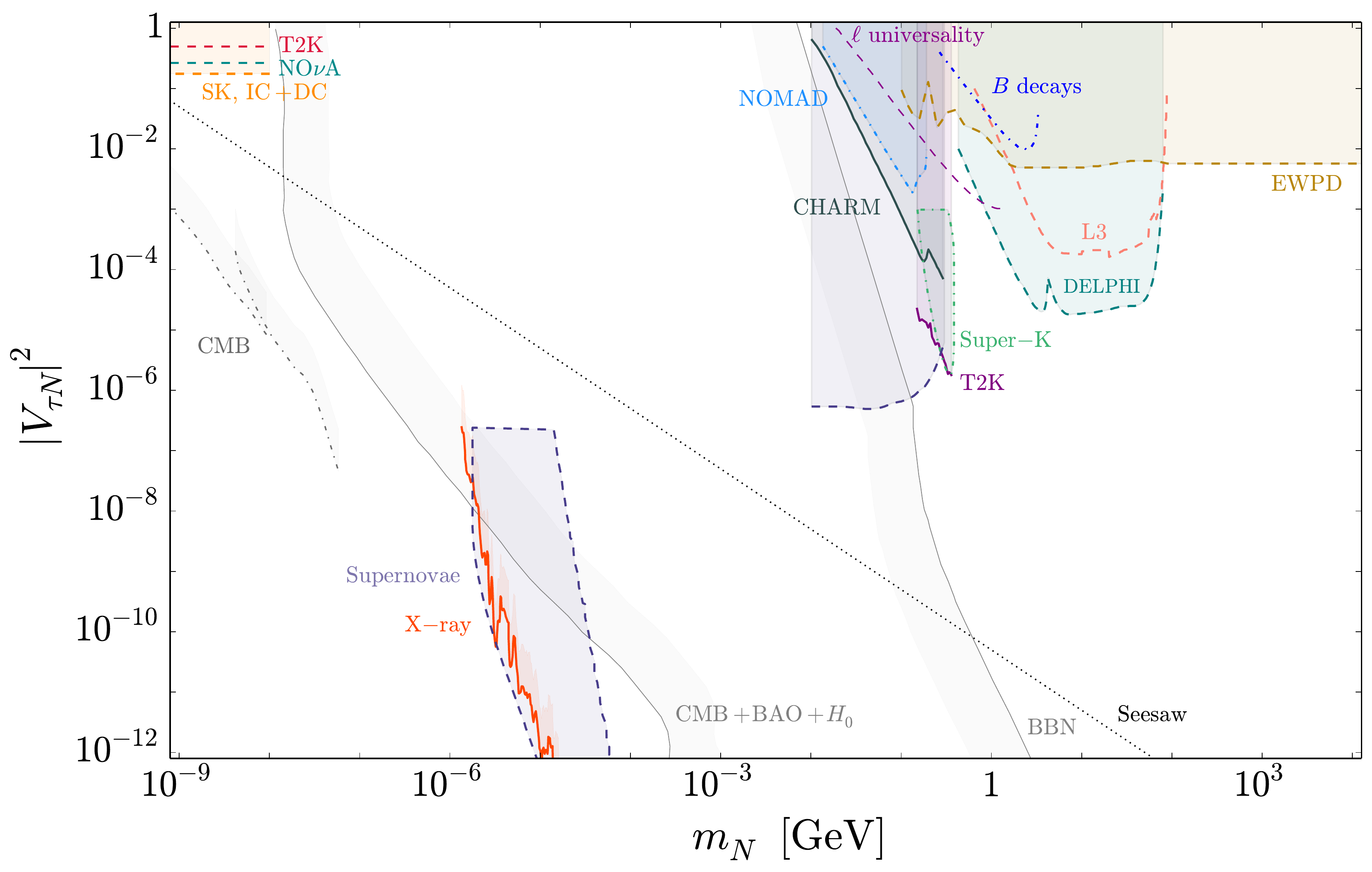}
		\caption{Constraints on the mass $m_N$ of the sterile neutrino and its squared mixing $|V_{\mu N}|^2$ with the muon neutrino (above) and $|V_{\tau N}|^2$ with the tau neutrino (below). The shaded regions are excluded by the searches listed in the appendix.}
		\label{fig:current_constraints_mutau}
	\end{figure}

	For completeness we plot in Fig.~\ref{fig:current_constraints_mutau} the constraints from a variety of experiments on the $m_{N}-|V_{\mu N}|^2$ and $m_{N}-|V_{\tau N}|^2$ parameter spaces (above and below respectively).
	
	Upper limits on the muon-sterile mixing strength have been placed above $m_{N}\sim1$ MeV by PSI~\cite{Daum:1987bg}, PIENU~\cite{Aguilar-Arevalo:2019owf}, KEK~\cite{Hayano:1982wu,Yamazaki:1984sj}, PS191~\cite{Bernardi:1987ek}, measurements of the muon decay spectrum~\cite{Shrock:1981wq}, Super-Kamiokande~\cite{Coloma:2019htx}, E949~\cite{Artamonov:2014urb}, NA62~\cite{Lazzeroni:2019}, T2K~\cite{Abe:2019kgx}, MicroBooNE~\cite{Abratenko:2019kez}, a variety of LNV meson decays~\cite{Atre:2009rg,Helo:2010cw,Kovalenko:2009td,Abada:2017jjx}, NuTeV~\cite{Vaitaitis:1999wq}, BEBC~\cite{CooperSarkar:1985nh}, FMMF~\cite{Gallas:1994xp}, NA3~\cite{Badier:1986xz}, CHARM~\cite{Vilain:1994vg}, LHCb~\cite{Aaij:2014aba}, Belle~\cite{Liventsev:2013zz}, ATLAS~\cite{Aad:2019kiz}, CMS~\cite{Sirunyan:2018mtv}, L3~\cite{Adriani:1992pq}, DELPHI~\cite{Abreu:1996pa} and electroweak precision data~\cite{Blennow:2016jkn, deBlas:2013gla, delAguila:2008pw, Antusch:2014woa}. In the light regime, i.e. at masses smaller than $m_{N}\sim100$ eV, limits have been placed by oscillation experiments such as IceCube \cite{TheIceCube:2016oqi,Aartsen:2017bap}, MINOS/MINOS+ \cite{Adamson:2017uda}, Super-Kamiokande \cite{Abe:2014gda}, NO$\nu$A \cite{Adamson:2017zcg}, CDHS \cite{Dydak:1983zq} and CCFR \cite{Stockdale:1984cg}. It should be noted that limits on the disappearance channel, i.e. active to sterile oscillation, tend towards a constant upper bound as a function of $\Delta m^2$ and hence $m_{N}$. This bound can in principle be extended to arbitrarily large $m_{N}$, covering the region between 100 eV and 1 MeV. Finally, limits have been set from considerations of supernovae \cite{Arguelles:2016uwb} and the non-observation of X-rays \cite{Ng:2019gch,Roach:2019ctw}. 
	
	Alternatively, upper limits on the tau-sterile mixing strength have been placed above $m_{N}\sim1$ MeV by NOMAD \cite{Astier:2001ck}, CHARM \cite{Orloff:2002de}, Super-Kamiokande, T2K, lepton universality and $B$ decays \cite{Cvetic:2017gkt}, L3, DELPHI and electroweak precision data. From oscillations upper bounds have been placed by IceCube \cite{TheIceCube:2016oqi,Aartsen:2017bap}, Super-Kamiokande \cite{Abe:2014gda} and NO$\nu$A \cite{Adamson:2017zcg}. The supernovae and X-ray constraints also apply for $|V_{\tau N}|^2$.

\bibliographystyle{JHEP}
\bibliography{hnl-0vbb}

\providecommand{\href}[2]{#2}\begingroup\raggedright\begin{thebibliography}{100}

\bibitem{Tanabashi:2018oca}
{\scshape Particle Data Group} collaboration, M.~Tanabashi et~al.,
  \emph{{Review of Particle Physics}},
  \href{http://dx.doi.org/10.1103/PhysRevD.98.030001}{\emph{Phys. Rev.} {\bf
  D98} (2018) 030001}.

\bibitem{Weinberg:1979sa}
S.~Weinberg, \emph{{Baryon and Lepton Nonconserving Processes}},
  \href{http://dx.doi.org/10.1103/PhysRevLett.43.1566}{\emph{Phys.Rev.Lett.}
  {\bf 43} (1979) 1566}.

\bibitem{Minkowski:1977sc}
P.~Minkowski, \emph{{$\mu \to e \gamma$ at a Rate of One Out of 1-Billion Muon
  Decays?}},
  \href{http://dx.doi.org/10.1016/0370-2693(77)90435-X}{\emph{Phys.Lett.} {\bf
  B67} (1977) 421}.

\bibitem{Mohapatra:1979ia}
R.~N. Mohapatra and G.~Senjanovi\'{c}, \emph{Neutrino mass and spontaneous
  parity nonconservation}, {\emph{Phys. Rev. Lett.} {\bf 44} (1980) 912}.

\bibitem{Gellmann:1980vs}
M.~Gell-Mann, P.~Ramond and R.~Slansky, \emph{{Complex Spinors and Unified
  Theories}}, {\emph{Conf. Proc.} {\bf C790927} (1979) 315--321},
  [\href{http://arxiv.org/abs/1306.4669}{{\tt 1306.4669}}].

\bibitem{Yanagida:1979as}
T.~Yanagida, \emph{{Horizontal Symmetry And Masses Of Neutrinos}},
  {\emph{Conf.Proc.} {\bf C7902131} (1979) 95}.

\bibitem{Schechter:1980gr}
J.~Schechter and J.~W.~F. Valle, \emph{Neutrino masses in $su(2) \times u(1)$
  theories}, {\emph{Phys. Rev.} {\bf D22} (1980) 2227}.

\bibitem{Aghanim:2018eyx}
{\scshape Planck} collaboration, N.~Aghanim et~al., \emph{{Planck 2018 results.
  VI. Cosmological parameters}},  \href{http://arxiv.org/abs/1807.06209}{{\tt
  1807.06209}}.

\bibitem{Pilaftsis:1991ug}
A.~Pilaftsis, \emph{{Radiatively induced neutrino masses and large Higgs
  neutrino couplings in the standard model with Majorana fields}},
  \href{http://dx.doi.org/10.1007/BF01482590}{\emph{Z.Phys.} {\bf C55} (1992)
  275}, [\href{http://arxiv.org/abs/hep-ph/9901206}{{\tt hep-ph/9901206}}].

\bibitem{Buchmuller:1991ce}
W.~Buchmuller, C.~Greub and P.~Minkowski, \emph{{Neutrino masses, neutral
  vector bosons and the scale of B-L breaking}},
  \href{http://dx.doi.org/10.1016/0370-2693(91)90952-M}{\emph{Phys.Lett.} {\bf
  B267} (1991) 395--399}.

\bibitem{Gluza:2002vs}
J.~Gluza, \emph{{On teraelectronvolt Majorana neutrinos}}, {\emph{Acta
  Phys.Polon.} {\bf B33} (2002) 1735--1746},
  [\href{http://arxiv.org/abs/hep-ph/0201002}{{\tt hep-ph/0201002}}].

\bibitem{Pilaftsis:2004xx}
A.~Pilaftsis, \emph{{Resonant tau-leptogenesis with observable lepton number
  violation}},
  \href{http://dx.doi.org/10.1103/PhysRevLett.95.081602}{\emph{Phys.Rev.Lett.}
  {\bf 95} (2005) 081602}, [\href{http://arxiv.org/abs/hep-ph/0408103}{{\tt
  hep-ph/0408103}}].

\bibitem{Kersten:2007vk}
J.~Kersten and A.~Y. Smirnov, \emph{{Right-Handed Neutrinos at CERN LHC and the
  Mechanism of Neutrino Mass Generation}},
  \href{http://dx.doi.org/10.1103/PhysRevD.76.073005}{\emph{Phys.Rev.} {\bf
  D76} (2007) 073005}, [\href{http://arxiv.org/abs/0705.3221}{{\tt
  0705.3221}}].

\bibitem{Xing:2009in}
Z.-z. Xing, \emph{{Naturalness and Testability of TeV Seesaw Mechanisms}},
  \href{http://dx.doi.org/10.1143/PTPS.180.112}{\emph{Prog.Theor.Phys.Suppl.}
  {\bf 180} (2009) 112}, [\href{http://arxiv.org/abs/0905.3903}{{\tt
  0905.3903}}].

\bibitem{Gavela:2009cd}
M.~Gavela, T.~Hambye, D.~Hernandez and P.~Hernandez, \emph{{Minimal Flavour
  Seesaw Models}},
  \href{http://dx.doi.org/10.1088/1126-6708/2009/09/038}{\emph{JHEP} {\bf 0909}
  (2009) 038}, [\href{http://arxiv.org/abs/0906.1461}{{\tt 0906.1461}}].

\bibitem{He:2009ua}
X.-G. He, S.~Oh, J.~Tandean and C.-C. Wen, \emph{{Large Mixing of Light and
  Heavy Neutrinos in Seesaw Models and the LHC}},
  \href{http://dx.doi.org/10.1103/PhysRevD.80.073012}{\emph{Phys.Rev.} {\bf
  D80} (2009) 073012}, [\href{http://arxiv.org/abs/0907.1607}{{\tt
  0907.1607}}].

\bibitem{Adhikari:2010yt}
R.~Adhikari and A.~Raychaudhuri, \emph{{Light neutrinos from massless texture
  and below TeV seesaw scale}},
  \href{http://dx.doi.org/10.1103/PhysRevD.84.033002}{\emph{Phys.Rev.} {\bf
  D84} (2011) 033002}, [\href{http://arxiv.org/abs/1004.5111}{{\tt
  1004.5111}}].

\bibitem{Ibarra:2010xw}
A.~Ibarra, E.~Molinaro and S.~Petcov, \emph{{TeV Scale See-Saw Mechanisms of
  Neutrino Mass Generation, the Majorana Nature of the Heavy Singlet Neutrinos
  and $(\beta\beta)_{0\nu}$-Decay}},
  \href{http://dx.doi.org/10.1007/JHEP09(2010)108}{\emph{JHEP} {\bf 1009}
  (2010) 108}, [\href{http://arxiv.org/abs/1007.2378}{{\tt 1007.2378}}].

\bibitem{Deppisch:2010fr}
F.~F. Deppisch and A.~Pilaftsis, \emph{{Lepton Flavour Violation and theta(13)
  in Minimal Resonant Leptogenesis}},
  \href{http://dx.doi.org/10.1103/PhysRevD.83.076007}{\emph{Phys.Rev.} {\bf
  D83} (2011) 076007}, [\href{http://arxiv.org/abs/1012.1834}{{\tt
  1012.1834}}].

\bibitem{Ibarra:2011xn}
A.~Ibarra, E.~Molinaro and S.~T. Petcov, \emph{{Low Energy Signatures of the
  TeV Scale See-Saw Mechanism}},
  \href{http://dx.doi.org/10.1103/PhysRevD.84.013005}{\emph{Phys.Rev.} {\bf
  D84} (2011) 013005}, [\href{http://arxiv.org/abs/1103.6217}{{\tt
  1103.6217}}].

\bibitem{Mitra:2011qr}
M.~Mitra, G.~Senjanovi\'{c} and F.~Vissani, \emph{{Neutrinoless Double Beta
  Decay and Heavy Sterile Neutrinos}},
  \href{http://dx.doi.org/10.1016/j.nuclphysb.2011.10.035}{\emph{Nucl.Phys.}
  {\bf B856} (2012) 26}, [\href{http://arxiv.org/abs/1108.0004}{{\tt
  1108.0004}}].

\bibitem{Shaposhnikov:2006nn}
M.~Shaposhnikov, \emph{{A Possible symmetry of the nuMSM}},
  \href{http://dx.doi.org/10.1016/j.nuclphysb.2006.11.003}{\emph{Nucl.Phys.}
  {\bf B763} (2007) 49}, [\href{http://arxiv.org/abs/hep-ph/0605047}{{\tt
  hep-ph/0605047}}].

\bibitem{Dev:2013oxa}
C.-H. Lee, P.~S.~B. Dev and R.~N. Mohapatra, \emph{{Natural TeV-scale
  left-right seesaw mechanism for neutrinos and experimental tests}},
  \href{http://dx.doi.org/10.1103/PhysRevD.88.093010}{\emph{Phys.Rev.} {\bf
  D88} (2013) 093010}, [\href{http://arxiv.org/abs/1309.0774}{{\tt
  1309.0774}}].

\bibitem{Chattopadhyay:2017zvs}
P.~Chattopadhyay and K.~M. Patel, \emph{{Discrete symmetries for electroweak
  natural type-I seesaw mechanism}},
  \href{http://dx.doi.org/10.1016/j.nuclphysb.2017.06.008}{\emph{Nucl. Phys.}
  {\bf B921} (2017) 487--506}, [\href{http://arxiv.org/abs/1703.09541}{{\tt
  1703.09541}}].

\bibitem{mohapatra:1986aw}
R.~N. Mohapatra, \emph{Mechanism for understanding small neutrino mass in
  superstring theories},
  \href{http://dx.doi.org/10.1103/PhysRevLett.56.561}{\emph{Phys. Rev. Lett.}
  {\bf 56} (1986) 561}.

\bibitem{Nandi:1985uh}
S.~Nandi and U.~Sarkar, \emph{{A Solution to the Neutrino Mass Problem in
  Superstring E6 Theory}},
  \href{http://dx.doi.org/10.1103/PhysRevLett.56.564}{\emph{Phys.Rev.Lett.}
  {\bf 56} (1986) 564}.

\bibitem{mohapatra:1986bd}
R.~N. Mohapatra and J.~W.~F. Valle, \emph{Neutrino mass and baryon-number
  nonconservation in superstring models}, {\emph{Phys. Rev.} {\bf D34} (1986)
  1642}.

\bibitem{wyler:1983dd}
D.~Wyler and L.~Wolfenstein, \emph{Massless neutrinos in left-right symmetric
  models}, {\emph{Nucl. Phys.} {\bf B218} (1983) 205}.

\bibitem{akhmedov:1995ip}
E.~K. Akhmedov, M.~Lindner, E.~Schnapka and J.~W.~F. Valle, \emph{Left-right
  symmetry breaking in njl approach}, {\emph{Phys. Lett.} {\bf B368} (1996)
  270--280}, [\href{http://arxiv.org/abs/hep-ph/9507275}{{\tt
  hep-ph/9507275}}].

\bibitem{akhmedov:1995vm}
E.~K. Akhmedov, M.~Lindner, E.~Schnapka and J.~W.~F. Valle, \emph{Dynamical
  left-right symmetry breaking}, {\emph{Phys. Rev.} {\bf D53} (1996)
  2752--2780}, [\href{http://arxiv.org/abs/hep-ph/9509255}{{\tt
  hep-ph/9509255}}].

\bibitem{malinsky:2005bi}
M.~Malinsky, J.~C. Romao and J.~W.~F. Valle, \emph{Novel supersymmetric so(10)
  seesaw mechanism}, {\emph{Phys. Rev. Lett.} {\bf 95} (2005) 161801}.

\bibitem{Ma:2009du}
E.~Ma, \emph{{Deciphering the Seesaw Nature of Neutrino Mass from Unitarity
  Violation}},
  \href{http://dx.doi.org/10.1142/S0217732309031776}{\emph{Mod.Phys.Lett.} {\bf
  A24} (2009) 2161--2165}, [\href{http://arxiv.org/abs/0904.1580}{{\tt
  0904.1580}}].

\bibitem{Dev:2012sg}
P.~S.~B. Dev and A.~Pilaftsis, \emph{{Minimal Radiative Neutrino Mass Mechanism
  for Inverse Seesaw Models}},
  \href{http://dx.doi.org/10.1103/PhysRevD.86.113001}{\emph{Phys.Rev.} {\bf
  D86} (2012) 113001}, [\href{http://arxiv.org/abs/1209.4051}{{\tt
  1209.4051}}].

\bibitem{Dev:2012bd}
P.~S.~B. Dev and A.~Pilaftsis, \emph{{Light and Superlight Sterile Neutrinos in
  the Minimal Radiative Inverse Seesaw Model}},
  \href{http://dx.doi.org/10.1103/PhysRevD.87.053007}{\emph{Phys.Rev.} {\bf
  D87} (2013) 053007}, [\href{http://arxiv.org/abs/1212.3808}{{\tt
  1212.3808}}].

\bibitem{Bray:2007ru}
S.~Bray, J.~S. Lee and A.~Pilaftsis, \emph{{Resonant CP violation due to heavy
  neutrinos at the LHC}},
  \href{http://dx.doi.org/10.1016/j.nuclphysb.2007.07.002}{\emph{Nucl.Phys.}
  {\bf B786} (2007) 95--118}, [\href{http://arxiv.org/abs/hep-ph/0702294}{{\tt
  hep-ph/0702294}}].

\bibitem{Gluza:2016qqv}
J.~Gluza, T.~Jelinski and R.~Szafron, \emph{{Lepton number violation and
  ‘Diracness’ of massive neutrinos composed of Majorana states}},
  \href{http://dx.doi.org/10.1103/PhysRevD.93.113017}{\emph{Phys. Rev.} {\bf
  D93} (2016) 113017}, [\href{http://arxiv.org/abs/1604.01388}{{\tt
  1604.01388}}].

\bibitem{Das:2017hmg}
A.~Das, P.~S.~B. Dev and R.~N. Mohapatra, \emph{{Same Sign versus Opposite Sign
  Dileptons as a Probe of Low Scale Seesaw Mechanisms}},
  \href{http://dx.doi.org/10.1103/PhysRevD.97.015018}{\emph{Phys. Rev.} {\bf
  D97} (2018) 015018}, [\href{http://arxiv.org/abs/1709.06553}{{\tt
  1709.06553}}].

\bibitem{Dev:2019rxh}
P.~S.~B. Dev, R.~N. Mohapatra and Y.~Zhang, \emph{{CP Violating Effects in
  Heavy Neutrino Oscillations: Implications for Colliders and Leptogenesis}},
  \href{http://arxiv.org/abs/1904.04787}{{\tt 1904.04787}}.

\bibitem{Atre:2009rg}
A.~Atre, T.~Han, S.~Pascoli and B.~Zhang, \emph{{The Search for Heavy Majorana
  Neutrinos}},
  \href{http://dx.doi.org/10.1088/1126-6708/2009/05/030}{\emph{JHEP} {\bf 0905}
  (2009) 030}, [\href{http://arxiv.org/abs/0901.3589}{{\tt 0901.3589}}].

\bibitem{Deppisch:2015qwa}
F.~F. Deppisch, P.~S.~B. Dev and A.~Pilaftsis, \emph{{Neutrinos and Collider
  Physics}}, \href{http://dx.doi.org/10.1088/1367-2630/17/7/075019}{\emph{New
  J. Phys.} {\bf 17} (2015) 075019},
  [\href{http://arxiv.org/abs/1502.06541}{{\tt 1502.06541}}].

\bibitem{deGouvea:2015euy}
A.~de~Gouvea and A.~Kobach, \emph{{Global Constraints on a Heavy Neutrino}},
  \href{http://dx.doi.org/10.1103/PhysRevD.93.033005}{\emph{Phys. Rev.} {\bf
  D93} (2016) 033005}, [\href{http://arxiv.org/abs/1511.00683}{{\tt
  1511.00683}}].

\bibitem{Chrzaszcz:2019inj}
M.~Chrzaszcz, M.~Drewes, T.~E. Gonzalo, J.~Harz, S.~Krishnamurthy and
  C.~Weniger, \emph{{A frequentist analysis of three right-handed neutrinos
  with GAMBIT}},  \href{http://arxiv.org/abs/1908.02302}{{\tt 1908.02302}}.

\bibitem{thooft:1979}
G.~t'Hooft, \emph{Lectures at Cargese Summer Inst. 1979}.
\newblock World Scientific, Singapore, 1982.

\bibitem{Grimus:2002nk}
W.~Grimus and L.~Lavoura, \emph{{One-loop corrections to the seesaw mechanism
  in the multi-Higgs-doublet standard model}},
  \href{http://dx.doi.org/10.1016/S0370-2693(02)02672-2}{\emph{Phys. Lett.}
  {\bf B546} (2002) 86--95}, [\href{http://arxiv.org/abs/hep-ph/0207229}{{\tt
  hep-ph/0207229}}].

\bibitem{Fernandez-Martinez:2015hxa}
E.~Fernandez-Martinez, J.~Hernandez-Garcia, J.~Lopez-Pavon and M.~Lucente,
  \emph{{Loop level constraints on Seesaw neutrino mixing}},
  \href{http://dx.doi.org/10.1007/JHEP10(2015)130}{\emph{JHEP} {\bf 10} (2015)
  130}, [\href{http://arxiv.org/abs/1508.03051}{{\tt 1508.03051}}].

\bibitem{Casas:2001sr}
J.~A. Casas and A.~Ibarra, \emph{Oscillating neutrinos and mu --> e, gamma},
  {\emph{Nucl. Phys.} {\bf B618} (2001) 171--204},
  [\href{http://arxiv.org/abs/hep-ph/0103065}{{\tt hep-ph/0103065}}].

\bibitem{Donini:2012tt}
A.~Donini, P.~Hernandez, J.~Lopez-Pavon, M.~Maltoni and T.~Schwetz, \emph{{The
  minimal 3+2 neutrino model versus oscillation anomalies}},
  \href{http://dx.doi.org/10.1007/JHEP07(2012)161}{\emph{JHEP} {\bf 1207}
  (2012) 161}, [\href{http://arxiv.org/abs/1205.5230}{{\tt 1205.5230}}].

\bibitem{Mertens:2018vuu}
{\scshape KATRIN} collaboration, S.~Mertens et~al., \emph{{A novel detector
  system for KATRIN to search for keV-scale sterile neutrinos}},
  \href{http://dx.doi.org/10.1088/1361-6471/ab12fe}{\emph{J. Phys.} {\bf G46}
  (2019) 065203}, [\href{http://arxiv.org/abs/1810.06711}{{\tt 1810.06711}}].

\bibitem{Adams:2013qkq}
{\scshape LBNE} collaboration, C.~Adams et~al., \emph{{The Long-Baseline
  Neutrino Experiment: Exploring Fundamental Symmetries of the Universe}},
  \href{http://arxiv.org/abs/1307.7335}{{\tt 1307.7335}}.

\bibitem{Hernandez:2018cgc}
P.~Hernández, J.~Jones-Pérez and O.~Suarez-Navarro, \emph{{Majorana vs
  Pseudo-Dirac Neutrinos at the ILC}},
  \href{http://dx.doi.org/10.1140/epjc/s10052-019-6728-1}{\emph{Eur. Phys. J.}
  {\bf C79} (2019) 220}, [\href{http://arxiv.org/abs/1810.07210}{{\tt
  1810.07210}}].

\bibitem{Chao:2009ef}
W.~Chao, Z.-g. Si, Y.-j. Zheng and S.~Zhou, \emph{{Testing the Realistic Seesaw
  Model with Two Heavy Majorana Neutrinos at the CERN Large Hadron Collider}},
  \href{http://dx.doi.org/10.1016/j.physletb.2009.11.059}{\emph{Phys. Lett.}
  {\bf B683} (2010) 26--32}, [\href{http://arxiv.org/abs/0907.0935}{{\tt
  0907.0935}}].

\bibitem{Dev:2015pga}
P.~S.~B. Dev and R.~N. Mohapatra, \emph{{Unified explanation of the $eejj$,
  diboson and dijet resonances at the LHC}},
  \href{http://dx.doi.org/10.1103/PhysRevLett.115.181803}{\emph{Phys. Rev.
  Lett.} {\bf 115} (2015) 181803}, [\href{http://arxiv.org/abs/1508.02277}{{\tt
  1508.02277}}].

\bibitem{Anamiati:2016uxp}
G.~Anamiati, M.~Hirsch and E.~Nardi, \emph{{Quasi-Dirac neutrinos at the LHC}},
  \href{http://dx.doi.org/10.1007/JHEP10(2016)010}{\emph{JHEP} {\bf 10} (2016)
  010}, [\href{http://arxiv.org/abs/1607.05641}{{\tt 1607.05641}}].

\bibitem{Cirelli:2004cz}
M.~Cirelli, G.~Marandella, A.~Strumia and F.~Vissani, \emph{{Probing
  oscillations into sterile neutrinos with cosmology, astrophysics and
  experiments}},
  \href{http://dx.doi.org/10.1016/j.nuclphysb.2004.11.056}{\emph{Nucl.Phys.}
  {\bf B708} (2005) 215--267}, [\href{http://arxiv.org/abs/hep-ph/0403158}{{\tt
  hep-ph/0403158}}].

\bibitem{Donini:2011jh}
A.~Donini, P.~Hernandez, J.~Lopez-Pavon and M.~Maltoni, \emph{{Minimal models
  with light sterile neutrinos}},
  \href{http://dx.doi.org/10.1007/JHEP07(2011)105}{\emph{JHEP} {\bf 1107}
  (2011) 105}, [\href{http://arxiv.org/abs/1106.0064}{{\tt 1106.0064}}].

\bibitem{Aad:2019kiz}
{\scshape ATLAS} collaboration, G.~Aad et~al., \emph{{Search for heavy neutral
  leptons in decays of $W$ bosons produced in 13 TeV $pp$ collisions using
  prompt and displaced signatures with the ATLAS detector}},
  \href{http://arxiv.org/abs/1905.09787}{{\tt 1905.09787}}.

\bibitem{Sirunyan:2018mtv}
{\scshape CMS} collaboration, A.~M. Sirunyan et~al., \emph{{Search for heavy
  neutral leptons in events with three charged leptons in proton-proton
  collisions at $\sqrt{s} =$ 13 TeV}},
  \href{http://dx.doi.org/10.1103/PhysRevLett.120.221801}{\emph{Phys. Rev.
  Lett.} {\bf 120} (2018) 221801}, [\href{http://arxiv.org/abs/1802.02965}{{\tt
  1802.02965}}].

\bibitem{Aad:2015xaa}
{\scshape ATLAS} collaboration, G.~Aad et~al., \emph{{Search for heavy Majorana
  neutrinos with the ATLAS detector in pp collisions at $ \sqrt{s}=8 $ TeV}},
  \href{http://dx.doi.org/10.1007/JHEP07(2015)162}{\emph{JHEP} {\bf 07} (2015)
  162}, [\href{http://arxiv.org/abs/1506.06020}{{\tt 1506.06020}}].

\bibitem{Sirunyan:2018xiv}
{\scshape CMS} collaboration, A.~M. Sirunyan et~al., \emph{{Search for heavy
  Majorana neutrinos in same-sign dilepton channels in proton-proton collisions
  at $ \sqrt{s}=13 $ TeV}},
  \href{http://dx.doi.org/10.1007/JHEP01(2019)122}{\emph{JHEP} {\bf 01} (2019)
  122}, [\href{http://arxiv.org/abs/1806.10905}{{\tt 1806.10905}}].

\bibitem{Benedikt:2018csr}
{\scshape FCC} collaboration, A.~Abada et~al., \emph{{FCC-hh: The Hadron
  Collider}}, \href{http://dx.doi.org/10.1140/epjst/e2019-900087-0}{\emph{Eur.
  Phys. J. ST} {\bf 228} (2019) 755--1107}.

\bibitem{Pascoli:2018heg}
S.~Pascoli, R.~Ruiz and C.~Weiland, \emph{{Heavy neutrinos with dynamic jet
  vetoes: multilepton searches at $ \sqrt{s}=14 $ , 27, and 100 TeV}},
  \href{http://dx.doi.org/10.1007/JHEP06(2019)049}{\emph{JHEP} {\bf 06} (2019)
  049}, [\href{http://arxiv.org/abs/1812.08750}{{\tt 1812.08750}}].

\bibitem{Das:2017zjc}
A.~Das, P.~S.~B. Dev and C.~S. Kim, \emph{{Constraining Sterile Neutrinos from
  Precision Higgs Data}},
  \href{http://dx.doi.org/10.1103/PhysRevD.95.115013}{\emph{Phys. Rev.} {\bf
  D95} (2017) 115013}, [\href{http://arxiv.org/abs/1704.00880}{{\tt
  1704.00880}}].

\bibitem{Adriani:1992pq}
{\scshape L3 Collaboration} collaboration, O.~Adriani et~al., \emph{{Search for
  isosinglet neutral heavy leptons in Z0 decays}},
  \href{http://dx.doi.org/10.1016/0370-2693(92)91579-X}{\emph{Phys.Lett.} {\bf
  B295} (1992) 371--382}.

\bibitem{Achard:2001qv}
{\scshape L3 Collaboration} collaboration, P.~Achard et~al., \emph{{Search for
  heavy isosinglet neutrino in $e^{+} e^{-}$ annihilation at LEP}},
  \href{http://dx.doi.org/10.1016/S0370-2693(01)00993-5}{\emph{Phys.Lett.} {\bf
  B517} (2001) 67--74}, [\href{http://arxiv.org/abs/hep-ex/0107014}{{\tt
  hep-ex/0107014}}].

\bibitem{Abreu:1996pa}
{\scshape DELPHI Collaboration} collaboration, P.~Abreu et~al., \emph{{Search
  for neutral heavy leptons produced in Z decays}},
  \href{http://dx.doi.org/10.1007/s002880050370,
  10.1007/s002880050370}{\emph{Z.Phys.} {\bf C74} (1997) 57--71}.

\bibitem{Banerjee:2015gca}
S.~Banerjee, P.~S.~B. Dev, A.~Ibarra, T.~Mandal and M.~Mitra, \emph{{Prospects
  of Heavy Neutrino Searches at Future Lepton Colliders}},
  \href{http://arxiv.org/abs/1503.05491}{{\tt 1503.05491}}.

\bibitem{Chakraborty:2018khw}
S.~Chakraborty, M.~Mitra and S.~Shil, \emph{{Fat Jet Signature of a Heavy
  Neutrino at Lepton Collider}},
  \href{http://dx.doi.org/10.1103/PhysRevD.100.015012}{\emph{Phys. Rev.} {\bf
  D100} (2019) 015012}, [\href{http://arxiv.org/abs/1810.08970}{{\tt
  1810.08970}}].

\bibitem{Das:2018usr}
A.~Das, S.~Jana, S.~Mandal and S.~Nandi, \emph{{Probing right handed neutrinos
  at the LHeC and lepton colliders using fat jet signatures}},
  \href{http://dx.doi.org/10.1103/PhysRevD.99.055030}{\emph{Phys. Rev.} {\bf
  D99} (2019) 055030}, [\href{http://arxiv.org/abs/1811.04291}{{\tt
  1811.04291}}].

\bibitem{Blondel:2014bra}
{\scshape FCC-ee Study Team} collaboration, A.~Blondel, E.~Gaverini, N.~Serra
  and M.~Shaposhnikov, \emph{{Search for Heavy Right Handed Neutrinos at the
  FCC-ee}},  \href{http://arxiv.org/abs/1411.5230}{{\tt 1411.5230}}.

\bibitem{Mondal:2016kof}
S.~Mondal and S.~K. Rai, \emph{{Probing the Heavy Neutrinos of Inverse Seesaw
  Model at the LHeC}},
  \href{http://dx.doi.org/10.1103/PhysRevD.94.033008}{\emph{Phys. Rev.} {\bf
  D94} (2016) 033008}, [\href{http://arxiv.org/abs/1605.04508}{{\tt
  1605.04508}}].

\bibitem{Antusch:2019eiz}
S.~Antusch, O.~Fischer and A.~Hammad, \emph{{Lepton-Trijet and Displaced Vertex
  Searches for Heavy Neutrinos at Future Electron-Proton Colliders}},
  \href{http://arxiv.org/abs/1908.02852}{{\tt 1908.02852}}.

\bibitem{Antusch:2016ejd}
S.~Antusch, E.~Cazzato and O.~Fischer, \emph{{Sterile neutrino searches at
  future $e^-e^+$, $pp$, and $e^-p$ colliders}},
  \href{http://dx.doi.org/10.1142/S0217751X17500786}{\emph{Int. J. Mod. Phys.}
  {\bf A32} (2017) 1750078}, [\href{http://arxiv.org/abs/1612.02728}{{\tt
  1612.02728}}].

\bibitem{Dercks:2018wum}
D.~Dercks, H.~K. Dreiner, M.~Hirsch and Z.~S. Wang, \emph{{Long-Lived Fermions
  at AL3X}}, \href{http://dx.doi.org/10.1103/PhysRevD.99.055020}{\emph{Phys.
  Rev.} {\bf D99} (2019) 055020}, [\href{http://arxiv.org/abs/1811.01995}{{\tt
  1811.01995}}].

\bibitem{Gligorov:2017nwh}
V.~V. Gligorov, S.~Knapen, M.~Papucci and D.~J. Robinson, \emph{{Searching for
  Long-lived Particles: A Compact Detector for Exotics at LHCb}},
  \href{http://dx.doi.org/10.1103/PhysRevD.97.015023}{\emph{Phys. Rev.} {\bf
  D97} (2018) 015023}, [\href{http://arxiv.org/abs/1708.09395}{{\tt
  1708.09395}}].

\bibitem{Feng:2017uoz}
J.~L. Feng, I.~Galon, F.~Kling and S.~Trojanowski, \emph{{ForwArd Search
  ExpeRiment at the LHC}},
  \href{http://dx.doi.org/10.1103/PhysRevD.97.035001}{\emph{Phys. Rev.} {\bf
  D97} (2018) 035001}, [\href{http://arxiv.org/abs/1708.09389}{{\tt
  1708.09389}}].

\bibitem{Chou:2016lxi}
J.~P. Chou, D.~Curtin and H.~J. Lubatti, \emph{{New Detectors to Explore the
  Lifetime Frontier}},
  \href{http://dx.doi.org/10.1016/j.physletb.2017.01.043}{\emph{Phys. Lett.}
  {\bf B767} (2017) 29--36}, [\href{http://arxiv.org/abs/1606.06298}{{\tt
  1606.06298}}].

\bibitem{Frank:2019pgk}
M.~Frank, M.~de~Montigny, P.-P.~A. Ouimet, J.~Pinfold, A.~Shaa and M.~Staelens,
  \emph{{Searching for Heavy Neutrinos with the MoEDAL-MAPP Detector at the
  LHC}},  \href{http://arxiv.org/abs/1909.05216}{{\tt 1909.05216}}.

\bibitem{Kling:2018wct}
F.~Kling and S.~Trojanowski, \emph{{Heavy Neutral Leptons at FASER}},
  \href{http://dx.doi.org/10.1103/PhysRevD.97.095016}{\emph{Phys. Rev.} {\bf
  D97} (2018) 095016}, [\href{http://arxiv.org/abs/1801.08947}{{\tt
  1801.08947}}].

\bibitem{Curtin:2018mvb}
D.~Curtin et~al., \emph{{Long-Lived Particles at the Energy Frontier: The
  MATHUSLA Physics Case}},
  \href{http://dx.doi.org/10.1088/1361-6633/ab28d6}{\emph{Rept. Prog. Phys.}
  {\bf 82} (2019) 116201}, [\href{http://arxiv.org/abs/1806.07396}{{\tt
  1806.07396}}].

\bibitem{Dev:2013wba}
P.~S.~B. Dev, A.~Pilaftsis and U.-K. Yang, \emph{{New Production Mechanism for
  Heavy Neutrinos at the LHC}},
  \href{http://dx.doi.org/10.1103/PhysRevLett.112.081801}{\emph{Phys.Rev.Lett.}
  {\bf 112} (2014) 081801}, [\href{http://arxiv.org/abs/1308.2209}{{\tt
  1308.2209}}].

\bibitem{Pilaftsis:1997dr}
A.~Pilaftsis, \emph{{Resonant CP violation induced by particle mixing in
  transition amplitudes}},
  \href{http://dx.doi.org/10.1016/S0550-3213(97)00469-0}{\emph{Nucl.Phys.} {\bf
  B504} (1997) 61--107}, [\href{http://arxiv.org/abs/hep-ph/9702393}{{\tt
  hep-ph/9702393}}].

\bibitem{Helo:2010cw}
J.~C. Helo, S.~Kovalenko and I.~Schmidt, \emph{{Sterile neutrinos in lepton
  number and lepton flavor violating decays}},
  \href{http://dx.doi.org/10.1016/j.nuclphysb.2011.07.020}{\emph{Nucl. Phys.}
  {\bf B853} (2011) 80--104}, [\href{http://arxiv.org/abs/1005.1607}{{\tt
  1005.1607}}].

\bibitem{Drewes:2019byd}
M.~Drewes, J.~Klarić and P.~Klose, \emph{{On Lepton Number Violation in Heavy
  Neutrino Decays at Colliders}},
  \href{http://dx.doi.org/10.1007/JHEP11(2019)032}{\emph{JHEP} {\bf 11} (2019)
  032}, [\href{http://arxiv.org/abs/1907.13034}{{\tt 1907.13034}}].

\bibitem{PhysRevD.84.052002}
{\scshape PIENU Collaboration} collaboration, M.~Aoki, M.~Blecher, D.~A.
  Bryman, S.~Chen, M.~Ding, L.~Doria et~al., \emph{Search for massive neutrinos
  in the decay $\ensuremath{\pi}\ensuremath{\rightarrow}e\ensuremath{\nu}$},
  \href{http://dx.doi.org/10.1103/PhysRevD.84.052002}{\emph{Phys. Rev. D} {\bf
  84} (Sep, 2011) 052002}.

\bibitem{Aguilar-Arevalo:2017vlf}
{\scshape PIENU} collaboration, A.~Aguilar-Arevalo et~al., \emph{{Improved
  search for heavy neutrinos in the decay $\pi\rightarrow e\nu$}},
  \href{http://dx.doi.org/10.1103/PhysRevD.97.072012}{\emph{Phys. Rev.} {\bf
  D97} (2018) 072012}, [\href{http://arxiv.org/abs/1712.03275}{{\tt
  1712.03275}}].

\bibitem{Bryman:2019ssi}
D.~A. Bryman and R.~Shrock, \emph{{Improved Constraints on Sterile Neutrinos in
  the MeV to GeV Mass Range}},  \href{http://arxiv.org/abs/1904.06787}{{\tt
  1904.06787}}.

\bibitem{Bryman:2019bjg}
D.~A. Bryman and R.~Shrock, \emph{{Constraints on Sterile Neutrinos in the MeV
  to GeV Mass Range}},  \href{http://arxiv.org/abs/1909.11198}{{\tt
  1909.11198}}.

\bibitem{CortinaGil:2017mqf}
{\scshape NA62} collaboration, E.~Cortina~Gil et~al., \emph{{Search for heavy
  neutral lepton production in $K^+$ decays}},
  \href{http://dx.doi.org/10.1016/j.physletb.2018.01.031}{\emph{Phys. Lett.}
  {\bf B778} (2018) 137--145}, [\href{http://arxiv.org/abs/1712.00297}{{\tt
  1712.00297}}].

\bibitem{Drewes:2018gkc}
M.~Drewes, J.~Hajer, J.~Klaric and G.~Lanfranchi, \emph{{NA62 sensitivity to
  heavy neutral leptons in the low scale seesaw model}},
  \href{http://dx.doi.org/10.1007/JHEP07(2018)105}{\emph{JHEP} {\bf 07} (2018)
  105}, [\href{http://arxiv.org/abs/1801.04207}{{\tt 1801.04207}}].

\bibitem{Abada:2016plb}
A.~Abada, D.~Becirevic, O.~Sumensari, C.~Weiland and R.~Zukanovich~Funchal,
  \emph{{Sterile neutrinos facing kaon physics experiments}},
  \href{http://dx.doi.org/10.1103/PhysRevD.95.075023}{\emph{Phys. Rev.} {\bf
  D95} (2017) 075023}, [\href{http://arxiv.org/abs/1612.04737}{{\tt
  1612.04737}}].

\bibitem{Abashian:2000cg}
A.~Abashian et~al., \emph{{The Belle Detector}},
  \href{http://dx.doi.org/10.1016/S0168-9002(01)02013-7}{\emph{Nucl. Instrum.
  Meth.} {\bf A479} (2002) 117--232}.

\bibitem{Liventsev:2013zz}
{\scshape Belle Collaboration} collaboration, D.~Liventsev et~al.,
  \emph{{Search for heavy neutrinos at Belle}},
  \href{http://dx.doi.org/10.1103/PhysRevD.87.071102}{\emph{Phys.Rev.} {\bf
  D87} (2013) 071102}, [\href{http://arxiv.org/abs/1301.1105}{{\tt
  1301.1105}}].

\bibitem{Badier:1986xz}
{\scshape NA3} collaboration, J.~Badier et~al., \emph{{Mass and Lifetime Limits
  on New Longlived Particles in 300-{GeV}/$c \pi^-$ Interactions}},
  \href{http://dx.doi.org/10.1007/BF01559588}{\emph{Z. Phys.} {\bf C31} (1986)
  21}.

\bibitem{Bergsma:1985is}
{\scshape CHARM} collaboration, F.~Bergsma et~al., \emph{{A Search for Decays
  of Heavy Neutrinos in the Mass Range 0.5-{GeV} to 2.8-{GeV}}},
  \href{http://dx.doi.org/10.1016/0370-2693(86)91601-1}{\emph{Phys. Lett.} {\bf
  166B} (1986) 473--478}.

\bibitem{Vilain:1994vg}
{\scshape CHARM II Collaboration} collaboration, P.~Vilain et~al.,
  \emph{{Search for heavy isosinglet neutrinos}},
  \href{http://dx.doi.org/10.1016/0370-2693(94)01422-9}{\emph{Phys.Lett.} {\bf
  B343} (1995) 453--458}.

\bibitem{Bernardi:1987ek}
G.~Bernardi, G.~Carugno, J.~Chauveau, F.~Dicarlo, M.~Dris et~al.,
  \emph{{FURTHER LIMITS ON HEAVY NEUTRINO COUPLINGS}},
  \href{http://dx.doi.org/10.1016/0370-2693(88)90563-1}{\emph{Phys.Lett.} {\bf
  B203} (1988) 332}.

\bibitem{Baranov:1992vq}
S.~Baranov, Y.~Batusov, A.~Borisov, S.~Bunyatov, V.~Y. Valuev et~al.,
  \emph{{Search for heavy neutrinos at the IHEP-JINR neutrino detector}},
  \href{http://dx.doi.org/10.1016/0370-2693(93)90405-7}{\emph{Phys.Lett.} {\bf
  B302} (1993) 336--340}.

\bibitem{Abe:2019kgx}
{\scshape T2K} collaboration, K.~Abe et~al., \emph{{Search for heavy neutrinos
  with the T2K near detector ND280}},
  \href{http://arxiv.org/abs/1902.07598}{{\tt 1902.07598}}.

\bibitem{Krasnov:2019kdc}
I.~Krasnov, \emph{{On DUNE prospects in the search for sterile neutrinos}},
  \href{http://arxiv.org/abs/1902.06099}{{\tt 1902.06099}}.

\bibitem{Ballett:2019bgd}
P.~Ballett, T.~Boschi and S.~Pascoli, \emph{{Heavy Neutral Leptons from
  low-scale seesaws at the DUNE Near Detector}},
  \href{http://arxiv.org/abs/1905.00284}{{\tt 1905.00284}}.

\bibitem{Alekhin:2015byh}
S.~Alekhin et~al., \emph{{A facility to Search for Hidden Particles at the CERN
  SPS: the SHiP physics case}},
  \href{http://dx.doi.org/10.1088/0034-4885/79/12/124201}{\emph{Rept. Prog.
  Phys.} {\bf 79} (2016) 124201}, [\href{http://arxiv.org/abs/1504.04855}{{\tt
  1504.04855}}].

\bibitem{SHiP:2018xqw}
{\scshape SHiP} collaboration, C.~Ahdida et~al., \emph{{Sensitivity of the SHiP
  experiment to Heavy Neutral Leptons}},
  \href{http://dx.doi.org/10.1007/JHEP04(2019)077}{\emph{JHEP} {\bf 04} (2019)
  077}, [\href{http://arxiv.org/abs/1811.00930}{{\tt 1811.00930}}].

\bibitem{Kovalenko:2009td}
S.~Kovalenko, Z.~Lu and I.~Schmidt, \emph{{Lepton Number Violating Processes
  Mediated by Majorana Neutrinos at Hadron Colliders}},
  \href{http://dx.doi.org/10.1103/PhysRevD.80.073014}{\emph{Phys.Rev.} {\bf
  D80} (2009) 073014}, [\href{http://arxiv.org/abs/0907.2533}{{\tt
  0907.2533}}].

\bibitem{Abada:2017jjx}
A.~Abada, V.~De~Romeri, M.~Lucente, A.~M. Teixeira and T.~Toma,
  \emph{{Effective Majorana mass matrix from tau and pseudoscalar meson lepton
  number violating decays}},
  \href{http://dx.doi.org/10.1007/JHEP02(2018)169}{\emph{JHEP} {\bf 02} (2018)
  169}, [\href{http://arxiv.org/abs/1712.03984}{{\tt 1712.03984}}].

\bibitem{Ablikim:2019gvd}
{\scshape BESIII} collaboration, M.~Ablikim et~al., \emph{{Search for heavy
  Majorana neutrino in lepton number violating decays of $D\to K \pi e^+
  e^+$}}, \href{http://dx.doi.org/10.1103/PhysRevD.99.112002}{\emph{Phys. Rev.}
  {\bf D99} (2019) 112002}, [\href{http://arxiv.org/abs/1902.02450}{{\tt
  1902.02450}}].

\bibitem{Chun:2019nwi}
E.~J. Chun, A.~Das, S.~Mandal, M.~Mitra and N.~Sinha, \emph{{Sensitivity of
  Lepton Number Violating Meson Decays in Different Experiments}},
  \href{http://arxiv.org/abs/1908.09562}{{\tt 1908.09562}}.

\bibitem{Shrock:1980vy}
R.~E. Shrock, \emph{{New Tests For, and Bounds On, Neutrino Masses and Lepton
  Mixing}},
  \href{http://dx.doi.org/10.1016/0370-2693(80)90235-X}{\emph{Phys.Lett.} {\bf
  B96} (1980) 159}.

\bibitem{Hiddemann:1995ce}
K.~H. Hiddemann, H.~Daniel and O.~Schwentker, \emph{{Limits on neutrino masses
  from the tritium beta spectrum}},
  \href{http://dx.doi.org/10.1088/0954-3899/21/5/008}{\emph{J. Phys.} {\bf G21}
  (1995) 639--650}.

\bibitem{Kraus:2012he}
C.~Kraus, A.~Singer, K.~Valerius and C.~Weinheimer, \emph{{Limit on sterile
  neutrino contribution from the Mainz Neutrino Mass Experiment}},
  \href{http://dx.doi.org/10.1140/epjc/s10052-013-2323-z}{\emph{Eur. Phys. J.}
  {\bf C73} (2013) 2323}, [\href{http://arxiv.org/abs/1210.4194}{{\tt
  1210.4194}}].

\bibitem{Belesev:2013cba}
A.~I. Belesev, A.~I. Berlev, E.~V. Geraskin, A.~A. Golubev, N.~A. Likhovid,
  A.~A. Nozik et~al., \emph{{The search for an additional neutrino mass
  eigenstate in the 2–100 eV region from ‘Troitsk nu-mass’ data: a
  detailed analysis}},
  \href{http://dx.doi.org/10.1088/0954-3899/41/1/015001}{\emph{J. Phys.} {\bf
  G41} (2014) 015001}, [\href{http://arxiv.org/abs/1307.5687}{{\tt
  1307.5687}}].

\bibitem{Abdurashitov:2017kka}
J.~N. Abdurashitov et~al., \emph{{First measeurements in search for keV-sterile
  neutrino in tritium beta-decay by Troitsk nu-mass experiment}},
  \href{http://dx.doi.org/10.1134/S0021364017120013}{\emph{Pisma Zh. Eksp.
  Teor. Fiz.} {\bf 105} (2017) 723--724},
  [\href{http://arxiv.org/abs/1703.10779}{{\tt 1703.10779}}].

\bibitem{PhysRevC.27.1175}
F.~P. Calaprice and D.~J. Millener, \emph{Heavy neutrinos and the beta spectra
  of $^{35}\mathrm{S}$, $^{18}\mathrm{F}$, and $^{19}\mathrm{Ne}$},
  \href{http://dx.doi.org/10.1103/PhysRevC.27.1175}{\emph{Phys. Rev. C} {\bf
  27} (Mar, 1983) 1175--1181}.

\bibitem{Holzschuh:2000nj}
E.~Holzschuh, L.~Palermo, H.~Stussi and P.~Wenk, \emph{{The beta-spectrum of
  S-35 and search for the admixture of heavy neutrinos}},
  \href{http://dx.doi.org/10.1016/S0370-2693(00)00476-7}{\emph{Phys. Lett.}
  {\bf B482} (2000) 1--9}.

\bibitem{Derbin:1997ut}
A.~V. Derbin, A.~I. Egorov, S.~V. Bakhlanov and V.~N. Muratova,
  \emph{{Measurement of the Ca-45 beta spectrum in search of deviations from
  the theoretical shape}}, \href{http://dx.doi.org/10.1134/1.567508}{\emph{JETP
  Lett.} {\bf 66} (1997) 88--92}.

\bibitem{Holzschuh:1999vy}
E.~Holzschuh, W.~Kundig, L.~Palermo, H.~Stussi and P.~Wenk, \emph{{Search for
  heavy neutrinos in the beta spectrum of Ni-63}},
  \href{http://dx.doi.org/10.1016/S0370-2693(99)00200-2}{\emph{Phys. Lett.}
  {\bf B451} (1999) 247--255}.

\bibitem{Schreckenbach:1983cg}
K.~Schreckenbach, G.~Colvin and F.~Von~Feilitzsch, \emph{{SEARCH FOR MIXING OF
  HEAVY NEUTRINOS IN THE BETA+ AND BETA- SPECTRA OF THE CU-64 DECAY}},
  \href{http://dx.doi.org/10.1016/0370-2693(83)90858-4}{\emph{Phys. Lett.} {\bf
  129B} (1983) 265--268}.

\bibitem{Derbin2018}
A.~V. Derbin, I.~S. Drachnev, I.~S. Lomskaya, V.~N. Muratova, N.~V. Pilipenko,
  D.~A. Semenov et~al., \emph{Search for a neutrino with a mass of 0.01--1.0
  mev in beta decays of 144ce--144pr nuclei},
  \href{http://dx.doi.org/10.1134/S0021364018200067}{\emph{JETP Letters} {\bf
  108} (Oct, 2018) 499--503}.

\bibitem{PhysRevLett.86.1978}
M.~Galeazzi, F.~Fontanelli, F.~Gatti and S.~Vitale, \emph{Limits on the
  existence of heavy neutrinos in the range 50--1000 ev from the study of the
  ${}^{187}\mathrm{Re}$ beta decay},
  \href{http://dx.doi.org/10.1103/PhysRevLett.86.1978}{\emph{Phys. Rev. Lett.}
  {\bf 86} (Mar, 2001) 1978--1981}.

\bibitem{Adhikari:2016bei}
M.~Drewes et~al., \emph{{A White Paper on keV Sterile Neutrino Dark Matter}},
  \href{http://dx.doi.org/10.1088/1475-7516/2017/01/025}{\emph{JCAP} {\bf 1701}
  (2017) 025}, [\href{http://arxiv.org/abs/1602.04816}{{\tt 1602.04816}}].

\bibitem{Derbin:1993wy}
A.~I. Derbin, A.~V. Chernyi, L.~A. Popeko, V.~N. Muratova, G.~A. Shishkina and
  S.~I. Bakhlanov, \emph{{Experiment on anti-neutrino scattering by electrons
  at a reactor of the Rovno nuclear power plant}}, {\emph{JETP Lett.} {\bf 57}
  (1993) 768--772}.

\bibitem{PhysRevD.52.1343}
C.~Hagner, M.~Altmann, F.~v. Feilitzsch, L.~Oberauer, Y.~Declais and
  E.~Kajfasz, \emph{Experimental search for the neutrino decay
  ${\ensuremath{\nu}}_{3}$\ensuremath{\rightarrow}${\ensuremath{\nu}}_{\mathit{j}}$+${\mathit{e}}^{+}$+${\mathit{e}}^{\mathrm{\ensuremath{-}}}$
  and limits on neutrino mixing},
  \href{http://dx.doi.org/10.1103/PhysRevD.52.1343}{\emph{Phys. Rev. D} {\bf
  52} (Aug, 1995) 1343--1352}.

\bibitem{PhysRevD.88.072010}
{\scshape Borexino Collaboration} collaboration, G.~Bellini, J.~Benziger,
  D.~Bick, G.~Bonfini, D.~Bravo, M.~Buizza~Avanzini et~al., \emph{New limits on
  heavy sterile neutrino mixing in $^{8}\mathrm{B}$ decay obtained with the
  borexino detector},
  \href{http://dx.doi.org/10.1103/PhysRevD.88.072010}{\emph{Phys. Rev. D} {\bf
  88} (Oct, 2013) 072010}.

\bibitem{deSalas:2017kay}
P.~F. de~Salas, D.~V. Forero, C.~A. Ternes, M.~Tortola and J.~W.~F. Valle,
  \emph{{Status of neutrino oscillations 2018: 3$\sigma$ hint for normal mass
  ordering and improved CP sensitivity}},
  \href{http://dx.doi.org/10.1016/j.physletb.2018.06.019}{\emph{Phys. Lett.}
  {\bf B782} (2018) 633--640}, [\href{http://arxiv.org/abs/1708.01186}{{\tt
  1708.01186}}].

\bibitem{Diaz:2019fwt}
A.~Diaz, C.~A. Argüelles, G.~H. Collin, J.~M. Conrad and M.~H. Shaevitz,
  \emph{{Where Are We With Light Sterile Neutrinos?}},
  \href{http://arxiv.org/abs/1906.00045}{{\tt 1906.00045}}.

\bibitem{Alekseev:2018efk}
{\scshape DANSS} collaboration, I.~Alekseev et~al., \emph{{Search for sterile
  neutrinos at the DANSS experiment}},
  \href{http://dx.doi.org/10.1016/j.physletb.2018.10.038}{\emph{Phys. Lett.}
  {\bf B787} (2018) 56--63}, [\href{http://arxiv.org/abs/1804.04046}{{\tt
  1804.04046}}].

\bibitem{Ko:2016owz}
{\scshape NEOS} collaboration, Y.~J. Ko et~al., \emph{{Sterile Neutrino Search
  at the NEOS Experiment}},
  \href{http://dx.doi.org/10.1103/PhysRevLett.118.121802}{\emph{Phys. Rev.
  Lett.} {\bf 118} (2017) 121802}, [\href{http://arxiv.org/abs/1610.05134}{{\tt
  1610.05134}}].

\bibitem{An:2016luf}
{\scshape Daya Bay} collaboration, F.~P. An et~al., \emph{{Improved Search for
  a Light Sterile Neutrino with the Full Configuration of the Daya Bay
  Experiment}},
  \href{http://dx.doi.org/10.1103/PhysRevLett.117.151802}{\emph{Phys. Rev.
  Lett.} {\bf 117} (2016) 151802}, [\href{http://arxiv.org/abs/1607.01174}{{\tt
  1607.01174}}].

\bibitem{Berryman:2019nvr}
J.~M. Berryman, \emph{{Constraining Sterile Neutrino Cosmology with Terrestrial
  Oscillation Experiments}},
  \href{http://dx.doi.org/10.1103/PhysRevD.100.023540}{\emph{Phys. Rev.} {\bf
  D100} (2019) 023540}, [\href{http://arxiv.org/abs/1905.03254}{{\tt
  1905.03254}}].

\bibitem{Barry:2011wb}
J.~Barry, W.~Rodejohann and H.~Zhang, \emph{{Light Sterile Neutrinos: Models
  and Phenomenology}},
  \href{http://dx.doi.org/10.1007/JHEP07(2011)091}{\emph{JHEP} {\bf 07} (2011)
  091}, [\href{http://arxiv.org/abs/1105.3911}{{\tt 1105.3911}}].

\bibitem{Abada:2017ieq}
A.~Abada, G.~Arcadi, V.~Domcke and M.~Lucente, \emph{{Neutrino masses,
  leptogenesis and dark matter from small lepton number violation?}},
  \href{http://dx.doi.org/10.1088/1475-7516/2017/12/024}{\emph{JCAP} {\bf 1712}
  (2017) 024}, [\href{http://arxiv.org/abs/1709.00415}{{\tt 1709.00415}}].

\bibitem{Ashenfelter:2018iov}
{\scshape PROSPECT} collaboration, J.~Ashenfelter et~al., \emph{{First search
  for short-baseline neutrino oscillations at HFIR with PROSPECT}},
  \href{http://dx.doi.org/10.1103/PhysRevLett.121.251802}{\emph{Phys. Rev.
  Lett.} {\bf 121} (2018) 251802}, [\href{http://arxiv.org/abs/1806.02784}{{\tt
  1806.02784}}].

\bibitem{Dentler:2018sju}
M.~Dentler, A.~Hernandez-Cabezudo, J.~Kopp, P.~A.~N. Machado, M.~Maltoni,
  I.~Martinez-Soler et~al., \emph{{Updated Global Analysis of Neutrino
  Oscillations in the Presence of eV-Scale Sterile Neutrinos}},
  \href{http://dx.doi.org/10.1007/JHEP08(2018)010}{\emph{JHEP} {\bf 08} (2018)
  010}, [\href{http://arxiv.org/abs/1803.10661}{{\tt 1803.10661}}].

\bibitem{Bolton:2019wta}
P.~D. Bolton and F.~F. Deppisch, \emph{{Probing nonstandard lepton number
  violating interactions in neutrino oscillations}},
  \href{http://dx.doi.org/10.1103/PhysRevD.99.115011}{\emph{Phys. Rev.} {\bf
  D99} (2019) 115011}, [\href{http://arxiv.org/abs/1903.06557}{{\tt
  1903.06557}}].

\bibitem{Abada:2007ux}
A.~Abada, C.~Biggio, F.~Bonnet, M.~B. Gavela and T.~Hambye, \emph{{Low energy
  effects of neutrino masses}},
  \href{http://dx.doi.org/10.1088/1126-6708/2007/12/061}{\emph{JHEP} {\bf 12}
  (2007) 061}, [\href{http://arxiv.org/abs/0707.4058}{{\tt 0707.4058}}].

\bibitem{Fernandez-Martinez:2016lgt}
E.~Fernandez-Martinez, J.~Hernandez-Garcia and J.~Lopez-Pavon, \emph{{Global
  constraints on heavy neutrino mixing}},
  \href{http://dx.doi.org/10.1007/JHEP08(2016)033}{\emph{JHEP} {\bf 08} (2016)
  033}, [\href{http://arxiv.org/abs/1605.08774}{{\tt 1605.08774}}].

\bibitem{Blennow:2016jkn}
M.~Blennow, P.~Coloma, E.~Fernandez-Martinez, J.~Hernandez-Garcia and
  J.~Lopez-Pavon, \emph{{Non-Unitarity, sterile neutrinos, and Non-Standard
  neutrino Interactions}},
  \href{http://dx.doi.org/10.1007/JHEP04(2017)153}{\emph{JHEP} {\bf 04} (2017)
  153}, [\href{http://arxiv.org/abs/1609.08637}{{\tt 1609.08637}}].

\bibitem{delAguila:2008pw}
F.~del Aguila, J.~de~Blas and M.~Perez-Victoria, \emph{{Effects of new leptons
  in Electroweak Precision Data}},
  \href{http://dx.doi.org/10.1103/PhysRevD.78.013010}{\emph{Phys.Rev.} {\bf
  D78} (2008) 013010}, [\href{http://arxiv.org/abs/0803.4008}{{\tt
  0803.4008}}].

\bibitem{Akhmedov:2013hec}
E.~Akhmedov, A.~Kartavtsev, M.~Lindner, L.~Michaels and J.~Smirnov,
  \emph{{Improving Electro-Weak Fits with TeV-scale Sterile Neutrinos}},
  \href{http://dx.doi.org/10.1007/JHEP05(2013)081}{\emph{JHEP} {\bf 1305}
  (2013) 081}, [\href{http://arxiv.org/abs/1302.1872}{{\tt 1302.1872}}].

\bibitem{deBlas:2013gla}
J.~de~Blas, \emph{{Electroweak limits on physics beyond the Standard Model}},
  \href{http://dx.doi.org/10.1051/epjconf/20136019008}{\emph{EPJ Web Conf.}
  {\bf 60} (2013) 19008}, [\href{http://arxiv.org/abs/1307.6173}{{\tt
  1307.6173}}].

\bibitem{Antusch:2014woa}
S.~Antusch and O.~Fischer, \emph{{Non-unitarity of the leptonic mixing matrix:
  Present bounds and future sensitivities}},
  \href{http://dx.doi.org/10.1007/JHEP10(2014)094}{\emph{JHEP} {\bf 1410}
  (2014) 94}, [\href{http://arxiv.org/abs/1407.6607}{{\tt 1407.6607}}].

\bibitem{Flieger:2019eor}
W.~Flieger, J.~Gluza and K.~Porwit, \emph{{New limits on neutrino non-standard
  mixings based on prescribed singular values}},
  \href{http://arxiv.org/abs/1910.01233}{{\tt 1910.01233}}.

\bibitem{Deppisch:2012vj}
F.~F. Deppisch, \emph{{Lepton Flavour Violation and Flavour Symmetries}},
  \href{http://dx.doi.org/10.1002/prop.201200126}{\emph{Fortsch.Phys.} {\bf 61}
  (2013) 622--644}, [\href{http://arxiv.org/abs/1206.5212}{{\tt 1206.5212}}].

\bibitem{Dohmen:1993mp}
{\scshape SINDRUM II Collaboration.} collaboration, C.~Dohmen et~al.,
  \emph{{Test of lepton flavor conservation in mu - e conversion on titanium}},
  \href{http://dx.doi.org/10.1016/0370-2693(93)91383-X}{\emph{Phys.Lett.} {\bf
  B317} (1993) 631--636}.

\bibitem{Abazajian:2012ys}
K.~N. Abazajian et~al., \emph{{Light Sterile Neutrinos: A White Paper}},
  \href{http://arxiv.org/abs/1204.5379}{{\tt 1204.5379}}.

\bibitem{Boyarsky:2009ix}
A.~Boyarsky, O.~Ruchayskiy and M.~Shaposhnikov, \emph{{The Role of sterile
  neutrinos in cosmology and astrophysics}},
  \href{http://dx.doi.org/10.1146/annurev.nucl.010909.083654}{\emph{Ann.Rev.Nucl.Part.Sci.}
  {\bf 59} (2009) 191--214}, [\href{http://arxiv.org/abs/0901.0011}{{\tt
  0901.0011}}].

\bibitem{Ruchayskiy:2012si}
O.~Ruchayskiy and A.~Ivashko, \emph{{Restrictions on the lifetime of sterile
  neutrinos from primordial nucleosynthesis}},
  \href{http://dx.doi.org/10.1088/1475-7516/2012/10/014}{\emph{JCAP} {\bf 1210}
  (2012) 014}, [\href{http://arxiv.org/abs/1202.2841}{{\tt 1202.2841}}].

\bibitem{Bezrukov:2009th}
F.~Bezrukov, H.~Hettmansperger and M.~Lindner, \emph{{keV sterile neutrino Dark
  Matter in gauge extensions of the Standard Model}},
  \href{http://dx.doi.org/10.1103/PhysRevD.81.085032}{\emph{Phys. Rev.} {\bf
  D81} (2010) 085032}, [\href{http://arxiv.org/abs/0912.4415}{{\tt
  0912.4415}}].

\bibitem{Nemevsek:2012cd}
M.~Nemevsek, G.~Senjanovic and Y.~Zhang, \emph{{Warm Dark Matter in Low Scale
  Left-Right Theory}},
  \href{http://dx.doi.org/10.1088/1475-7516/2012/07/006}{\emph{JCAP} {\bf 1207}
  (2012) 006}, [\href{http://arxiv.org/abs/1205.0844}{{\tt 1205.0844}}].

\bibitem{El-Zant:2013nta}
A.~El-Zant, S.~Khalil and A.~Sil, \emph{{Warm dark matter in a $B-L$ inverse
  seesaw scenario}},
  \href{http://dx.doi.org/10.1103/PhysRevD.91.035030}{\emph{Phys. Rev.} {\bf
  D91} (2015) 035030}, [\href{http://arxiv.org/abs/1308.0836}{{\tt
  1308.0836}}].

\bibitem{Biswas:2018iny}
A.~Biswas, D.~Borah and D.~Nanda, \emph{{keV Neutrino Dark Matter in a Fast
  Expanding Universe}},
  \href{http://dx.doi.org/10.1016/j.physletb.2018.10.012}{\emph{Phys. Lett.}
  {\bf B786} (2018) 364--372}, [\href{http://arxiv.org/abs/1809.03519}{{\tt
  1809.03519}}].

\bibitem{Vincent:2014rja}
A.~C. Vincent, E.~F. Martinez, P.~Hernández, M.~Lattanzi and O.~Mena,
  \emph{{Revisiting cosmological bounds on sterile neutrinos}},
  \href{http://dx.doi.org/10.1088/1475-7516/2015/04/006}{\emph{JCAP} {\bf 1504}
  (2015) 006}, [\href{http://arxiv.org/abs/1408.1956}{{\tt 1408.1956}}].

\bibitem{Abada:2014zra}
A.~Abada, G.~Arcadi and M.~Lucente, \emph{{Dark Matter in the minimal Inverse
  Seesaw mechanism}},
  \href{http://dx.doi.org/10.1088/1475-7516/2014/10/001}{\emph{JCAP} {\bf 1410}
  (2014) 001}, [\href{http://arxiv.org/abs/1406.6556}{{\tt 1406.6556}}].

\bibitem{Abazajian:2017tcc}
K.~N. Abazajian, \emph{{Sterile neutrinos in cosmology}},
  \href{http://dx.doi.org/10.1016/j.physrep.2017.10.003}{\emph{Phys. Rept.}
  {\bf 711-712} (2017) 1--28}, [\href{http://arxiv.org/abs/1705.01837}{{\tt
  1705.01837}}].

\bibitem{Shi:1998km}
X.-D. Shi and G.~M. Fuller, \emph{{A New dark matter candidate: Nonthermal
  sterile neutrinos}},
  \href{http://dx.doi.org/10.1103/PhysRevLett.82.2832}{\emph{Phys. Rev. Lett.}
  {\bf 82} (1999) 2832--2835},
  [\href{http://arxiv.org/abs/astro-ph/9810076}{{\tt astro-ph/9810076}}].

\bibitem{Dodelson:1993je}
S.~Dodelson and L.~M. Widrow, \emph{{Sterile-neutrinos as dark matter}},
  \href{http://dx.doi.org/10.1103/PhysRevLett.72.17}{\emph{Phys. Rev. Lett.}
  {\bf 72} (1994) 17--20}, [\href{http://arxiv.org/abs/hep-ph/9303287}{{\tt
  hep-ph/9303287}}].

\bibitem{Bulbul:2014sua}
E.~Bulbul et~al., \emph{{Detection of An Unidentified Emission Line in the
  Stacked X-ray spectrum of Galaxy Clusters}},
  \href{http://dx.doi.org/10.1088/0004-637X/789/1/13}{\emph{Astrophys.J.} {\bf
  789} (2014) 13}, [\href{http://arxiv.org/abs/1402.2301}{{\tt 1402.2301}}].

\bibitem{Boyarsky:2014jta}
A.~Boyarsky, O.~Ruchayskiy, D.~Iakubovskyi and J.~Franse, \emph{{Unidentified
  Line in X-Ray Spectra of the Andromeda Galaxy and Perseus Galaxy Cluster}},
  \href{http://dx.doi.org/10.1103/PhysRevLett.113.251301}{\emph{Phys.Rev.Lett.}
  {\bf 113} (2014) 251301}, [\href{http://arxiv.org/abs/1402.4119}{{\tt
  1402.4119}}].

\bibitem{Boyarsky:2014ska}
A.~Boyarsky, J.~Franse, D.~Iakubovskyi and O.~Ruchayskiy, \emph{{Checking the
  dark matter origin of 3.53~keV line with the Milky Way center}},
  \href{http://arxiv.org/abs/1408.2503}{{\tt 1408.2503}}.

\bibitem{Ng:2019gch}
K.~C.~Y. Ng, B.~M. Roach, K.~Perez, J.~F. Beacom, S.~Horiuchi, R.~Krivonos
  et~al., \emph{{New Constraints on Sterile Neutrino Dark Matter from $NuSTAR$
  M31 Observations}},
  \href{http://dx.doi.org/10.1103/PhysRevD.99.083005}{\emph{Phys. Rev.} {\bf
  D99} (2019) 083005}, [\href{http://arxiv.org/abs/1901.01262}{{\tt
  1901.01262}}].

\bibitem{Roach:2019ctw}
B.~M. Roach, K.~C.~Y. Ng, K.~Perez, J.~F. Beacom, S.~Horiuchi, R.~Krivonos
  et~al., \emph{{NuSTAR Tests of Sterile-Neutrino Dark Matter: New Galactic
  Bulge Observations and Combined Impact}},
  \href{http://arxiv.org/abs/1908.09037}{{\tt 1908.09037}}.

\bibitem{Neronov:2015kca}
A.~Neronov and D.~Malyshev, \emph{{Toward a full test of the $\nu$MSM sterile
  neutrino dark matter model with Athena}},
  \href{http://dx.doi.org/10.1103/PhysRevD.93.063518}{\emph{Phys. Rev.} {\bf
  D93} (2016) 063518}, [\href{http://arxiv.org/abs/1509.02758}{{\tt
  1509.02758}}].

\bibitem{hirata:1987hu}
{\scshape Kamiokande-II collaboration} collaboration, K.~Hirata et~al.,
  \emph{Observation of a neutrino burst from the supernova sn1987a},
  {\emph{Phys. Rev. Lett.} {\bf 58} (1987) 1490--1493}.

\bibitem{bionta:1987qt}
R.~M. Bionta et~al., \emph{Observation of a neutrino burst in coincidence with
  supernova sn 1987a in the large magellanic cloud}, {\emph{Phys. Rev. Lett.}
  {\bf 58} (1987) 1494}.

\bibitem{Kainulainen:1990bn}
K.~Kainulainen, J.~Maalampi and J.~T. Peltoniemi, \emph{{Inert neutrinos in
  supernovae}},
  \href{http://dx.doi.org/10.1016/0550-3213(91)90354-Z}{\emph{Nucl. Phys.} {\bf
  B358} (1991) 435--446}.

\bibitem{Shi:1993ee}
X.~Shi and G.~Sigl, \emph{{A Type II supernovae constraint on electron-neutrino
  - sterile-neutrino mixing}},
  \href{http://dx.doi.org/10.1016/0370-2693(94)90233-X,
  10.1016/0370-2693(94)91232-7}{\emph{Phys. Lett.} {\bf B323} (1994) 360--366},
  [\href{http://arxiv.org/abs/hep-ph/9312247}{{\tt hep-ph/9312247}}].

\bibitem{Nunokawa:1997ct}
H.~Nunokawa, J.~T. Peltoniemi, A.~Rossi and J.~W.~F. Valle, \emph{{Supernova
  bounds on resonant active sterile neutrino conversions}},
  \href{http://dx.doi.org/10.1103/PhysRevD.56.1704}{\emph{Phys. Rev.} {\bf D56}
  (1997) 1704--1713}, [\href{http://arxiv.org/abs/hep-ph/9702372}{{\tt
  hep-ph/9702372}}].

\bibitem{Hidaka:2006sg}
J.~Hidaka and G.~M. Fuller, \emph{{Dark matter sterile neutrinos in stellar
  collapse: Alteration of energy/lepton number transport and a mechanism for
  supernova explosion enhancement}},
  \href{http://dx.doi.org/10.1103/PhysRevD.74.125015}{\emph{Phys. Rev.} {\bf
  D74} (2006) 125015}, [\href{http://arxiv.org/abs/astro-ph/0609425}{{\tt
  astro-ph/0609425}}].

\bibitem{Hidaka:2007se}
J.~Hidaka and G.~M. Fuller, \emph{{Sterile Neutrino-Enhanced Supernova
  Explosions}}, \href{http://dx.doi.org/10.1103/PhysRevD.76.083516}{\emph{Phys.
  Rev.} {\bf D76} (2007) 083516}, [\href{http://arxiv.org/abs/0706.3886}{{\tt
  0706.3886}}].

\bibitem{Tamborra:2011is}
I.~Tamborra, G.~G. Raffelt, L.~Hudepohl and H.-T. Janka, \emph{{Impact of
  eV-mass sterile neutrinos on neutrino-driven supernova outflows}},
  \href{http://dx.doi.org/10.1088/1475-7516/2012/01/013}{\emph{JCAP} {\bf 1201}
  (2012) 013}, [\href{http://arxiv.org/abs/1110.2104}{{\tt 1110.2104}}].

\bibitem{Warren:2014qza}
M.~L. Warren, M.~Meixner, G.~Mathews, J.~Hidaka and T.~Kajino, \emph{{Sterile
  neutrino oscillations in core-collapse supernovae}},
  \href{http://dx.doi.org/10.1103/PhysRevD.90.103007}{\emph{Phys. Rev.} {\bf
  D90} (2014) 103007}, [\href{http://arxiv.org/abs/1405.6101}{{\tt
  1405.6101}}].

\bibitem{Fuller:2009zz}
G.~M. Fuller, A.~Kusenko and K.~Petraki, \emph{{Heavy sterile neutrinos and
  supernova explosions}},
  \href{http://dx.doi.org/10.1016/j.physletb.2008.11.016}{\emph{Phys. Lett.}
  {\bf B670} (2009) 281--284}, [\href{http://arxiv.org/abs/0806.4273}{{\tt
  0806.4273}}].

\bibitem{Raffelt:2011nc}
G.~G. Raffelt and S.~Zhou, \emph{{Supernova bound on keV-mass sterile neutrinos
  reexamined}}, \href{http://dx.doi.org/10.1103/PhysRevD.83.093014}{\emph{Phys.
  Rev.} {\bf D83} (2011) 093014}, [\href{http://arxiv.org/abs/1102.5124}{{\tt
  1102.5124}}].

\bibitem{Arguelles:2016uwb}
C.~A. Argüelles, V.~Brdar and J.~Kopp, \emph{{Production of keV Sterile
  Neutrinos in Supernovae: New Constraints and Gamma Ray Observables}},
  \href{http://dx.doi.org/10.1103/PhysRevD.99.043012}{\emph{Phys. Rev.} {\bf
  D99} (2019) 043012}, [\href{http://arxiv.org/abs/1605.00654}{{\tt
  1605.00654}}].

\bibitem{Oberauer:1993yr}
L.~Oberauer, C.~Hagner, G.~Raffelt and E.~Rieger, \emph{{Supernova bounds on
  neutrino radiative decays}},
  \href{http://dx.doi.org/10.1016/0927-6505(93)90004-W}{\emph{Astropart. Phys.}
  {\bf 1} (1993) 377--386}.

\bibitem{Gariazzo:2015rra}
S.~Gariazzo, C.~Giunti, M.~Laveder, Y.~F. Li and E.~M. Zavanin, \emph{{Light
  sterile neutrinos}},
  \href{http://dx.doi.org/10.1088/0954-3899/43/3/033001}{\emph{J. Phys.} {\bf
  G43} (2016) 033001}, [\href{http://arxiv.org/abs/1507.08204}{{\tt
  1507.08204}}].

\bibitem{Asaka:2006nq}
T.~Asaka, M.~Laine and M.~Shaposhnikov, \emph{{Lightest sterile neutrino
  abundance within the nuMSM}},
  \href{http://dx.doi.org/10.1088/1126-6708/2007/01/091,
  10.1007/JHEP02(2015)028}{\emph{JHEP} {\bf 01} (2007) 091},
  [\href{http://arxiv.org/abs/hep-ph/0612182}{{\tt hep-ph/0612182}}].

\bibitem{Dolgov:2003sg}
A.~D. Dolgov and F.~L. Villante, \emph{{BBN bounds on active sterile neutrino
  mixing}},
  \href{http://dx.doi.org/10.1016/j.nuclphysb.2003.11.031}{\emph{Nucl. Phys.}
  {\bf B679} (2004) 261--298}, [\href{http://arxiv.org/abs/hep-ph/0308083}{{\tt
  hep-ph/0308083}}].

\bibitem{Hannestad:2015tea}
S.~Hannestad, R.~S. Hansen, T.~Tram and Y.~Y.~Y. Wong, \emph{{Active-sterile
  neutrino oscillations in the early Universe with full collision terms}},
  \href{http://dx.doi.org/10.1088/1475-7516/2015/08/019}{\emph{JCAP} {\bf 1508}
  (2015) 019}, [\href{http://arxiv.org/abs/1506.05266}{{\tt 1506.05266}}].

\bibitem{Mirizzi:2013gnd}
A.~Mirizzi, G.~Mangano, N.~Saviano, E.~Borriello, C.~Giunti, G.~Miele et~al.,
  \emph{{The strongest bounds on active-sterile neutrino mixing after Planck
  data}}, \href{http://dx.doi.org/10.1016/j.physletb.2013.08.015}{\emph{Phys.
  Lett.} {\bf B726} (2013) 8--14}, [\href{http://arxiv.org/abs/1303.5368}{{\tt
  1303.5368}}].

\bibitem{Bridle:2016isd}
S.~Bridle, J.~Elvin-Poole, J.~Evans, S.~Fernandez, P.~Guzowski and
  S.~Soldner-Rembold, \emph{{A Combined View of Sterile-Neutrino Constraints
  from CMB and Neutrino Oscillation Measurements}},
  \href{http://dx.doi.org/10.1016/j.physletb.2016.11.050}{\emph{Phys. Lett.}
  {\bf B764} (2017) 322--327}, [\href{http://arxiv.org/abs/1607.00032}{{\tt
  1607.00032}}].

\bibitem{LopezPavon:2012zg}
J.~Lopez-Pavon, S.~Pascoli and C.-f. Wong, \emph{{Can heavy neutrinos dominate
  neutrinoless double beta decay?}},
  \href{http://dx.doi.org/10.1103/PhysRevD.87.093007}{\emph{Phys. Rev.} {\bf
  D87} (2013) 093007}, [\href{http://arxiv.org/abs/1209.5342}{{\tt
  1209.5342}}].

\bibitem{Awasthi:2013ff}
R.~L. Awasthi, M.~K. Parida and S.~Patra, \emph{{Neutrino masses, dominant
  neutrinoless double beta decay, and observable lepton flavor violation in
  left-right models and SO(10) grand unification with low mass $ W_R, Z_R$
  bosons}}, \href{http://dx.doi.org/10.1007/JHEP08(2013)122}{\emph{JHEP} {\bf
  1308} (2013) 122}, [\href{http://arxiv.org/abs/1302.0672}{{\tt 1302.0672}}].

\bibitem{Abada:2018qok}
A.~Abada, A.~Hernandez-Cabezudo and X.~Marcano, \emph{{Beta and Neutrinoless
  Double Beta Decays with KeV Sterile Fermions}},
  \href{http://dx.doi.org/10.1007/JHEP01(2019)041}{\emph{JHEP} {\bf 01} (2019)
  041}, [\href{http://arxiv.org/abs/1807.01331}{{\tt 1807.01331}}].

\bibitem{Tello:2010am}
V.~Tello, M.~Nemevsek, F.~Nesti, G.~Senjanovic and F.~Vissani,
  \emph{{Left-Right Symmetry: from LHC to Neutrinoless Double Beta Decay}},
  \href{http://dx.doi.org/10.1103/PhysRevLett.106.151801}{\emph{Phys.Rev.Lett.}
  {\bf 106} (2011) 151801}, [\href{http://arxiv.org/abs/1011.3522}{{\tt
  1011.3522}}].

\bibitem{Chakrabortty:2012mh}
J.~Chakrabortty, H.~Z. Devi, S.~Goswami and S.~Patra, \emph{{Neutrinoless
  double-$\beta$ decay in TeV scale Left-Right symmetric models}},
  \href{http://dx.doi.org/10.1007/JHEP08(2012)008}{\emph{JHEP} {\bf 1208}
  (2012) 008}, [\href{http://arxiv.org/abs/1204.2527}{{\tt 1204.2527}}].

\bibitem{Barry:2013xxa}
J.~Barry and W.~Rodejohann, \emph{{Lepton number and flavour violation in
  TeV-scale left-right symmetric theories with large left-right mixing}},
  \href{http://dx.doi.org/10.1007/JHEP09(2013)153}{\emph{JHEP} {\bf 1309}
  (2013) 153}, [\href{http://arxiv.org/abs/1303.6324}{{\tt 1303.6324}}].

\bibitem{Dev:2013vxa}
P.~S.~B. Dev, S.~Goswami, M.~Mitra and W.~Rodejohann, \emph{{Constraining
  Neutrino Mass from Neutrinoless Double Beta Decay}},
  \href{http://dx.doi.org/10.1103/PhysRevD.88.091301}{\emph{Phys.Rev.} {\bf
  D88} (2013) 091301}, [\href{http://arxiv.org/abs/1305.0056}{{\tt
  1305.0056}}].

\bibitem{Huang:2013kma}
W.-C. Huang and J.~Lopez-Pavon, \emph{{On neutrinoless double beta decay in the
  minimal left-right symmetric model}},
  \href{http://dx.doi.org/10.1140/epjc/s10052-014-2853-z}{\emph{Eur.Phys.J.}
  {\bf C74} (2014) 2853}, [\href{http://arxiv.org/abs/1310.0265}{{\tt
  1310.0265}}].

\bibitem{Stefanik:2015twa}
D.~Stefanik, R.~Dvornicky, F.~Simkovic and P.~Vogel, \emph{{Reexamining the
  light neutrino exchange mechanism of the $0\nu\beta\beta$ decay with left-
  and right-handed leptonic and hadronic currents}},
  \href{http://dx.doi.org/10.1103/PhysRevC.92.055502}{\emph{Phys. Rev.} {\bf
  C92} (2015) 055502}, [\href{http://arxiv.org/abs/1506.07145}{{\tt
  1506.07145}}].

\bibitem{Horoi:2015gdv}
M.~Horoi and A.~Neacsu, \emph{{Analysis of mechanisms that could contribute to
  neutrinoless double-beta decay}},
  \href{http://dx.doi.org/10.1103/PhysRevD.93.113014}{\emph{Phys. Rev.} {\bf
  D93} (2016) 113014}, [\href{http://arxiv.org/abs/1511.00670}{{\tt
  1511.00670}}].

\bibitem{Deppisch:2017vne}
F.~F. Deppisch, C.~Hati, S.~Patra, P.~Pritimita and U.~Sarkar,
  \emph{{Neutrinoless double beta decay in left-right symmetric models with a
  universal seesaw mechanism}},
  \href{http://dx.doi.org/10.1103/PhysRevD.97.035005}{\emph{Phys. Rev.} {\bf
  D97} (2018) 035005}, [\href{http://arxiv.org/abs/1701.02107}{{\tt
  1701.02107}}].

\bibitem{Bolton:2019bou}
P.~D. Bolton, F.~F. Deppisch, C.~Hati, S.~Patra and U.~Sarkar,
  \emph{{Alternative formulation of left-right symmetry with $B-L$ conservation
  and purely Dirac neutrinos}},
  \href{http://dx.doi.org/10.1103/PhysRevD.100.035013}{\emph{Phys. Rev.} {\bf
  D100} (2019) 035013}, [\href{http://arxiv.org/abs/1902.05802}{{\tt
  1902.05802}}].

\bibitem{P_s_1999}
H.~Päs, M.~Hirsch, H.~Klapdor-Kleingrothaus and S.~Kovalenko, \emph{Towards a
  superformula for neutrinoless double beta decay},
  \href{http://dx.doi.org/10.1016/s0370-2693(99)00330-5}{\emph{Physics Letters
  B} {\bf 453} (May, 1999) 194–198}.

\bibitem{Drewes:2019mhg}
M.~Drewes, \emph{{On the Minimal Mixing of Heavy Neutrinos}},
  \href{http://arxiv.org/abs/1904.11959}{{\tt 1904.11959}}.

\bibitem{Benes:2005hn}
P.~Benes, A.~Faessler, F.~Simkovic and S.~Kovalenko, \emph{{Sterile neutrinos
  in neutrinoless double beta decay}},
  \href{http://dx.doi.org/10.1103/PhysRevD.71.077901}{\emph{Phys.Rev.} {\bf
  D71} (2005) 077901}, [\href{http://arxiv.org/abs/hep-ph/0501295}{{\tt
  hep-ph/0501295}}].

\bibitem{Kotila:2012zza}
J.~Kotila and F.~Iachello, \emph{{Phase space factors for double-$\beta$
  decay}}, \href{http://dx.doi.org/10.1103/PhysRevC.85.034316}{\emph{Phys.Rev.}
  {\bf C85} (2012) 034316}, [\href{http://arxiv.org/abs/1209.5722}{{\tt
  1209.5722}}].

\bibitem{Stoica:2013lka}
S.~Stoica and M.~Mirea, \emph{{New calculations for phase space factors
  involved in double-$\beta$ decay}},
  \href{http://dx.doi.org/10.1103/PhysRevC.88.037303}{\emph{Phys. Rev.} {\bf
  C88} (2013) 037303}, [\href{http://arxiv.org/abs/1307.0290}{{\tt
  1307.0290}}].

\bibitem{Mirea:2014dza}
M.~Mirea, T.~Pahomi and S.~Stoica, \emph{{Phase Space Factors for Double Beta
  Decay: an up-date}},  \href{http://arxiv.org/abs/1411.5506}{{\tt 1411.5506}}.

\bibitem{Faessler:2014kka}
A.~Faessler, M.~Gonzalez, S.~Kovalenko and F.~Simkovic, \emph{{Arbitrary mass
  Majorana neutrinos in neutrinoless double beta decay}},
  \href{http://dx.doi.org/10.1103/PhysRevD.90.096010}{\emph{Phys.Rev.} {\bf
  D90} (2014) 096010}, [\href{http://arxiv.org/abs/1408.6077}{{\tt
  1408.6077}}].

\bibitem{Hybvarinen:2015}
J.~Hyv\"arinen and J.~Suhonen, \emph{Nuclear matrix elements for
  $0\ensuremath{\nu}\ensuremath{\beta}\ensuremath{\beta}$ decays with light or
  heavy majorana-neutrino exchange},
  \href{http://dx.doi.org/10.1103/PhysRevC.91.024613}{\emph{Phys. Rev. C} {\bf
  91} (Feb, 2015) 024613}.

\bibitem{Barea:2015zfa}
J.~Barea, J.~Kotila and F.~Iachello, \emph{{Limits on sterile neutrino
  contributions to neutrinoless double beta decay}},
  \href{http://dx.doi.org/10.1103/PhysRevD.92.093001}{\emph{Phys. Rev.} {\bf
  D92} (2015) 093001}, [\href{http://arxiv.org/abs/1509.01925}{{\tt
  1509.01925}}].

\bibitem{Blennow:2010th}
M.~Blennow, E.~Fernandez-Martinez, J.~Lopez-Pavon and J.~Menendez,
  \emph{{Neutrinoless double beta decay in seesaw models}},
  \href{http://dx.doi.org/10.1007/JHEP07(2010)096}{\emph{JHEP} {\bf 07} (2010)
  096}, [\href{http://arxiv.org/abs/1005.3240}{{\tt 1005.3240}}].

\bibitem{Barea:2013bz}
J.~Barea, J.~Kotila and F.~Iachello, \emph{{Nuclear matrix elements for
  double-$\beta$ decay}},
  \href{http://dx.doi.org/10.1103/PhysRevC.87.014315}{\emph{Phys. Rev.} {\bf
  C87} (2013) 014315}, [\href{http://arxiv.org/abs/1301.4203}{{\tt
  1301.4203}}].

\bibitem{Graf:2018ozy}
L.~Graf, F.~F. Deppisch, F.~Iachello and J.~Kotila, \emph{{Short-Range
  Neutrinoless Double Beta Decay Mechanisms}},
  \href{http://dx.doi.org/10.1103/PhysRevD.98.095023}{\emph{Phys. Rev.} {\bf
  D98} (2018) 095023}, [\href{http://arxiv.org/abs/1806.06058}{{\tt
  1806.06058}}].

\bibitem{Engel:2016xgb}
J.~Engel and J.~Menéndez, \emph{{Status and Future of Nuclear Matrix Elements
  for Neutrinoless Double-Beta Decay: A Review}},
  \href{http://dx.doi.org/10.1088/1361-6633/aa5bc5}{\emph{Rept. Prog. Phys.}
  {\bf 80} (2017) 046301}, [\href{http://arxiv.org/abs/1610.06548}{{\tt
  1610.06548}}].

\bibitem{Wiringa:1994wb}
R.~B. Wiringa, V.~G.~J. Stoks and R.~Schiavilla, \emph{{An Accurate
  nucleon-nucleon potential with charge independence breaking}},
  \href{http://dx.doi.org/10.1103/PhysRevC.51.38}{\emph{Phys. Rev.} {\bf C51}
  (1995) 38--51}, [\href{http://arxiv.org/abs/nucl-th/9408016}{{\tt
  nucl-th/9408016}}].

\bibitem{Machleidt:2000ge}
R.~Machleidt, \emph{{The High precision, charge dependent Bonn nucleon-nucleon
  potential (CD-Bonn)}},
  \href{http://dx.doi.org/10.1103/PhysRevC.63.024001}{\emph{Phys. Rev.} {\bf
  C63} (2001) 024001}, [\href{http://arxiv.org/abs/nucl-th/0006014}{{\tt
  nucl-th/0006014}}].

\bibitem{Machleidt:2011zz}
R.~Machleidt and D.~R. Entem, \emph{{Chiral effective field theory and nuclear
  forces}}, \href{http://dx.doi.org/10.1016/j.physrep.2011.02.001}{\emph{Phys.
  Rept.} {\bf 503} (2011) 1--75}, [\href{http://arxiv.org/abs/1105.2919}{{\tt
  1105.2919}}].

\bibitem{Hergert:2015awm}
H.~Hergert, S.~K. Bogner, T.~D. Morris, A.~Schwenk and K.~Tsukiyama, \emph{{The
  In-Medium Similarity Renormalization Group: A Novel Ab Initio Method for
  Nuclei}}, \href{http://dx.doi.org/10.1016/j.physrep.2015.12.007}{\emph{Phys.
  Rept.} {\bf 621} (2016) 165--222},
  [\href{http://arxiv.org/abs/1512.06956}{{\tt 1512.06956}}].

\bibitem{Cirigliano:2019vdj}
V.~Cirigliano, W.~Dekens, J.~De~Vries, M.~L. Graesser, E.~Mereghetti,
  S.~Pastore et~al., \emph{{A renormalized approach to neutrinoless double-beta
  decay}},  \href{http://arxiv.org/abs/1907.11254}{{\tt 1907.11254}}.

\bibitem{Detmold:2018zan}
W.~Detmold and D.~Murphy, \emph{{Nuclear Matrix Elements for Neutrinoless
  Double Beta Decay from Lattice QCD}},
  \href{http://dx.doi.org/10.22323/1.334.0262}{\emph{PoS} {\bf LATTICE2018}
  (2019) 262}, [\href{http://arxiv.org/abs/1811.05554}{{\tt 1811.05554}}].

\bibitem{DellOro:2016tmg}
S.~Dell'Oro, S.~Marcocci, M.~Viel and F.~Vissani, \emph{{Neutrinoless double
  beta decay: 2015 review}},
  \href{http://dx.doi.org/10.1155/2016/2162659}{\emph{Adv. High Energy Phys.}
  {\bf 2016} (2016) 2162659}, [\href{http://arxiv.org/abs/1601.07512}{{\tt
  1601.07512}}].

\bibitem{KamLAND-Zen:2016pfg}
{\scshape KamLAND-Zen} collaboration, A.~Gando et~al., \emph{{Search for
  Majorana Neutrinos near the Inverted Mass Hierarchy Region with
  KamLAND-Zen}}, \href{http://dx.doi.org/10.1103/PhysRevLett.117.109903,
  10.1103/PhysRevLett.117.082503}{\emph{Phys. Rev. Lett.} {\bf 117} (2016)
  082503}, [\href{http://arxiv.org/abs/1605.02889}{{\tt 1605.02889}}].

\bibitem{Agostini:2018tnm}
{\scshape GERDA} collaboration, M.~Agostini et~al., \emph{{Improved Limit on
  Neutrinoless Double-$\beta$ Decay of $^{76}$Ge from GERDA Phase II}},
  \href{http://dx.doi.org/10.1103/PhysRevLett.120.132503}{\emph{Phys. Rev.
  Lett.} {\bf 120} (2018) 132503}, [\href{http://arxiv.org/abs/1803.11100}{{\tt
  1803.11100}}].

\bibitem{Chen:2016qcd}
X.~Chen et~al., \emph{{PandaX-III: Searching for neutrinoless double beta decay
  with high pressure$^{136}$Xe gas time projection chambers}},
  \href{http://dx.doi.org/10.1007/s11433-017-9028-0}{\emph{Sci. China Phys.
  Mech. Astron.} {\bf 60} (2017) 061011},
  [\href{http://arxiv.org/abs/1610.08883}{{\tt 1610.08883}}].

\bibitem{Albert:2017hjq}
{\scshape nEXO} collaboration, J.~B. Albert et~al., \emph{{Sensitivity and
  Discovery Potential of nEXO to Neutrinoless Double Beta Decay}},
  \href{http://dx.doi.org/10.1103/PhysRevC.97.065503}{\emph{Phys. Rev.} {\bf
  C97} (2018) 065503}, [\href{http://arxiv.org/abs/1710.05075}{{\tt
  1710.05075}}].

\bibitem{Abgrall:2017syy}
{\scshape LEGEND} collaboration, N.~Abgrall et~al., \emph{{The Large Enriched
  Germanium Experiment for Neutrinoless Double Beta Decay (LEGEND)}},
  \href{http://dx.doi.org/10.1063/1.5007652}{\emph{AIP Conf. Proc.} {\bf 1894}
  (2017) 020027}, [\href{http://arxiv.org/abs/1709.01980}{{\tt 1709.01980}}].

\bibitem{Wang:2015raa}
{\scshape CUPID} collaboration, G.~Wang et~al., \emph{{CUPID: CUORE (Cryogenic
  Underground Observatory for Rare Events) Upgrade with Particle
  IDentification}},  \href{http://arxiv.org/abs/1504.03599}{{\tt 1504.03599}}.

\bibitem{Daum:1987bg}
M.~Daum, B.~Jost, R.~Marshall, R.~Minehart, W.~Stephens et~al., \emph{{Search
  for Admixtures of Massive Neutrinos in the Decay $\pi^+ \to \mu^+$
  Neutrino}},
  \href{http://dx.doi.org/10.1103/PhysRevD.36.2624}{\emph{Phys.Rev.} {\bf D36}
  (1987) 2624}.

\bibitem{Aguilar-Arevalo:2019owf}
{\scshape PIENU} collaboration, A.~Aguilar-Arevalo et~al., \emph{{Search for
  heavy neutrinos in $\pi \to \mu\nu$ decay}},
  \href{http://dx.doi.org/10.1016/j.physletb.2019.134980}{\emph{Phys. Lett.}
  {\bf B798} (2019) 134980}, [\href{http://arxiv.org/abs/1904.03269}{{\tt
  1904.03269}}].

\bibitem{Hayano:1982wu}
R.~Hayano, T.~Taniguchi, T.~Yamanaka, T.~Tanimori, R.~Enomoto et~al.,
  \emph{{HEAVY NEUTRINO SEARCH USING K(mu2) DECAY}},
  \href{http://dx.doi.org/10.1103/PhysRevLett.49.1305}{\emph{Phys.Rev.Lett.}
  {\bf 49} (1982) 1305}.

\bibitem{Yamazaki:1984sj}
T.~Yamazaki, T.~Ishikawa, Y.~Akiba, M.~Iwasaki, K.~Tanaka et~al., \emph{{Search
  for Heavy Neutrinos in Kaon Decay}}, {\emph{Conf.Proc.} {\bf C840719} (1984)
  I.262}.

\bibitem{Shrock:1981wq}
R.~E. Shrock, \emph{{General Theory of Weak Processes Involving Neutrinos. 2.
  Pure Leptonic Decays}},
  \href{http://dx.doi.org/10.1103/PhysRevD.24.1275}{\emph{Phys. Rev.} {\bf D24}
  (1981) 1275}.

\bibitem{Coloma:2019htx}
P.~Coloma, P.~Hernández, V.~Muñoz and I.~Shoemaker, \emph{{New constraints on
  Heavy Neutral Leptons from Super-Kamiokande data}},
  \href{http://arxiv.org/abs/1911.09129}{{\tt 1911.09129}}.

\bibitem{Artamonov:2014urb}
{\scshape E949} collaboration, A.~Artamonov et~al., \emph{{Search for heavy
  neutrinos in $K^+\to\mu^+\nu_H$ decays}},
  \href{http://dx.doi.org/10.1103/PhysRevD.91.059903,
  10.1103/PhysRevD.91.052001}{\emph{Phys.Rev.} {\bf D91} (2015) 052001},
  [\href{http://arxiv.org/abs/1411.3963}{{\tt 1411.3963}}].

\bibitem{Lazzeroni:2019}
C.~Lazzeroni and E.~Goudzovski, \emph{{'KAON 2019 experimental summary' and
  'Exotic searches at the NA62 experiment at CERN', Talks at the International
  Conference on Kaon Physics 2019}}, .

\bibitem{Abratenko:2019kez}
{\scshape MicroBooNE} collaboration, P.~Abratenko et~al., \emph{{Search for
  heavy neutral leptons decaying into muon-pion pairs in the MicroBooNE
  detector}},  \href{http://arxiv.org/abs/1911.10545}{{\tt 1911.10545}}.

\bibitem{Vaitaitis:1999wq}
{\scshape NuTeV Collaboration, E815 Collaboration} collaboration, A.~Vaitaitis
  et~al., \emph{{Search for neutral heavy leptons in a high-energy neutrino
  beam}},
  \href{http://dx.doi.org/10.1103/PhysRevLett.83.4943}{\emph{Phys.Rev.Lett.}
  {\bf 83} (1999) 4943--4946}, [\href{http://arxiv.org/abs/hep-ex/9908011}{{\tt
  hep-ex/9908011}}].

\bibitem{CooperSarkar:1985nh}
{\scshape WA66 Collaboration} collaboration, A.~M. Cooper-Sarkar et~al.,
  \emph{{Search for Heavy Neutrino Decays in the {BEBC} Beam Dump Experiment}},
  \href{http://dx.doi.org/10.1016/0370-2693(85)91493-5}{\emph{Phys.Lett.} {\bf
  B160} (1985) 207}.

\bibitem{Gallas:1994xp}
{\scshape FMMF Collaboration} collaboration, E.~Gallas et~al., \emph{{Search
  for neutral weakly interacting massive particles in the Fermilab Tevatron
  wide band neutrino beam}},
  \href{http://dx.doi.org/10.1103/PhysRevD.52.6}{\emph{Phys.Rev.} {\bf D52}
  (1995) 6--14}.

\bibitem{Aaij:2014aba}
{\scshape LHCb collaboration} collaboration, R.~Aaij et~al., \emph{{Search for
  Majorana neutrinos in $B^- \to \pi^+\mu^-\mu^-$ decays}},
  \href{http://dx.doi.org/10.1103/PhysRevLett.112.131802}{\emph{Phys.Rev.Lett.}
  {\bf 112} (2014) 131802}, [\href{http://arxiv.org/abs/1401.5361}{{\tt
  1401.5361}}].

\bibitem{TheIceCube:2016oqi}
{\scshape IceCube} collaboration, M.~G. Aartsen et~al., \emph{{Searches for
  Sterile Neutrinos with the IceCube Detector}},
  \href{http://dx.doi.org/10.1103/PhysRevLett.117.071801}{\emph{Phys. Rev.
  Lett.} {\bf 117} (2016) 071801}, [\href{http://arxiv.org/abs/1605.01990}{{\tt
  1605.01990}}].

\bibitem{Aartsen:2017bap}
{\scshape IceCube} collaboration, M.~G. Aartsen et~al., \emph{{Search for
  sterile neutrino mixing using three years of IceCube DeepCore data}},
  \href{http://dx.doi.org/10.1103/PhysRevD.95.112002}{\emph{Phys. Rev.} {\bf
  D95} (2017) 112002}, [\href{http://arxiv.org/abs/1702.05160}{{\tt
  1702.05160}}].

\bibitem{Adamson:2017uda}
{\scshape MINOS+} collaboration, P.~Adamson et~al., \emph{{Search for sterile
  neutrinos in MINOS and MINOS+ using a two-detector fit}},
  \href{http://dx.doi.org/10.1103/PhysRevLett.122.091803}{\emph{Phys. Rev.
  Lett.} {\bf 122} (2019) 091803}, [\href{http://arxiv.org/abs/1710.06488}{{\tt
  1710.06488}}].

\bibitem{Abe:2014gda}
{\scshape Super-Kamiokande} collaboration, K.~Abe et~al., \emph{{Limits on
  sterile neutrino mixing using atmospheric neutrinos in Super-Kamiokande}},
  \href{http://dx.doi.org/10.1103/PhysRevD.91.052019}{\emph{Phys. Rev.} {\bf
  D91} (2015) 052019}, [\href{http://arxiv.org/abs/1410.2008}{{\tt
  1410.2008}}].

\bibitem{Adamson:2017zcg}
{\scshape NOvA} collaboration, P.~Adamson et~al., \emph{{Search for
  active-sterile neutrino mixing using neutral-current interactions in NOvA}},
  \href{http://dx.doi.org/10.1103/PhysRevD.96.072006}{\emph{Phys. Rev.} {\bf
  D96} (2017) 072006}, [\href{http://arxiv.org/abs/1706.04592}{{\tt
  1706.04592}}].

\bibitem{Dydak:1983zq}
F.~Dydak et~al., \emph{{A Search for Muon-neutrino Oscillations in the Delta
  m**2 Range 0.3-eV**2 to 90-eV**2}},
  \href{http://dx.doi.org/10.1016/0370-2693(84)90688-9}{\emph{Phys. Lett.} {\bf
  134B} (1984) 281}.

\bibitem{Stockdale:1984cg}
I.~E. Stockdale et~al., \emph{{Limits on Muon Neutrino Oscillations in the Mass
  Range 55-eV**2 < Delta m**2 < 800-eV**2}},
  \href{http://dx.doi.org/10.1103/PhysRevLett.52.1384}{\emph{Phys. Rev. Lett.}
  {\bf 52} (1984) 1384}.

\bibitem{Astier:2001ck}
{\scshape NOMAD Collaboration} collaboration, P.~Astier et~al., \emph{{Search
  for heavy neutrinos mixing with tau neutrinos}},
  \href{http://dx.doi.org/10.1016/S0370-2693(01)00362-8}{\emph{Phys.Lett.} {\bf
  B506} (2001) 27--38}, [\href{http://arxiv.org/abs/hep-ex/0101041}{{\tt
  hep-ex/0101041}}].

\bibitem{Orloff:2002de}
J.~Orloff, A.~N. Rozanov and C.~Santoni, \emph{{Limits on the mixing of tau
  neutrino to heavy neutrinos}},
  \href{http://dx.doi.org/10.1016/S0370-2693(02)02769-7}{\emph{Phys.Lett.} {\bf
  B550} (2002) 8--15}, [\href{http://arxiv.org/abs/hep-ph/0208075}{{\tt
  hep-ph/0208075}}].

\bibitem{Cvetic:2017gkt}
G.~Cvetič, F.~Halzen, C.~S. Kim and S.~Oh, \emph{{Anomalies in (semi)-leptonic
  $B$ decays $B^{\pm} \to \tau^{\pm} \nu$, $B^{\pm} \to D \tau^{\pm} \nu$ and
  $B^{\pm} \to D^* \tau^{\pm} \nu$, and possible resolution with sterile
  neutrino}},
  \href{http://dx.doi.org/10.1088/1674-1137/41/11/113102}{\emph{Chin. Phys.}
  {\bf C41} (2017) 113102}, [\href{http://arxiv.org/abs/1702.04335}{{\tt
  1702.04335}}].

\end{thebibliography}\endgroup
\end{document}